\documentclass[12pt]{article}
\usepackage{fancyhdr}

\usepackage{mathrsfs}
\usepackage[T1]{fontenc}
\usepackage{mathpazo}
\usepackage{setspace}
\usepackage{amsfonts}
\usepackage{amssymb}
\usepackage{amsmath}
\usepackage{epsfig}
\usepackage{latexsym}
\usepackage{color}
\usepackage{graphicx}
\usepackage{nicefrac}
\usepackage[latin1]{inputenc}
\usepackage{slashed}
\usepackage{multirow}
\usepackage{comment}
\usepackage{cancel}
\usepackage{soul}
\usepackage{hyperref}
\usepackage[nosort]{cite}
\usepackage{datetime}

\usepackage{braket}

\usepackage{tikz}
\usepackage{capt-of}
\usetikzlibrary{decorations.markings}

\usepackage[titletoc]{appendix}
\usepackage[normalem]{ulem}

\def\hybrid{\topmargin -20pt    \oddsidemargin 0pt
        \headheight 0pt \headsep 0pt
        \textwidth 6.25in       
        \textheight 9.25in       
        \marginparwidth .875in
        \parskip 5pt plus 1pt   \jot = 1.5ex}

\hybrid

\def\baselinestretch{1.2}

\catcode`\@=11

\def\marginnote#1{}
\newcount\hour
\newcount\minute
\newtoks\amorpm
\hour=\time\divide\hour by60
\minute=\time{\multiply\hour by60 \global\advance\minute by-\hour}
\edef\standardtime{{\ifnum\hour<12 \global\amorpm={am}%
        \else\global\amorpm={pm}\advance\hour by-12 \fi
        \ifnum\hour=0 \hour=12 \fi
        \number\hour:\ifnum\minute<10 0\fi\number\minute\the\amorpm}}
\edef\militarytime{\number\hour:\ifnum\minute<10 0\fi\number\minute}

\def\draftlabel#1{{\@bsphack\if@filesw {\let\thepage\relax
   \xdef\@gtempa{\write\@auxout{\string
      \newlabel{#1}{{\@currentlabel}{\thepage}}}}}\@gtempa
   \if@nobreak \ifvmode\nobreak\fi\fi\fi\@esphack}
        \gdef\@eqnlabel{#1}}
\def\@eqnlabel{}
\def\@vacuum{}
\def\draftmarginnote#1{\marginpar{\raggedright\scriptsize\tt#1}}

\def\draft{\oddsidemargin -.5truein
        \def\@oddfoot{\sl preliminary draft \hfil
        \rm\thepage\hfil\sl\today\quad\militarytime}
        \let\@evenfoot\@oddfoot \overfullrule 3pt
        \let\label=\draftlabel
        \let\marginnote=\draftmarginnote
   \def\@eqnnum{(\theequation)\rlap{\kern\marginparsep\tt\@eqnlabel}%
\global\let\@eqnlabel\@vacuum}  }

\def\preprint{\twocolumn\sloppy\flushbottom\parindent 2em
        \leftmargini 2em\leftmarginv .5em\leftmarginvi .5em
        \oddsidemargin -.5in    \evensidemargin -.5in
        \columnsep .4in \footheight 0pt
        \textwidth 10.in        \topmargin  -.4in
        \headheight 12pt \topskip .4in
        \textheight 6.9in \footskip 0pt
        \def\@oddhead{\thepage\hfil\addtocounter{page}{1}\thepage}
        \let\@evenhead\@oddhead \def\@oddfoot{} \def\@evenfoot{} }

\def\numberbysection{\@addtoreset{equation}{section}
        \def\theequation{\thesection.\arabic{equation}}}

\def\underline#1{\relax\ifmmode\@@underline#1\else
        $\@@underline{\hbox{#1}}$\relax\fi}

\def\titlepage{\@restonecolfalse\if@twocolumn\@restonecoltrue\onecolumn
     \else \newpage \fi \thispagestyle{empty}\c@page\z@
        \def\thefootnote{\fnsymbol{footnote}} }

\def\endtitlepage{\if@restonecol\twocolumn \else \newpage \fi
        \def\thefootnote{\arabic{footnote}}
        \setcounter{footnote}{0}}  

\catcode`@=12
\relax

\def\figcap{\section*{Figure Captions\markboth
        {FIGURECAPTIONS}{FIGURECAPTIONS}}\list
        {Figure \arabic{enumi}:\hfill}{\settowidth\labelwidth{Figure
999:}
        \leftmargin\labelwidth
        \advance\leftmargin\labelsep\usecounter{enumi}}}
 \relax
\def\tablecap{\section*{Table Captions\markboth
        {TABLECAPTIONS}{TABLECAPTIONS}}\list
        {Table \arabic{enumi}:\hfill}{\settowidth\labelwidth{Table
999:}
        \leftmargin\labelwidth
        \advance\leftmargin\labelsep\usecounter{enumi}}}
 \relax
\def\reflist{\section*{References\markboth
        {REFLIST}{REFLIST}}\list
        {[\arabic{enumi}]\hfill}{\settowidth\labelwidth{[999]}
        \leftmargin\labelwidth
        \advance\leftmargin\labelsep\usecounter{enumi}}}
 \relax
\makeatletter
\newcounter{pubctr}
\def\publist{\@ifnextchar[{\@publist}{\@@publist}}
\def\@publist[#1]{\list
        {[\arabic{pubctr}]\hfill}{\settowidth\labelwidth{[999]}
        \leftmargin\labelwidth
        \advance\leftmargin\labelsep
        \@nmbrlisttrue\def\@listctr{pubctr}
        \setcounter{pubctr}{#1}\addtocounter{pubctr}{-1}}}
\def\@@publist{\list
        {[\arabic{pubctr}]\hfill}{\settowidth\labelwidth{[999]}
        \leftmargin\labelwidth
        \advance\leftmargin\labelsep
        \@nmbrlisttrue\def\@listctr{pubctr}}}
 \relax
\makeatother
\newskip\humongous \humongous=0pt plus 1000pt minus 1000pt

\newif\ifdtup

\relax

\def\be{\begin{equation}}
\def\ee{\end{equation}}
\def\ba{\begin{eqnarray}}
\def\ea{\end{eqnarray}}

\def\del{\partial}

\def\a{\alpha}

\def\b{\beta}

\def\g{\gamma}
\def\G{\Gamma}
\def\d{\delta}
\def\D{\Delta}
\def\e{\epsilon}

\def\m{\mu}
\def\n{\nu}

\def\Om{\Omega}
\def\l{\lambda}

\def\s{\sigma}

\def\no{\noindent}

\def\qq{\qquad}

\def\IR{\relax{\rm I\kern-.18em R}}

\def \z { {\bar z} }

\def \w { {\bar w} }

\def \ha {{1\over 2}}

\def \ov {\over}

\def\IR{\relax{\rm I\kern-.18em R}}
\def\IL{\relax{\rm I\kern-.18em L}}

\def\inv{^{\raise.15ex\hbox{${\scriptscriptstyle -}$}\kern-.05em 1}}

\def\Tr{{\rm Tr}}

\begin{document}

\renewcommand{\theequation}{\thesection.\arabic{equation}}
\csname @addtoreset\endcsname{equation}{section}

\newcommand{\beq}{\begin{equation}}
\newcommand{\eeq}[1]{\label{#1}\end{equation}}
\newcommand{\ber}{\begin{equation}}
\newcommand{\eer}[1]{\label{#1}\end{equation}}
\newcommand{\eqn}[1]{(\ref{#1})}
\begin{titlepage}
\begin{center}

\renewcommand*{\thefootnote}{\arabic{footnote}}


${}$
\vskip .2 in

\vskip .4cm

{\large\bf
$\lambda$-deformations in the upper-half plane}

\vskip 0.4in

{\bf Konstantinos Sfetsos}\footnote{E-mail:~ksfetsos@phys.uoa.gr} \hskip .2cm
and \hskip .2cm  {\bf Konstantinos Siampos}\footnote{E-mail:~konstantinos.siampos@phys.uoa.gr}
\vskip 0.16in

 {\em
Department of Nuclear and Particle Physics,\\
Faculty of Physics, National and Kapodistrian University of Athens,\\
Athens 15784, Greece\\
}

\vskip 0.12in


\vskip .5in
\end{center}

\centerline{\bf Abstract }

\no
We formulate $\lambda$-deformed $\sigma$-models as QFTs in the upper-half plane. For different boundary conditions we compute correlation functions of currents and primary operators, exactly in the deformation parameter $\lambda$ and for large values of the level  $k$ of the underlying WZW model.
To perform our computations we use either conformal perturbation theory in association with Cardy's doubling trick,  as well as meromorphicity arguments and a non-perturbative symmetry in the parameter space $(\lambda,k)$, or standard QFT techniques based on
the free field expansion of the $\sigma$-model action, with the free fields obeying appropriate boundary conditions.
Both methods have their own advantages yielding consistent and rich, compared to those in the absence of a boundary,
complementary results. We pay particular attention, albeit not exclusively, to integrability preserving boundary conditions.

\vskip .4in
\noindent
\end{titlepage}
\vfill
\eject

\newpage

\tableofcontents

\noindent

\def\baselinestretch{1.2}
\baselineskip 20 pt
\noindent


\setcounter{equation}{0}

\section{Introduction }

Theories with boundaries are of tremendous interest
and have attracted  over the years a lot of attention from physical and mathematical view points.
Boundary conditions have played a fundamental r\^ole in the very development of string theory, since treating them carefully led to the
discovery of D-branes~\cite{Dai:1989ua,Leigh:1989jq,Horava:1989ga,Green:1991et,Bachas:1995kx}, which in turn boosted our understanding of non-perturbative aspects of
string theory~\cite{Hull:1994ys,Witten:1995ex} and also was instrumental in precisely formulating
the AdS/CFT correspondence~\cite{Maldacena:1997re,Gubser:1998bc,Witten:1998qj,Witten:1998zw}.

\no
Quantum field theories (QFTs), in particular conformal field theories (CFTs), in spaces with boundaries have several additional applications, besides string theory, in condensed matter and statistical mechanics systems, determining at critical points the surface correlation functions and the corresponding critical exponents~\cite{Cardy:1984bb,McAvity:1995zd}.
The aim of this paper is to investigate the response of such boundary CFTs, to finite deformations.
In particular, we will investigate this issue, mostly focusing on quantum and integrability aspects,
in the context of the $\l$-deformed models of~\cite{Sfetsos:2013wia}.

\no
In a CFT with plane boundary, there is still a remnant of the conformal group which consists of those
transformations leaving the boundary surface intact~\cite{Cardy:1984bb}. For example, let us consider a scalar operator
$\Phi(x,y)$ of scaling dimension $\D$ and place the boundary at $y=0$. 
The one-point correlation function of such an operator is 
constrained by demanding translation invariance and also covariance under special conformal transformation parallel to the surface.
It turns out that it is generically non-vanishing, in contrast to the full plane result, and it is given up to a proportionality constant by~\cite{McAvity:1995zd}
\be
\label{Osborn.1pt}
\langle \Phi(x,y)\rangle_\text{uhp}\propto\frac{1}{y^{\D}}\,,
\ee
where uhp stands for the upper-half plane.
Similarly, the two-point correlation function of two such operators is given by~\cite{Cardy:1984bb,McAvity:1995zd}
\be
\label{Cardy.Osborn}
\langle\Phi_1(x_1,y_1)\Phi_2(x_2,y_2)\rangle_\text{uhp}=
\frac{\psi\left(\zeta\right)}{y_1^{\D_1}y_2^{\D_2}} ,\qquad \zeta=\frac{(x_1-x_2)^2+y_1^2+y_2^2}{y_1y_2}\,,
\ee
where $\psi(\zeta)$ is an undetermined function,
which depends on the specific theory for given boundary conditions.
The above result applies in any dimension, 
with the appropriate realization of $(x_1-x_2)^2$ as the norm of the corresponding vector, 
as only translational and special conformal transformations were involved.
 In the two-dimensional case \eqref{Osborn.1pt} and \eqref{Cardy.Osborn} apply for operators with
equal weights for the holomorphic and anti-holomorphic sectors $(h,h)$, with $\D=2 h$.
Despite the fact that the operators may have unequal scaling dimension, their two-point function is generically non-vanishing, in contrast to the case without a boundary.
The fact that even a two-point correlation function can be determined up to an unknown function and not up to a normalization constant as it is the case with no boundaries, suggests that these theories and their deformations thereof, have a much richer structure.

\no
The $\s$-models we will be focusing on, are known generically as $\l$-deformations~\cite{Sfetsos:2013wia,Hollowood:2014rla,Hollowood:2014qma,Georgiou:2016urf,Georgiou:2017jfi,Sfetsos:2017sep,Sfetsos:2015nya,Georgiou:2018hpd,Georgiou:2018gpe,Driezen:2019ykp}.
They describe deformations of CFTs corresponding to WZW~\cite{Witten:1983ar} or gauged WZW
theories~\cite{Karabali:1988au,Karabali:1989dk,DiVecchia:1984df,DiVecchia:1984ksr} by current or parafermionic bilinears, respectively.
More concretely, in the present work we will use the prototype $\l$-deformed action~\cite{Sfetsos:2013wia}\footnote{
\label{licon}
The worldsheet light-cone coordinates $\s^\pm$ and $(\tau,\sigma)$ are given by
\begin{equation*}
\s^\pm=\tau\pm\s\,,\quad\del_\tau=\del_++\del_-\,,\quad\del_\s=\del_+-\del_-\,.
\end{equation*}}
\be
\label{lambda.action}
S_{k,\l}(g)=S_\text{WZW,k}(g)+\frac{k}{\pi}\int\text{d}^2\s\, R^a_+\Big(\l^{-1}\mathbb{1}-D^T\Big)_{ab}^{-1}L^b_-\,,
\ee
where $S_\text{WZW,k}(g)$ is the WZW action for a group element $g$ of a compact semisimple group $G$ and
the right-/left-invariant one-forms as well  as the adjoint action are given by
\be
\label{panatha}
R^a_+=-i\,\text{Tr}\big(t_a\del_+gg^{-1}\big)\,,\quad L^a_-=-i\,\text{Tr}\big(t_ag^{-1}\del_-g\big)\,,\quad D_{ab}=\Tr\big(t_agt_bg^{-1}\big)\ ,
\ee
where $t_a$'s form a basis of the Lie algebra $\frak{g}$ of $G$, normalized as $\text{Tr}(t_at_b)=\d_{ab}$ and 
obeying the Lie algebra $[t_a,t_b]=if_{abc}\,t_c$. The structure constants $f_{abc}$ are real, the Killing metric is 
$\d_{ab}$, with $a=1,\dots,\text{d}_G$, $\text{d}_G$ being the dimension of the group $G$.
The constant $k$ is a positive integer which is taken to be large enough so that the curvature
of the metric corresponding to the $\s$-model  \eqn{lambda.action} is small. The physical range of the parameter $\l$ is the interval $[0,1)$ and in that range the model is non-singular. The action \eqref{lambda.action} is the effective description of the non-Abelian Thirring model
\be
\label{Thirring.action}
S(g)=S_\text{WZW,k}(g)+\l\frac{k}{\pi}\int\text{d}^2\s\,R^a_+L^a_-\, ,
\ee
to which it reduces for small values of $\l$. In addition, \eqn{lambda.action} enjoys a non-perturbative symmetry
involving flipping the sign of $k$, inverting $\l$ and the group element $g$~\cite{Itsios:2014lca}.
This symmetry is not manifest in \eqn{Thirring.action} and for that action arises only after employing
path integral arguments~\cite{Kutasov:1989aw}.
Moreover, \eqn{lambda.action} is integrable~\cite{Sfetsos:2013wia} which is not a property of \eqn{Thirring.action} as well.
Furthermore, \eqn{lambda.action} has well behaved zoom-in limits around $\l=1$ and $\l=-1$, that is the non-Abelian T-dual of the principal chiral model and the pseudo chiral model, respectively~\cite{Sfetsos:2013wia,Georgiou:2016iom}.
Apart from the aforementioned symmetry the above actions share the same $\b$-function~\cite{Kutasov:1989dt,Itsios:2014lca,Sfetsos:2014jfa,Gerganov:2000mt} and symmetry algebra~\cite{Sfetsos:2013wia,Balog:1993es,Georgiou:2015nka,Georgiou:2016iom}.
In fact, \eqref{lambda.action} is the all-loop in $\l$ effective action of \eqref{Thirring.action}, for large values of $k$.
Its simplicity and the fact that it is an all order action in the deformation parameter $\l$,
makes it a rear example in the literature.

\no
A lot a progress has been made in recent years in understanding several quantum aspects of the above actions from complementary view points and methods. These range from low order conformal perturbation theory, symmetry arguments based on the aforementioned non-perturbative symmetry and demanding meromorphicity near $\l=\pm1$ (correlated with the limit $k\to\infty$) as well as gravitational methods.
These efforts allowed the exact determination of two- and three-point functions of current, composite current-bilinear and primary fields and of their anomalous dimensions ~\cite{Georgiou:2015nka,Georgiou:2016iom,Georgiou:2019jcf,Sagkrioti:2020mkw}.
A rather distinct approach in performing computations in this context was put forward in~\cite{Georgiou:2019aon}. This approach utilizes the expansion of \eqref{lambda.action} around the free field point, thus manifestly taking into account the exact dependence of the interaction vertices  on the deformation parameter $\l$. The above approach was further extended for $\l$-deformations based on coset CFTs  in~\cite{Georgiou:2020bpx}.

Given the progress outlined above it is important to extend the analysis of $\l$-deformed $\s$-models as QFTs in spaces with boundaries,
the simplest case being in the upper-half plane. As argued above based on the structure of \eqn{Cardy.Osborn},
we expect a much richer structure arising in the study of quantum aspects of \eqn{lambda.action} in spaces with boundaries than in the
absence of them. The immediate extra ingredient is dealing with boundary conditions, in particular distinguishing them according to whether or not they preserve integrability.
We note that integrability preserving boundaries conditions were exposed for the $\l$-deformed action \eqn{lambda.action}
in~\cite{Driezen:2018glg}, nicely extending previous works for consistent boundary conditions for WZW model, in particular that of~\cite{Alekseev:1998mc,Felder:1999ka,Stanciu:1999id,Figueroa-OFarrill:1999cmq}.

The plan of the paper is as follows:
In Section~\ref{CFTuhp},  we review and apply the results of~\cite{Cardy:1984bb} to evaluate single and composite
current correlation functions on the upper-half plane at the conformal point, paving the way for the more complicated equations in the subsequent sections.
In addition, we also evaluate the one-point correlation function of primary fields.
In Section~\ref{CFTpertuhp}, we move away from the conformal point and we compute current and primary field
correlation functions  exactly in $\l$ and to leading order in $\nicefrac1k$,  using conformal perturbation.
In Section~\ref{freefields.uhp}, we perform similar computations using the free field expansion as a basis and standard QFT techniques.
The end results are in agreement and complementary to those in Section~\ref{CFTpertuhp}.
In Section~\ref{concl}, we present our conclusions and future directions of this work.
Appendix~\ref{integrals.quiver}, stands as a quiver for the various integrals on the upper-half which are needed in 
Sections~\ref{CFTpertuhp},~\ref{freefields.uhp} of the present work.
Finally, Appendices~\ref{2loop.append} and~\ref{1loop.append}, provide computational details for Section~\ref{CFTpertuhp}.

\section{CFTs in the upper-half plane}
\label{CFTuhp}

In this section we discuss CFT correlation functions in the upper-half plane using conventional QFT methods as well as
Cardy's doubling trick~\cite{Cardy:1984bb}, for a review see also section 11.2.1. in~\cite{CFT.book},
which greatly facilitates the relevant computations. To enhance the pedagogical component of the paper
we present elementary examples with free fields and non-Abelian currents.

\subsection{Cardy's doubling trick}

Consider an operator $\Phi^{(h,\bar h)}(z,\z)$ with conformal weights $h$ and $\bar h$ for the 
holomorphic and anti-holomorphic sectors, respectively. We are interested in  correlation functions involving this operator in the upper-half plane (uhp)
of the form
\be
\label{dCtrick0}
\langle \Phi^{(h,\bar h)}(z,\z) \cdots\rangle_{\rm uhp}\ ,
\ee
where the ellipsis denote all other operators in the correlator.
Assuming that there is no flow of energy across the boundary that is $\bar T(\z)=T(z)$, when $\z= z$,
Cardy's doubling trick~\cite{Cardy:1984bb}
amounts in realizing the above correlator by one defined in the \emph{entire plane}
\be
\label{dCtrick1}
\langle \Phi^{(h,\bar h)}(z,\z)  \bar \Phi^{(\bar h,h)}(\z,z) \cdots\rangle_{\rm hol}\  ,
\ee
where we have inserted the conjugate operator  $\bar\Phi^{(\bar h,h)}(\z,z)$ into the
correlator at the mirror image point with respect to the real axis, i.e. at $(\z,z)$ and with the conformal weights interchanged.
The $\z$-dependence ($z$-dependence) in the inserted conjugate operator should be understood as a holomorphic (an antiholomorphic)
coordinate which takes values in the lower-half (upper-half) plane.
In addition, we restrict the evaluation of
the correlator in the holomorphic sector, which in \eqref{dCtrick1} is indicated by the subscript.
Of course, all other operators in the correlator are treated in a similar manner. We emphasize that the replacement of 
\eqn{dCtrick0} by \eqn{dCtrick1} is at the level of Ward identities, see Eq.(4.10) in~\cite{Cardy:1984bb},
leaving room for integration constants to be specified for the appropriate boundary conditions to be  obeyed. 

\no
As an illustration we evaluate the one-point function of the operator $\Phi^{(h,\bar h)}(z,\z)$
\begin{equation}
\label{1pt.double}
\langle \Phi^{(h,\bar h)}(z,\z)\rangle_\text{uhp}=
\langle \Phi^{(h,\bar h)}(z,\z) \Phi^{(\bar h,h)}(\z,z)\rangle_\text{hol}=\frac{A_\Phi\,\d_{h,\bar h}}{(z-\z)^{2h}}\,,
\end{equation}
where $A_\Phi$ is a constant. This is consistent with the general result \eqref{Osborn.1pt} with $\D=2h$ and $z\to x+i y$.

\no
As an application of particular interest consider the case with $\bar h=0$. Then
$\Phi^{(h,0)}(z,\z)$ is essentially $\z$-independent so that we may denote it by  $\Phi^{h}(z)$ and its conjugate field
by $\bar \Phi^{h}(z)$. Then
\be
\label{dCtrick2}
\langle \Phi^{h}(z) \cdots\rangle_{\rm uhp} = \langle \Phi^{h}(z)  \bar\Phi^{h}( z) \cdots\rangle_{\rm hol}
= \langle \Phi^{h}(z) \cdots\rangle\ ,
\ee
where in the last step we have omitted $\bar\Phi^{h}(z)$ since being a purely anti-holomorphic operator
it does not contribute to the  holomorphic sector of the correlator. For the same reason we have removed the subscript.

\no
An equally important application is when $h=0$. Then
$\Phi^{(0,\bar h)}(z,\z)$ is essentially $z$-independent so that we may denote it by  $\bar \Phi^{\bar h}(\z)$ and its conjugate field  by $ \Phi^{\bar h}(\z)$. Then
\be
\label{dCtrick3}
\langle \bar \Phi^{\bar h}(\z) \cdots\rangle_{\rm uhp} = \langle \bar \Phi^{\bar h}(\z)  \Phi^{\bar h}(\z)  \cdots\rangle_{\rm hol}
= \langle \Phi^{\bar h}(\z) \cdots\rangle\ ,
\ee
where in the last step we have omitted $\bar\Phi^{\bar h}(\z)$ since, as before,
 being a purely anti-holomorphic operator it does not contribute to the
holomorphic correlator. As a result of the above analysis purely holomorphic operators are left in correlation function as they
are not accompanied by their conjugates. Similarly, purely anti-holomorphic operators are just replaced by their
corresponding holomorphic conjugates. Hence, quite generally we have that
\be
\label{dCtrick4}
\langle  \Phi^{h}(z)  \bar \Phi^{\bar h}(\z) \dots \rangle_{\rm uhp}=  \langle \Phi^{h}(z)  \Phi^{\bar h}(\z)  \dots \rangle\ .
\ee
Let us recapitulate, using Cardy's doubling trick one reduces the evaluation of conformal correlation functions in the upper-half plane to correlation functions in the entire plane, as they obey the same Ward identities, which nevertheless should encode the original boundary conditions upon appropriate choice of the integration constants.

\subsection{Correlation functions of Abelian currents}
\label{subsection 2.2}

We first discuss the appropriate boundary conditions for a single free field $\phi$ in the Minkowski as well as in the
Euclidean regimes. In the case where the world-sheet $S$ has a non-trivial boundary $\del S$ on the real axis, possible boundary conditions are
\be
\label{dbcf}
{\rm Dirichlet\ b.c.}:\quad \phi^a|_{ \del S}=0\  \Rightarrow\  \del_\tau\phi^a|_{ \del S} = 0 \ \Leftrightarrow\
(\del_+\phi^a +\del_-\phi^a)|_{ \del S} = 0
\ee
and
\be
\label{nbcb}
{\rm Neumann\ b.c.}:\quad \del_\s\phi^a|_{ \del S} = 0 \ \Leftrightarrow\
(\del_+\phi^a -\del_-\phi^a)|_{ \del S} = 0 \ .
\ee
In order to pass to the Euclidean regime we perform the analytic continuation
\be
\label{analy}
\begin{split}
& \tau\to -i t\ ,\quad z= t+i \s\ ,\quad \z= t-i \s\ ,
\\
&
\s^+ \to - i z\ ,\quad \s^- \to - i \z\ ,\quad   \del_+\to i \del\ ,\quad \del_-\to i \bar \del\ ,
\\
& \del_\tau \to i (\del  +\bar \del)\ ,\quad  \del_\s \to i (\del  -\bar \del)\ .
\end{split}
\ee
The holomorphic and anti-holomorphic derivatives of  $\phi$ are Virasoro primary fields of
corresponding dimension one. For them we will use the notation
\be
\label{noott}
j(z) = \del \phi(z,\z)\ ,\qq \bar j(\z) = \bar\del \phi(z,\z)\ .
\ee
In order to perform field theory computations using Wick's theorem, we need to determine the basic two-point function
\be
G(z,w)= \langle \phi(z,\z) \phi(w,\w)\rangle_{\text{uhp}}\  .
\ee
In the following we restrict our domain $S$ to the upper-half plane.
Hence, the boundary $\del S$ is the real axis where the Dirichlet and Neumann boundary conditions are set.
In the  Euclidean regime this corresponds to set $\z=z$, yielding
\be
\label{dii}
{\rm Dirichlet\ b.c.:}\qq  \phi(z,\z)|_{\z=z}=0\ \Rightarrow\ (\del +\bar \del)  G|_{ \z=z} = 0\
\ee
and that
\be
\label{nee}
{\rm Neumann\ b.c.:}\qq   (\del -\bar \del)  G |_{ \z=z} = 0\ .
\ee
The solution to the two-point function with the above two different boundary conditions
can be easily presented in a unified way which also includes the case in which the theory is
defined in the entire plane. It reads that
\be
\label{fifi}
\langle \phi(z,\z) \phi(w,\w)\rangle_{\rm uhp} = -\ln {|z-w|^2  -\e\ln |z-\bar w|^2}\ 
\ee
where
\be
\e=0 , + 1, -1 \ ,\quad {\rm for\ the\ entire\ plane, \  Neumann\ b.c., Dirichlet\ b.c.}\
\ee
and obeys the usual Green-function equation
\be
\label{olympi}
\del\bar \del G = -\pi \d^{(2)}(z-w)
\ee
\no
and in our conventions $\d^{(2)}(z)=\d(t)\d(\s)$.
In the following we will use $x_i$ and $\bar x_i$ for the Euclidean world-sheet coordinates as this facilitates the more complicated
calculations performed later in the paper when more than two fields are involved.
From the above we find the following propagators for the currents
\be
\label{free.two.point}
\begin{split}
&\langle j(x_1) j(x_2) \rangle_\text{uhp}= -{1\ov x_{12}^2} + \e \pi\, \d^{(2)}(x_1-\bar x_2) \,,
\\
&\langle j(x_1)\bar j(\bar x_2)\rangle_\text{uhp} = -\, \frac{\e}{(x_1-\bar x_2)^2}+\pi\, \d^{(2)}(x_{12})\,,
\end{split}
\ee
where $x_{12}= x_1-x_2$. The terms involving $\d$-functions arise from
the fact that  $\displaystyle \del {1\ov \z} = \bar \del {1\ov z}= \pi \d^{(2)}(z)$.
We may ignore the $\d$-function in the first correlator above since $ x_1$ and $\bar x_2$ are located at different parts of the complex plane.

\no
We would like to realize the above two-point functions using CFT methods. The {\it holomorphic conformal block} of four operators with conformal weights $h_i$, $i=1,2,3,4$ takes the form~\cite{Polyakov:1970xd}
\be
\label{four2}
G^{(4)}=x_{13}^{h_2+h_4}x_{24}^{h_1+h_3}x_{12}^{-h_1-h_2}x_{23}^{-h_2-h_3}x_{34}^{-h_3-h_4}x_{14}^{-h_1-h_4}F(\xi)\,,\qquad
\xi=\frac{x_{12}x_{34}}{x_{13}x_{24}}\,,
\ee
where the function $F(\xi)$ is to be determined.
According to Cardy's doubling trick~\cite{Cardy:1984bb}, this can be used, as we do below, to evaluate two-point functions in which two
of the four operators and their locations are the conjugate ones, that is $x_{3,4}=\bar x_{1,2}$ 
and $h_{3,4}=\bar h_{1,2}$ yielding
\be
\begin{split}
\label{2pt.Cardy.double}
G^{(2)}_{\rm uhp}= & \left(x_1-\bar x_1\right)^{h_2+\bar h_2}\left(x_2-\bar x_2\right)^{h_1+\bar h_1}
\left(x_1-\bar x_2\right)^{-h_1-\bar h_2}(x_2-\bar x_1)^{-h_2-\bar h_1}
\\
& \times x_{12}^{-h_1-h_2}\bar x_{12}^{-\bar h_1-\bar h_2}F(\xi)\,,
\end{split}
\ee
with
\be
\label{comparison.xi}
\xi=-\frac{|x_{12}|^2}{(x_1-\bar x_1)(\bar x_2- x_2)}\leqslant0\,,\qquad
1-\xi=\frac{|x_1-\bar x_2|^2}{(x_1-\bar x_1)(\bar x_2- x_2)}\geqslant0
\ee
and the function $F(\xi)$ incorporates the boundary conditions~\cite{Cardy:1984bb}.
A comment is in order concerning the relationship of the two-point functions \eqref{2pt.Cardy.double}
with \eqref{Cardy.Osborn}. Their explicit connection reads
\be
\label{2pt.gen.rel}
\psi(\zeta)=\frac{4^{h_1+h_2}}{(\zeta^2-4)^{h_1+h_2}}F(\xi)\,,\quad \zeta=2(1-2\xi)\geqslant2\,,\quad \D_{1,2}=2h_{1,2}=2\bar h_{1,2}\,.
\ee

\paragraph{The $\langle jj\rangle_\text{uhp}$ two-point function:}
The two-point function evaluated on the upper-half plane is
given by \eqref{free.two.point}. According to Cardy's doubling trick this is given by the
two-point function \eqref{2pt.Cardy.double} in which the holomorphic conformal weights are
$h_{1,2}=1$  and $\bar h_{1,2}=0$.
In order to reproduce \eqref{free.two.point} we have to choose
\be
F(\xi)=\xi-1\,,
\ee
where the invariant ratio $\xi$ was defined in \eqref{comparison.xi}.

\paragraph{The $\langle j\bar j\rangle_\text{uhp}$ two-point function:} The two-point function is given in \eqref{free.two.point}.
According to Cardy's doubling trick this is given by the
two-point function \eqref{2pt.Cardy.double} in which the holomorphic conformal weights are
$h_1=\bar h_2=1$  and $ h_2=\bar h_1=0$.
The above choice leads to
\be
\langle j(x_1)\bar j(\bar x_2)\rangle_\text{uhp}=\frac{1}{(x_1-\bar x_2)^2}\frac{F(\xi)}{\xi}\,,
\ee
where the parameter $\xi$ is given in \eqref{comparison.xi}.
It can be shown that
\be
\d^{(2)}(x_{12})=-\frac{1}{\pi(x_1-\bar x_2)^2}\d(\xi)
\ee
and therefore \eqref{free.two.point} takes the form
\be
\frac{F(\xi)}{\xi}=-\e-\d(\xi)\,.
\ee
The parameter $\e$ should be thought an integration constant in the Ward identities in the upper-half plane, such that the boundary conditions 
\eqref{dii} and \eqref{nee} are obeyed. Equivalently we should include the factor of $\e$ when applying Cardy's doubling trick 
\be
\label{current.free.Cardy}
\bar j(\bar x)\quad \Longrightarrow \quad \e\, j(\bar x)\,,\quad \e=\pm 1\,,
\ee
in the corresponding correlator evaluated in the entire plane.
As an elementary example let us again consider the two-point function \eqref{free.two.point}
\be
\langle j(x_1)\bar j(\bar x_2)\rangle_\text{uhp} =\e \langle j(x_1) j(\bar x_2)\rangle=
-\, {\e\ov (x_1-\bar x_2)^2}\,,
\ee
where we have omitted the contact term.
Finally, we emphasize that \eqref{current.free.Cardy} becomes handy in computing higher
point correlators.

\paragraph{Composite current-bilinear operator:}

Let us consider the composite current-bilinear ${\cal O}(x,\bar x)=j(x)\bar j(\bar x)$ of conformal weight $(1,1)$. 
Its one-point function on the upper-half plane can be 
obtained through \eqref{free.two.point}
\be
\langle{\cal O}(x,\bar x)\rangle_\text{uhp}=-\frac{\e}{(x-\bar x)^2}\,,
\ee
in agreement with \eqref{1pt.double} with $A_\Phi=-\epsilon$.
Its two-point function
$\langle{\cal O}(x_1,\bar x_1){\cal O}(x_2,\bar x_2)\rangle_\text{uhp}$ on the 
upper-half plane can be evaluated using Wick contraction and \eqref{free.two.point}
\be
\label{2pt.comp.Abel}
\begin{split}
&\langle{\cal O}(x_1,\bar x_1){\cal O}(x_2,\bar x_2)\rangle_\text{uhp}=
\langle j(x_1)\bar j(\bar x_1)j(x_2)\bar j(\bar x_2)\rangle_\text{uhp}\\
&\qq =\frac{1}{|x_{12}|^4}+\frac{\e^2}{|x_1-\bar x_2|^4}+\frac{\e^2}{(x_1-\bar x_1)^2(x_2-\bar x_2)^2}\, ,
\end{split}
\ee
where we have omitted contact terms.
Alternatively, we may obtain the same result through Cardy's doubling trick
\be
\langle j(x_1)\bar j(\bar x_1)j(x_2)\bar j(\bar x_2)\rangle_\text{uhp}=
\langle j(x_1) j(\bar x_1)j(x_2) j(\bar x_2)\rangle\,.
\ee
Indeed, this fits in the general expression of the two-point function as obtained from the holomorphic conformal block
 two-point function \eqref{2pt.Cardy.double} with $h_{1,2}=\bar h_{1,2}=1$ after choosing
\be
\label{F.Abel}
F(\xi)=(1-\xi)^2+\e^2\xi^2(1+(1-\xi)^2)\,,
\ee
where the $\xi$ is given by \eqref{comparison.xi}. The two-point function \eqref{2pt.comp.Abel}, fits in the generic form of \eqref{Cardy.Osborn}
\begin{equation}
\psi(\zeta)=\frac{\e^2}{16}+\frac{1}{(\zeta-2)^2}+\frac{\e^2}{(\zeta+2)^2}\,,\quad \zeta=2(1-2\xi)\geqslant2\,,\quad \D_{1,2}=2\,,
\end{equation}
which is also in agreement with \eqref{2pt.gen.rel} and \eqref{F.Abel}.

\subsection{Non-Abelian currents}

We now turn our attention to non-Abelian currents $J_a(x)$
and $\bar J_a(\bar x)$ satisfying the usual operator product expansion
\be
\label{KM.OPE}
J_a(x_1)J_b(x_2)=\frac{\d_{ab}}{x_{12}^2}+\frac{f_{abc}}{\sqrt{k}\,x_{12}}J_c(x_2)\,,\quad
J_a(x_1)\bar J_b(\bar x_2)=0\,,
\ee
where here the structure constants $f_{abc}$ are taken to be imaginary.  A similar algebra is also obeyed by the $\bar  J_a(\bar x)$'s.
Concerning the allowed boundary conditions at $\bar x= x$, the one below
\be
\label{KM.bnr}
J_a(x)=\bar J_a(\bar x)\,,
\ee
is consistent, at it preserves the form of \eqref{KM.OPE} for the holomorphic and the anti-holomorphic currents, 
up to inner or outer automorphisms of the algebra also preserving the Killing metric $\d_{ab}$.
In fact, this boundary condition describes D-branes whose world-volume is described in terms of the conjugacy classes of the group $G$~\cite{Alekseev:1998mc,Felder:1999ka,Stanciu:1999id,Figueroa-OFarrill:1999cmq}.
Note that this is not the case for the boundary condition $J_a(x)=-\bar J_a(\bar x),$ at $\bar x= x$, as there is an explicit obstruction from the non-Abelian term in the current algebra \eqref{KM.OPE}, contrary to the Abelian case. Hence, the current algebra is not preserved. It turns out that this boundary condition describes D-branes whose world-volume includes coset spaces~\cite{Stanciu:1999id} and we will not be interested in these boundary conditions in the present paper.

\no
Using Cardy's doubling trick we find that the one-point correlation function vanishes
\be
\langle J_a(x)\rangle_\text{uhp}=\langle J_a(x)\rangle=0\,,
\ee
since the only vector structure,
is the trace of the structure constant $f_{abc}$ with the Killing metric $\d_{ab}$ and this vanishes for the semisimple group $G$ under consideration.

\no
Moving on to the two-point function of the Kac--Moody currents the discussion is analogue to the free field case.
Hence, using Cardy's doubling trick we find that
\be
\label{2pt.CFT}
\langle J_a(x_1)  J_b(x_2)\rangle_\text{uhp}=\frac{\d_{ab}}{x_{12}^2}\,,\quad
\langle J_a(x_1)\bar J_b(\bar x_2)\rangle_\text{uhp}=\frac{\d_{ab}}{(x_1-\bar x_2)^2}\,.
\ee

\no
Also, the three-point functions are given by
\be
\label{3pt.CFT}
\begin{split}
&\langle J_a(x_1)  J_b(x_2) J_c(x_3)\rangle_\text{uhp}=\frac{f_{abc}}{\sqrt{k}\,x_{12}x_{23}x_{13}}\,,\\
&\langle J_a(x_1)  J_b(x_2)\bar J_c(\bar x_3)\rangle_\text{uhp}=\frac{f_{abc}}{\sqrt{k}\,x_{12}(x_1-\bar x_3)(x_2-\bar x_3)}\,.
\end{split}
\ee
Note that the above correlators \eqref{2pt.CFT} and \eqref{3pt.CFT} are consistent with the boundary condition \eqref{KM.bnr}.

\no
Let us consider the composite current-bilinear ${\cal O}(x,\bar x)= J_a(x)\bar  J_a(\bar x)$ which has conformal weight $(1,1)$.
Its one-point function on the upper-half plane is read from \eqref{2pt.CFT}
\be
\langle{\cal O}(x,\bar x)\rangle_\text{uhp}=\frac{\text{d}_G}{(x-\bar x)^2}\,,
\ee
in agreement with \eqref{1pt.double} with $A_\Phi=\text{d}_G$.
We may evaluate its two-point function on the upper-half plane using the above prescription
as well as the four-point function of currents on the full plane
\begin{equation}
\begin{split}
& \langle  J_{a_1}( z_1)  J_{a_2}( z_2)  J_{a_3}( z_3)  J_{a_4}( z_4)\rangle=
\frac{1}{k}\left(\frac{f_{a_1a_3e}f_{a_2a_4e}}{ z_{12} z_{13} z_{24} z_{34}}
-\frac{f_{a_1a_4e}f_{a_2a_3e}}{ z_{12} z_{14} z_{23} z_{34}}\right)
\\
&\qq\qq\qq
+\frac{\delta_{a_1a_2}\delta_{a_3a_4}}{ z_{12}^2 z_{34}^2}
+\frac{\delta_{a_1a_3}\delta_{a_2a_4}}{ z_{13}^2 z_{24}^2}
+\frac{\delta_{a_1a_4}\delta_{a_2a_3}}{ z_{14}^2 z_{23}^2}\,.
\end{split}
\end{equation}
We find that
\be
\label{2pointOCFT}
\begin{split}
&\langle{\cal O}(x_1,\bar x_1){\cal O}(x_2,\bar x_2)\rangle_\text{uhp}\\
&\qquad =\langle  J_a(x_1)\bar  J_a(\bar x_1) J_b(x_2)\bar  J_b(\bar x_2)\rangle_\text{uhp}
=\langle  J_a(x_1)  J_b(x_2)  J_a(\bar x_1)  J_b(\bar x_2)\rangle\\
&\qquad =\frac{\text{d}_G}{|x_{12}|^4}+\frac{\text{d}_G}{|x_1-\bar x_2|^4}+\frac{\text{d}_G^2}{(x_1-\bar x_1)^2(x_2-\bar x_2)^2}
+\frac{c_G}{k} \frac{\text{d}_G}{|x_{12}|^2|x_1-\bar x_2|^2}\,,
\end{split}
\ee
where we note the explicit $\nicefrac{c_G}{k}$ dependence,
with $c_G$ being the quadratic Casimir in the adjoint representation, defined by $f_{acd} f_{bcd}= - c_G \d_{ab}$. This is a
positive number for the compact Lie groups under consideration.
Moreover, the result takes the form of \eqref{2pt.Cardy.double} with
\be
\label{F.Non.Abel}
F(\xi)=\text{d}_G\left(1-\left(2+\frac{c_G}{k}\right)\xi(1-\xi)+\text{d}_G\,\xi^2(1-\xi)^2\right)\,
\ee
and the invariant ratio $\xi$ is given by \eqref{comparison.xi}. 
The two-point function \eqref{2pointOCFT}, fits in the generic form of \eqref{Cardy.Osborn}
\begin{equation}
\small
\psi(\zeta)=\frac{\text{d}_G^2}{16}+\frac{\text{d}_G}{(\zeta-2)^2}+\frac{\text{d}_G}{(\zeta+2)^2}+\frac{c_G\text{d}_G}{k(\zeta^2-4)}\,,
\quad \zeta=2(1-2\xi)\geqslant2\,,\quad \D_{1,2}=2\, ,
\end{equation}
which is also in agreement with \eqref{2pt.gen.rel} and \eqref{F.Non.Abel}.
As a further check, the above expression degenerates to \eqref{2pt.comp.Abel} in the Abelian (single field) limit in which $c_G=0$ and $\text{d}_G=1$.

\no
Before concluding this subsection let us comment on the limiting case when one
of the two operators in \eqref{2pointOCFT} approaches the boundary. 
In that case the corresponding two-point function reads
\be
\label{bB2pt}
\langle{\cal O}(x_1,\bar x_1)T(t_2)\rangle_\text{uhp}=\frac{\text{d}_G}{|x_1-t_2|^4}\,,
\ee
where the operator $T(t)$ is defined via the usual normal ordering procedure
\be
T(t)=\frac{k}{2k+c_G}\lim_{\s\to0^+}\left({\cal O}(x,\bar x)-\frac{\text{d}_G}{(x-\bar x)^2}\right)
\ee
and $x=t+i\s$. The two-point correlation function \eqref{bB2pt} corresponds to a non-vanishing bulk-boundary 
operator product expansion and once we turn on the bulk perturbation, that is driven 
by the composite current-bilinear ${\cal O}(x,\bar x)$ -- see Eq.\eqref{Thirring.action}, it induces an RG flow on the coupling 
of the boundary operator $T(t)$~\cite{Fredenhagen:2006dn}.
In the present work we will be interested in evaluating bulk correlators on the upper-half plane and away from conformal point, which are unaffected 
by the presence of the boundary fields for the boundary conditions of \eqref{KM.bnr}, see~\cite{Fredenhagen:2006dn}.\footnote{\label{annulus}
Let us note that in higher-genus diagrams, the bulk RG flows and correlators are also modified 
due to a Fishler--Susskind type of mechanism~\cite{Fischler:1986ci,Fischler:1986tb}. The RG flows were explicitly analyzed for the annulus case in~\cite{Keller:2007nd}.}

\subsection{Primary fields}
\label{pr2.4}

We now turn our attention to affine primary fields $\Phi_{i,i'}(x,\bar x)$, transforming in the irreducible representations $R$ and $R'$,
under the action of the currents $J_a$ and $\bar J_a$. In terms of the Hermitian matrices $t_a$ and $\tilde t_a$ and using the notation of~\cite{Georgiou:2016iom} we have that
\be
\label{J.on.primaries}
\begin{split}
&J_a(x_1)\Phi_{i,i'}(x_2,\bar x_2)=-\frac{\left(t_a\right)_{ij}}{\sqrt{k}}\frac{\Phi_{j,i'}(x_2,\bar x_2)}{x_{12}}\,,\\
&\bar J_a(\bar x_1)\Phi_{i,i'}(x_2,\bar x_2)=\frac{\left(\tilde t_a\right)_{j'i'}}{\sqrt{k}}\frac{\Phi_{i,j'}(x_2,\bar x_2)}{\bar x_{12}}\,,
\end{split}
\ee
where $i=1,2,\dots,\text{dimR}$ and $i'=1,2,\dots,\text{dimR'}$. The signs on the right hand sides are
such that the representation matrices obey the same Lie algebra, i.e. $[t_a,t_b]=f_{abc} t_c$.

\no
These fields are also Virasoro primaries with holomorphic and anti-holomorphic
conformal dimension~\cite{Knizhnik:1984nr}
\be
h_R=\frac{c_R}{2k+c_G}\,,\qquad h_{R'}=\frac{c_{R'}}{2k+c_G}\,,
\ee
where $c_R$ and $c_{R'}$ are the quadratic Casimir operators in the representations $R$ and $R'$ respectively which are defined as
\be
\label{Casimir.RRp}
\left(t_at_a\right)_{ij}=c_R\d_{ij}\, ,\quad \left(\tilde t_a\tilde t_a\right)_{i'j'}=c_{R'}\d_{i'j'}
\ee
and we also note that for the adjoint representation $(t_a)_{bc}= (\tilde t_a)_{bc}= -f_{abc}$.

\no
Using Cardy's doubling trick we may evaluate the one-point function of a general affine primary field $\Phi_{i,i'}(x,\bar x)$,
transforming as described above. In the whole plane such a one-point function vanishes identically.
We need the conjugate field $\bar \Phi_{i',i}(\bar x,x)$, which transforms under $J_a$ in \eqn{J.on.primaries}
in the conjugate to $R'$ representation with matrices $-\tilde t^*_a$. Similarly,
under $\bar J_a$ in \eqn{J.on.primaries} the same field transforms  in the conjugate to $R$ representation
with matrices $- t^*_a$. Therefore, the one-point function is non-vanishing if and only if the representations $R$ and $R'$ are identical so that 
$\tilde t_a =t_a$ and equals to\footnote{
\label{norm.1pt} We have ignored an overall coefficient encoding the boundary conditions for the primary field. This can be obtained from Verlinde's formula by projecting Cardy's boundary states~\cite{Cardy:1989ir} on the Ishibashi ones, see for example Eq.(4.20)~\cite{Schomerus:2002dc} for the  $\text{su}(2)$ case. Rather unexpectedly this coefficient can be derived via a semi-classical analysis of the Born--Infeld action up to a shift of the level of the current algebra~\cite{Bachas:2000ik,Bachas:2001id}. The ignored normalization is affecting all higher-point correlators and in addition carries over information for the boundary conditions obeyed by the primary fields. However, we are not 
interested in these issues in the current work.}
\be
\label{onepoint.CFT}
\langle \Phi_{i,i'}(x,\bar x)\rangle_\text{uhp}=\langle \Phi_{i,i'}(x,\bar x)\bar \Phi_{i',i}(\bar x,x)\rangle_\text{hol}=
\frac{\d_{ii'}}{(x-\bar x)^{2h_R}}\,,
\ee
in agreement with \eqref{1pt.double}, where $h=h_R$ and $A_\Phi=1$.

\section{CFT perturbation in the upper-half plane}

\label{CFTpertuhp}

In this section as well as in the accompanying Appendices~\ref{2loop.append},~\ref{1loop.append}
we employ conformal perturbation and compute two-point correlation functions of currents and composite current-bilinear, 
three-point functions of currents and one-point function of primaries in the upper-half plane and
beyond the conformal point. These computations will be performed in Euclidean signature.

\no
As a reminder we note that, the structure constants of the Lie algebra in this section and also in the 
accompanying Appendices~\ref{2loop.append},~\ref{1loop.append} are taken to be imaginary.

\no
In the Euclidean regime~\eqref{analy}, the WZW perturbation~\eqref{Thirring.action} when considered on the upper-half plane contributes in the path integral as 
$-\frac{\l}{\pi}\int_S\text{d}^2z\, J^a(z,\z)\bar J^a(z,\z)$. 
Consider a set of generic fields $\{{\cal A}_1(x_1,\bar x_1), {\cal A}_2(x_2,\bar x_2),\cdots\}$, then to order
$\l^n$ their correlation function takes the form
\be
\small
\begin{split}
&\langle {\cal A}_1(x_1,\bar x_1){\cal A}_2(x_2,\bar x_2)\cdots\rangle^{(n)}_\text{uhp}=\frac{1}{n!}\left(-\frac{\l}{\pi}\right)^n
\int_S\text{d}^2z_{1\cdots n}\langle J_{a_1}(z_1,\z_1)\cdots J_{a_n}(z_n,\z_n)\, ,
\\
&\bar J_{a_1}(z_1,\z_1)\cdots\bar J_{a_n}(z_n,\z_n){\cal A}_1(x_1,\bar x_1){\cal A}_2(x_2,\bar x_2)\cdots\rangle_\text{uhp}\,,
\end{split}
\ee
where $\text{d}^2z_{1\cdots n}=\text{d}^2z_1\cdots\text{d}^2z_n$.
The subscript $S$ in the integral denotes the domain of integration, which is the upper-half plane $S=\{\text{Im}z\geqslant0\}$.
To evaluate the above integral we follow the regularization prescription described in~\cite{Georgiou:2016iom},
but in the present work we also allow internal points to coincide with external ones.
In other words, we keep all $\delta$-functions (except those corresponding to external points) and we introduce a short-distance
regulator $\varepsilon$ whenever an integral diverges.
In this way we do not need to worry about keeping the order of integrations intact.
In addition, if an integral diverges for large distances we restrict the integration within the radius $R$ of a half-disc located at the upper-half plane.

\no
For our computations we need the basic integral
\be
\label{J1.basic.inttext}
\int_S\frac{\text{d}^2z}{(z-x_1)(\z-\bar x_2)}=\pi\ln\frac{R(\bar x_2- x_1)}{|x_{12}|^2}-\frac{i\pi^2}{2}\,,
\ee
evaluated in Appendix~\ref{J1.basic.int}, as well as the integrals
\be
\begin{split}
&\int_S\frac{\text{d}^2z}{(z-x_1)^2(\z-\bar x_2)}=\pi\left(\frac{1}{x_1-\bar x_2}-\frac{1}{x_{12}}\right)\,,
\\
&\int_S\frac{\text{d}^2z}{(z-x_1)(\z-\bar x_2)^2}=-\pi\left(\frac{1}{x_1-\bar x_2}-\frac{1}{\bar x_{12}}\right)\,,
\\
&\int_S\frac{\text{d}^2z}{(z-x_1)^2(\z-\bar x_2)^2}=\frac{\pi}{(x_1-\bar x_2)^2}+\pi^2\d^{(2)}(x_{12})\,,
\end{split}
\ee
obtained by taking appropriate derivatives of \eqref{J1.basic.inttext}.

\subsection{Three-point current correlation functions}

We turn our attention to correlators of the form $\langle J J J\rangle$ and $\langle J J\bar J\rangle$ up
to
$\nicefrac{\l}{\sqrt{k}}$ in the conformal perturbation theory. This is enough to determine their full $\l$-dependence by employing the symmetry in coupling space~\cite{Itsios:2014lca}
\be
\label{nonpsym}
\l\to\l^{-1}\ ,\qquad k\to-k\,,
\ee
valid for $k\gg 1$, as well as the non-Abelian limit (correlated $\l\to 1, k\to \infty$), the pseudo-dual
limit (correlated $\l\to -1, k\to \infty$) and also by matching with the perturbative result up to order $\nicefrac{\l}{\sqrt{k}}$ we are about to compute. This technique has been already fruitful in evaluating the same correlators in the
entire plane~\cite{Georgiou:2015nka,Georgiou:2016iom}, for unequal levels as well~\cite{Georgiou:2016zyo}.

\subsubsection{The three-point function $\langle J J J\rangle$}

We will show that the three-point function $\langle J J J\rangle$ to order $\nicefrac{\l}{\sqrt{k}}$ takes the
following form
\ba
\label{JJJ.CFT.pert}
&&\langle   J_a(x_1)  J_b(x_2)  J_c(x_3)\rangle^{(\l)}_{\text{uhp}}=\frac{f_{abc}}{\sqrt{k}\,x_{12}x_{13}x_{23}}+\frac{\l\,f_{abc}}{\sqrt{k}}\left(\frac{1}{x_{23}^2}\left(\frac{1}{x_1-\bar x_2}+\frac{1}{\bar x_3-x_1}\right)\right.
\nonumber
\\
&&\qq\  \left.+\frac{\pi}{x_{12}^2}\left(\frac{1}{\bar x_2-x_3}+\frac{1}{x_3-\bar x_1}\right)+\frac{\pi}{x_{13}^2}\left(\frac{1}{x_2-\bar x_3}+\frac{1}{\bar x_1-x_2}\right)\right)\,.
\ea
From the above result, the symmetry \eqref{nonpsym} and regularity at the non-Abelian and pseudo-dual limits  we obtain the full $\l$-dependence of the couplings and hence of the correlator
(the steps are identical to those in \cite{Georgiou:2016iom} for the same correlator)
\be
\label{JJJe}
 \boxed{
\begin{split}
&\langle J_a(x_1) J_b(x_2) J_c(x_3)\rangle^{(\l),\text{exact}}_{\text{uhp}}
 = {1+\l+\l^2\ov \sqrt{k(1-\l)(1+\l)^3}}\, { f_{abc}\ov x_{12} x_{13} x_{23}}
\\
&\quad
+ {\l \ov \sqrt{k(1-\l)(1+\l)^3}}\, f_{abc} \bigg({1\ov x_{23}^2 (x_1-\bar x_2)} -
{1\ov x_{13}^2 (x_2-\bar x_1)}  +\text{cyclic in 1,2,3}\bigg)\ .
\end{split}
}
\ee
In what follows we shall work out the details in deriving \eqref{JJJ.CFT.pert},
namely the conformal result \eqref{JJJ.CFT} as well as the one-loop correction to it \eqref{JJJ.CFT.pert1}.

\paragraph{Conformal result:} At the conformal point the correlator equals to \eqref{3pt.CFT}. We repeat it here for the reader's convenience
\be
\label{JJJ.CFT}
\langle   J_a(x_1)  J_b(x_2)   J_c(x_3)\rangle_\text{uhp}
=\frac{f_{abc}}{\sqrt{k}\,x_{12}x_{13}x_{23}}\,.
\ee

\paragraph{One-loop:} To the above result we add the contribution of order $\nicefrac{\l}{\sqrt{k}}$ which reads
\be
\label{JJJ.oneloop}
\langle  J_a(x_1)  J_b(x_2)  J_c(x_3) \rangle^{(1)}_\text{uhp}=
-\frac{\l}{\pi}\int_S\text{d}^2z\,\langle   J_a(x_1)  J_b(x_2)  J_c(x_3)  J_d(z)\bar  J_d(\z)\rangle_\text{uhp}\,.
\ee
To evaluate the above five-point function we first use Cardy's doubling trick, in particular \eqref{dCtrick4} and then
we perform  the corresponding contractions using \eqref{KM.OPE}. We have that
\be
\label{JJJnonab}
\begin{split}
\langle   J_a &(x_1)   J_b(x_2)  J_c(x_3)  J_d(z)\bar  J_d(\z)\rangle_\text{uhp}=
\langle   J_a(x_1)  J_b(x_2)  J_c(x_3)  J_d(z)  J_d(\z)\rangle\\
&=\frac{\langle J_b(x_2) J_c(x_3)  J_a(\z)\rangle}{(x_1-z)^2}+\frac{\langle J_b(x_2) J_c(x_3)  J_a(z)\rangle}{(x_1-\z)^2}\\
&+\frac{f_{abe}\langle  J_e(x_2) J_c(x_3) J_d(z)  J_d(\z)\rangle}{\sqrt{k}\,x_{12}}
+\frac{f_{ace}\langle  J_b(x_2) J_e(x_3) J_d(z)  J_d(\z)\rangle}{\sqrt{k}\,x_{13}}\\
&+\frac{f_{ade}\langle J_b(x_2) J_c(x_3) J_e(z)  J_d(\z)\rangle}{\sqrt{k}\,(x_1-z)}
+\frac{f_{ade}\langle J_b(x_2) J_c(x_3) J_d(z)  J_e(\z)\rangle}{\sqrt{k}\,(x_1-\z)}\\
&=\frac{f_{abc}}{\sqrt{k}(x_1-z)^2x_{23}(x_2-\z)(x_3-\z)}+\frac{f_{abc}}{\sqrt{k}(x_1-\z)^2x_{23}(x_2-z)(x_3-z)}\\
&+\frac{f_{abc}}{\sqrt{k}\,x_{12}}\left(\frac{\text{d}_G}{x_{23}^2(z-\z)^2}+\frac{1}{(z-x_2)^2(\z-x_3)^2}+\frac{1}{(z-x_3)^2(\z-x_2)^2}\right)\\
&-\frac{f_{abc}}{\sqrt{k}\,x_{13}}\left(\frac{\text{d}_G}{x_{23}^2(z-\z)^2}+\frac{1}{(z-x_2)^2(\z-x_3)^2}+\frac{1}{(z-x_3)^2(\z-x_2)^2}\right)\\
&-\frac{f_{abc}}{\sqrt{k}(x_1-z)}\left(\frac{1}{(z-x_2)^2(\z-x_3)^2}-\frac{1}{(z-x_3)^2(\z-x_2)^2}\right)\\
&+\frac{f_{abc}}{\sqrt{k}(x_1-\z)}\left(\frac{1}{(z-x_2)^2(\z-x_3)^2}-\frac{1}{(z-x_3)^2(\z-x_2)^2}\right)\,.
\end{split}
\ee
Then we insert this expression into \eqref{JJJ.oneloop} and compute the corresponding integrals. We will also dismiss terms
corresponding to bubble diagrams which it turns out to be proportional to the group dimension $\text{d}_G$.

\no
First we start with the integral arising from the first term after the last equality above. It is given by
\be
\frac{1}{x_{23}}\int_S\frac{\text{d}^2z}{(x_1-z)^2(x_2-\z)(x_3-\z)}=
\frac{1}{x_{23}^2}\del_{x_1}\int_S\frac{\text{d}^2z}{z-x_1}\left(\frac{1}{\z-x_2}-\frac{1}{\z-x_3}\right)=0\,,
\ee
where we have used \eqref{ix1x2}.

\no
The integral arising from the following term is
\be
\begin{split}
&\frac{1}{x_{23}}\int_S\frac{\text{d}^2z}{(x_1-\z)^2(x_2-z)(x_3-z)}=
\frac{1}{x_{23}^2}\del_{x_1}\int_S\frac{\text{d}^2z}{\z-x_1}\left(\frac{1}{z-x_2}-\frac{1}{z-x_3}\right)\\
&\qq =\frac{\pi}{x_{23}^2}\left(\frac{1}{x_1-\bar x_3}-\frac{1}{x_1-\bar x_2}\right)\,,
\end{split}
\ee
where we have used again \eqref{ix1x2}.
The terms in the following second and third lines after the equality (apart from the bubble diagrams)
vanish as it can easily seen using once again \eqref{ix1x2}.

\no
The terms in the next fourth line equal to
\be
\begin{split}
\int_S\frac{\text{d}^2z}{z-x_1}& \left(\frac{1}{(z-x_2)^2(\z-x_3)^2}-\frac{1}{(z-x_3)^2(\z-x_2)^2}\right)=I_{321}-I_{231}\\
&=\frac{\pi}{x_{12}^2}\left(\frac{1}{x_3-\bar x_2}-\frac{1}{x_3-\bar x_1}\right)-\frac{\pi}{x_{13}^2}\left(\frac{1}{x_2-\bar x_3}-\frac{1}{x_2-\bar x_1}\right)\,,
\end{split}
\ee
using the results of Appendix~\ref{I123.appendix}.
The terms in the fifth line are vanishing as they are related to the integral $J_{123}$
\be
\int_S\frac{\text{d}^2z}{x_1-\z}\left(\frac{1}{(z-x_2)^2(\z-x_3)^2}-\frac{1}{(z-x_3)^2(\z-x_2)^2}\right)=-J_{321}+J_{231}=0\,,
\ee
using the results of Appendix~\ref{J123.appendix}.

\no
Inserting the above into \eqref{JJJ.oneloop} we find that
\be
\label{JJJ.CFT.pert1}
\begin{split}
\langle   J_a(x_1) &  J_b(x_2)  J_c(x_3)\rangle^{(1)}_{\text{uhp}}=\frac{\l\,f_{abc}}{\sqrt{k}}\left(\frac{1}{x_{23}^2}\left(\frac{1}{x_1-\bar x_2}+\frac{1}{\bar x_3-x_1}\right)\right.\\
&\left.+\frac{\pi}{x_{12}^2}\left(\frac{1}{\bar x_2-x_3}+\frac{1}{x_3-\bar x_1}\right)+\frac{\pi}{x_{13}^2}\left(\frac{1}{x_2-\bar x_3}+\frac{1}{\bar x_1-x_2}\right)\right)\ .
\end{split}
\ee
Adding the latter with the conformal result \eqref{JJJ.CFT}, we reach \eqref{JJJ.CFT.pert}.

\subsubsection{The three-point function $\langle J J\bar J\rangle$}

We will show that the three-point function $\langle J J\bar J\rangle$ to order $\nicefrac{\l}{\sqrt{k}}$ takes the
following form
\be
\small
\label{JJJbar.oneloop.end}
\begin{split}
&\langle   J_a(x_1)  J_b(x_2)\bar  J_c(\bar x_3)\rangle^{(\l)}_\text{uhp}=\frac{f_{abc}}{\sqrt{k}\,x_{12}(x_1-\bar x_3)(x_2-\bar x_3)}\\
&+\frac{\l\,f_{abc}}{\sqrt{k}}\left(\frac{\bar x_{12}}{x_{12}^2\bar x_{13}\bar x_{23}}+\frac{1}{(x_1-\bar x_3)^2}\left(\frac{1}{\bar x_1-x_2}+\frac{1}{x_{23}}\right)
+\frac{1}{(x_2-\bar x_3)^2}\left(\frac{1}{x_1-\bar x_2}-\frac{1}{x_{13}}\right)\right)\, .
\end{split}
\ee
As in the previous subsection, this result, the symmetry \eqref{nonpsym} and regularity at the non-Abelian and pseudo-dual limits,
allows for the full $\l$-dependence of the couplings and hence of the correlator
\be
\boxed{
\label{jjjb}
\begin{split}
\langle   J_a(x_1) & J_b(x_2)\bar  J_c(\bar x_3)\rangle^{(\l),\text{exact}}_\text{uhp}
 = {1+\l+\l^2\ov \sqrt{k(1-\l)(1+\l)^3}}\frac{f_{abc}}{x_{12}(x_1-\bar x_3)(x_2-\bar x_3)}\\
&+\frac{\l\,f_{abc}}{\sqrt{k(1-\l)(1+\l)^3}}\left(\frac{\bar x_{12}}{x_{12}^2\bar x_{13}\bar x_{23}}+
\frac{1}{(x_1-\bar x_3)^2}\left(\frac{1}{\bar x_1-x_2}+\frac{1}{x_{23}}\right)\right.\\
&\left.+\frac{1}{(x_2-\bar x_3)^2}\left(\frac{1}{x_1-\bar x_2}-\frac{1}{x_{13}}\right)\right)\, .
\end{split}
}
\ee
Before we proceed, we note the above correlator and \eqn{JJJe}
are consistent with the Dirichlet boundary condition at $\bar x_3= x_3$, that is
\be
\label{bc.JJJ.JJJbar}
{\big\langle}  J_a(x_1) J_b(x_2)
\Big( J_c(x_3)-\bar J_c(\bar x_3)\Big){\big\rangle}^{(\l)}_\text{uhp}\Big{|}_{\bar x_3= x_3}=0\,.
\ee

\no
In what follows we work out the details in deriving \eqref{JJJbar.oneloop.end},
that is the conformal result \eqref{JJJbar.CFT} and the one-loop contribution \eqref{JJJbar.oneloop1}.

\paragraph{Conformal result:}  At the conformal point the correlator equals to \eqref{3pt.CFT} and is repeated here for the reader's convenience
\be
\label{JJJbar.CFT}
\langle   J_a(x_1)  J_b(x_2)  \bar  J_c(\bar x_3)\rangle_\text{uhp}
=\frac{f_{abc}}{\sqrt{k}\,x_{12}(x_1-\bar x_3)(x_2-\bar x_3)}\,.
\ee
\paragraph{One-loop:} The contribution to the above result of order $\nicefrac{\l}{\sqrt{k}}$ reads
\be
\label{JJJbar.oneloop}
\langle  J_a(x_1)  J_b(x_2) \bar  J_c(\bar x_3) \rangle^{(1)}_\text{uhp}=
-\frac{\l}{\pi}\int_S\text{d}^2z\langle   J_a(x_1)  J_b(x_2) \bar  J_c(\bar x_3)  J_d(z)\bar  J_d(\z)\rangle_\text{uhp}\,.
\ee
We evaluate the above five-point function via Cardy's doubling trick and \eqref{KM.OPE}
\be
\small
\begin{split}
&\langle   J_a(x_1)  J_b(x_2) \bar  J_c(\bar x_3)  J_d(z)\bar  J_d(\z)\rangle_\text{uhp}=
\langle   J_a(x_1)  J_b(x_2)  J_c(\bar x_3)  J_d(z)  J_d(\z)\rangle\\
&=\frac{\langle J_b(x_2) J_c(\bar x_3)  J_a(\z)\rangle}{(x_1-z)^2}+\frac{\langle J_b(x_2) J_c(\bar x_3)  J_a(z)\rangle}{(x_1-\z)^2}\\
&+\frac{f_{abe}\langle J_e(x_2) J_c(\bar x_3) J_d(z)\bar  J_d(\z)\rangle}{\sqrt{k}\,x_{12}}
+\frac{f_{ace}\langle J_b(x_2) J_e(\bar x_3) J_d(z)  J_d(\z)\rangle}{\sqrt{k}\,(x_1-\bar x_3)}\\
&+\frac{f_{ade}\langle J_b(x_2) J_c(\bar x_3) J_e(z)  J_d(\z)\rangle}{\sqrt{k}\,(x_1-z)}
+\frac{f_{ade}\langle J_b(x_2) J_c(\bar x_3) J_d(z)  J_e(\z)\rangle}{\sqrt{k}\,(x_1-\z)}\\
&=\frac{f_{abc}}{\sqrt{k}(x_1-z)^2(x_2-\bar x_3)(x_2-\z)(\bar x_3-\z)}+\frac{f_{abc}}{\sqrt{k}(x_1-\z)^2(x_2-\bar x_3)(x_2-z)(\bar x_3-z)}\\
&+\frac{f_{abc}}{\sqrt{k}\,x_{12}}\left(\frac{\text{d}_G}{(x_2-\bar x_3)^2(z-\z)^2}+\frac{1}{(z-x_2)^2(\z-\bar x_3)^2}+\frac{1}{(z-\bar x_3)^2(\z-x_2)^2}\right)\\
&-\frac{f_{abc}}{\sqrt{k}\,(x_1-\bar x_3)}\left(\frac{\text{d}_G}{(x_2-\bar x_3)^2(z-\z)^2}+\frac{1}{(z-x_2)^2(\z-\bar x_3)^2}+\frac{1}{(z-\bar x_3)^2(\z-x_2)^2}\right)\\
&-\frac{f_{abc}}{\sqrt{k}(x_1-z)}\left(\frac{1}{(z-x_2)^2(\z-\bar x_3)^2}-\frac{1}{(z-\bar x_3)^2(\z-x_2)^2}\right)\\
&+\frac{f_{abc}}{\sqrt{k}(x_1-\z)}\left(\frac{1}{(z-x_2)^2(\z-\bar x_3)^2}-\frac{1}{(z-\bar x_3)^2(\z-x_2)^2}\right)\,,
\end{split}
\ee
then we insert it into \eqref{JJJbar.oneloop} and compute the corresponding integrals, dismissing bubble diagrams terms.

\no
The first term of the first line after the last equality is
\be
\small
\begin{split}
&\int_S\frac{\text{d}^2z}{(x_1-z)^2(x_2-\bar x_3)(x_2-\z)(\bar x_3-\z)}=
\frac{1}{(x_2-\bar x_3)^2}\del_{x_1}\int_S\frac{\text{d}^2z}{z-x_1}\left(\frac{1}{\z-x_2}-\frac{1}{\z-\bar x_3}\right)\\
&=\frac{\pi}{(x_2-\bar x_3)^2}\left(\frac{1}{x_{13}}-\frac{1}{x_1-\bar x_3}\right)\, ,
\end{split}
\ee
where we have used the integral \eqref{int.I1x1x2} and the first of \eqref{JJ.int.text}.
For the second term of the same line we find in a similar manner that
\be
\int_S\frac{\text{d}^2z}{(x_1-\z)^2(x_2-\bar x_3)(x_2-z)(\bar x_3-z)}=\frac{\pi}{(x_2-\bar x_3)^2}\left(\frac{1}{x_1-\bar x_3}-\frac{1}{x_1-\bar x_2}\right)\,.
\ee
In addition, the terms in the second line, after the last equality, cancel among themselves as it can easily see using \eqref{JJ.int.text0}.
This is also true for the following third line as well.

\no
The last two lines which can be evaluated using the integrals of \eqref{JJJbar.int}, specifically in order of
appearance $A_{123}, E_{123}, B_{123}$ and $F_{123}$. Employing all the above into \eqref{JJJbar.oneloop} we find
\be
\small
\label{JJJbar.oneloop1}
\begin{split}
&\langle   J_a(x_1)  J_b(x_2)\bar  J_c(\bar x_3)\rangle^{(1)}_\text{uhp}=\\
&\frac{\l\,f_{abc}}{\sqrt{k}}\left(\frac{\bar x_{12}}{x_{12}^2\bar x_{13}\bar x_{23}}+\frac{1}{(x_1-\bar x_3)^2}\left(\frac{1}{\bar x_1-x_2}+\frac{1}{x_{23}}\right)
+\frac{1}{(x_2-\bar x_3)^2}\left(\frac{1}{x_1-\bar x_2}-\frac{1}{x_{13}}\right)\right)\,
\end{split}
\ee
and adding the latter with the conformal result \eqref{JJJbar.CFT}, we reach \eqref{JJJbar.oneloop.end}.

\subsection{Two-point current correlation functions}

We will compute the two-point functions $\langle J J\rangle$ and $\langle J\bar J\rangle$, beyond the
conformal point and read the corresponding anomalous dimension. Our results will be exact in  the parameter $\l$ and up to order
$\nicefrac1k$ in the large $k$ expansion. We already note that these correlators have a much richer structure than the corresponding ones
for the entire plane~\cite{Georgiou:2015nka}.

\subsubsection{The two-point function $\langle J J\rangle$}
\label{JJ.oneloop.main}

We will show that the two-point function $\langle J J\rangle$
to order $\nicefrac{\l^2}{k}$ takes the following form
\be
\label{JJ.2FULL}
\begin{split}
&\langle   J_a(x_1)  J_b(x_2)\rangle^{(\l)}_\text{uhp}
\\
&=\frac{\d_{ab}}{x_{12}^2}\left(1-\frac{c_G}{k}\l^2 \frac{\xi}{1-\xi}-\frac{c_G}{k}\l(1-2\l)\ln(1-\xi)+
\frac{c_G}{k} \l^2\ln\frac{\varepsilon^2}{|x_{12}|^2}\right)\,,
\end{split}
\ee
where $\varepsilon$ is a short-distance cut-off and we have ignored a factor $1+\l^2$ inside the parenthesis 
which only
affects the overall normalization of the correlator.
From the logarithmic term of the two-point function \eqref{JJ.2FULL}, we can extract the anomalous dimension
of the current operator to order $\nicefrac{\l^2}{k}$.
From \eqref{JJ.2FULL} and the non-perturbative symmetry \eqref{nonpsym} as
well as regularity at the non-Abelian and pseudo-dual limits at $\l=\pm 1$ we get the exact dependence of the correlator in $\l$ to order $\nicefrac1k$
\be
\boxed{
\begin{split}
\label{JJ.FULL}
&\langle   J_a(x_1)  J_b(x_2)\rangle^{^{(\l),\text{exact}}}_\text{uhp}
=\frac{\d_{ab}}{x_{12}^2}\left(1-\g_ J\frac{\xi}{1-\xi}
+\d_J\ln(1-\xi)
+\g_J\ln\frac{\varepsilon^2}{|x_{12}|^2}\right)\,,
\end{split}
}
\ee
where the parameters $\g_J$ and the  $\d_J$ are given by
\be
\label{anom.J.exact}
\g_J=\frac{c_G\l^2}{k(1-\l)(1+\l)^3}\geqslant 0\,,\quad
\d_J=-\frac{c_G\l(1+\l^2)}{k(1-\l)(1+\l)^3}\, .
\ee
The anomalous dimension of the operator $ J_a$ is identified with the parameter $\g_J$ and matches the whole plane result, see Eq.(2.5) in~\cite{Georgiou:2015nka}. This agreement should have been expected
on physical grounds as the anomalous dimension is determined by the short-distance behavior and 
in the upper-half plane the presence of the boundary has no effect (see footnote~\ref{annulus}).

\no
In what follows we shall work out the details in proving \eqref{JJ.2FULL}.

\paragraph{Conformal result:}
 At the conformal point the correlator equals to \eqref{2pt.CFT}, repeated here for the reader's convenience
\be
\label{JJ.CFT.solo}
\langle   J_a(x_1)  J_b(x_2)\rangle_\text{uhp}=\frac{\d_{ab}}{x_{12}^2}\,.
\ee

\paragraph{One-loop:}
Turning on the deformation parameter, the
one-loop contribution to the two-point function reads
\be
\label{JJ.oneloop}
\langle   J_a(x_1)  J_b(x_2)\rangle^{(1)}_\text{uhp}=
-\frac{\l}{\pi}\int_S\text{d}^2z\langle   J_a(x_1)  J_b(x_2)  J_c(z)\bar  J_c(\z)\rangle_\text{uhp}\,.
\ee
Using Cardy's doubling trick, we can evaluate the above four point function as
\be
\begin{split}
\langle   J_a(x_1)&  J_b(x_2)  J_c(z)\bar  J_c(\z)\rangle_\text{uhp}=
\langle   J_a(x_1)  J_b(x_2)  J_c(z)  J_c(\z)\rangle\\
&=\frac{\d_{ab}\,\text{d}_G}{x_{12}^2(z-\bar z)^2}+\frac{\d_{ab}}{(x_1-z)^2(x_2-\z)^2}+\frac{\d_{ab}}{(x_1-\z)^2(x_2-z)^2}\\
&\quad +\frac{c_G\d_{ab}}{k(z-x_1)(z-x_2)(\z-x_1)(\z-x_2)}\,.
\end{split}
\ee
Inserting the above correlation function into \eqref{JJ.oneloop} we find that its first term is a bubble diagram, hence it is dismissed.
Moving on to the second and third terms we find that  these vanish since
\be
I_{1|1}(x_1;x_2)=\int_S\frac{\text{d}^2z}{(z-x_1)^2(\z-x_2)^2}=0\,,
\ee
which can be easily shown using the results of Appendix~\ref{I1x1x2.sect}.
So, we are left with the following integral
\be
\langle   J_a(x_1)  J_b(x_2)\rangle^{(1)}_\text{uhp}=
-\frac{\l}{\pi}\frac{c_G\d_{ab}}{k}\int_S\frac{\text{d}^2z}{(z-x_1)(z-x_2)(\z-x_1)(\z-x_2)}\,.
\ee
To evaluate it we use twice the identity
\be
\label{identity.split}
\frac{1}{(z-x_1)(z-x_2)}=\frac{1}{x_{12}}\left(\frac{1}{z-x_1}-\frac{1}{z-x_2}\right)\,
\ee
and the integral
\be
\label{int.I1x1x2}
I_1(x_1;x_2)=\int_S\frac{\text{d}^2z}{(z-x_1)(\z-x_2)}=\pi\ln\frac{R}{\bar x_1-x_2}+\frac{i\pi^2}{2}\,,
\ee
evaluated in Appendix~\ref{I1x1x2.sect}, where
$R$ is a large cut-off radius of a half-disc located at the upper-half plane.
Employing the above we can easily find that
\be
\label{JJ.lego.int}
\int_S\frac{\text{d}^2z}{(z-x_1)(z-x_2)(\z-x_1)(\z-x_2)}=\frac{\pi}{x_{12}^2}\ln\frac{|x_1-\bar x_2|^2}{(x_1-\bar x_1)
(\bar x_2- x_2)}\, .
\ee
So the one-loop contribution reads
\be
\label{JJ.1loop.solo}
\langle   J_a(x_1)  J_b(x_2)\rangle^{(1)}_\text{uhp}=-\l\frac{c_G\d_{ab}}{k}\frac{1}{x_{12}^2}
\ln(1-\xi)\,,
\ee
in terms of the invariant ratio \eqref{comparison.xi}.
At this point we will present the one-loop result since we will need it when we establish the mixed
correlator $\langle J\bar J\rangle$ below.
Adding up \eqref{JJ.CFT.solo} and \eqref{JJ.1loop.solo} we find that to order $\nicefrac{\l}{k}$ it reads
\be
\boxed{
\label{JJ.1FULL}
\langle   J_a(x_1)  J_b(x_2)\rangle^{(\l)}_\text{uhp}
=\frac{\d_{ab}}{x_{12}^2}\left(1-\frac{c_G}{k}\l\ln(1-\xi)\right)\,.
}
\ee

\paragraph{Two-loop:} Moving to the two-loop order in $\l$, we need to evaluate
\be
\label{JJ.2loop}
\langle  J_a(x_1)  J_b(x_2)\rangle^{(2)}_\text{uhp}=\frac{\l^2}{2\pi^2}\int_S\text{d}^2z_{12}
\langle  J_a(x_1)  J_b(x_2) J_c(z_1)\bar  J_c(\z_1) J_d(z_2)\bar  J_d(\z_2)\rangle_\text{uhp}\,.
\ee
After a rather long but straightforward computation
which is sketched in Appendix~\ref{2loop.append} we reach the end result \eqref{JJ.2loop.solo.App} that we repeat also here
\be
\label{JJ.2loop.solo}
\langle  J_a(x_1)  J_b(x_2)\rangle^{(2)}_\text{uhp}=
\l^2\left(1-\frac{c_G}{k}\frac{\xi}{1-\xi}\right)\frac{\d_{ab}}{x_{12}^2}+
\frac{c_G\l^2}{k}\frac{\d_{ab}}{x_{12}^2}\ln\frac{(1-\xi)^2\varepsilon^2}{|x_{12}|^2}\,,
\ee
where again $\varepsilon$ is a short distance cut-off.

\no
Finally, the end result can be read by adding up the CFT, one-loop
and two-loop results, that is Eqs.\eqref{JJ.CFT.solo}, \eqref{JJ.1loop.solo} and \eqref{JJ.2loop.solo} respectively,
we find \eqref{JJ.2FULL}.

\subsubsection{The two-point function $\langle J\bar J\rangle$}
\label{JJbar.oneloop.main}

We will show that the two-point function $\langle J\bar J\rangle$ to order $\nicefrac{\l}{k}$ takes the following form
\be
\boxed{
\label{JbarJ.oneloop0}
\langle   J_a(x_1)\bar  J_b(\bar x_2)\rangle^{(\l)}_\text{uhp}=\frac{\d_{ab}}{(x_1-\bar x_2)^2}
\left(1-\l\frac{c_G}{k}\ln(-\xi)\right)\,. }
\ee
Before we proceed with the various contributions leading to this result, we note that, this correlator and \eqref{JJ.1FULL} are
 consistent with the Dirichlet boundary condition at $\bar x_2= x_2$, that is
\be
\label{bc.JJ.JbarJ}
{\big\langle}  J_a(x_1)
\Big( J_b(x_2)-\bar J_b(\bar x_2)\Big){\big\rangle}^{(\l)}_\text{uhp}\Big{|}_{\bar x_2= x_2}=0\,.
\ee

\no
In what follows we shall work out the details in proving \eqref{JbarJ.oneloop0}.

\paragraph{Conformal result:}
At the conformal point the correlator equals to \eqref{2pt.CFT}, which we repeat here as well
\be
\label{JJbar.CFT}
\langle   J_a(x_1)\bar  J_b(x_2)\rangle_\text{uhp}=\frac{\d_{ab}}{(x_1-\bar x_2)^2}\,.
\ee
\paragraph{One-loop:}
Turning on the deformation parameter $\l$, the $\langle J\bar J\rangle$ correlator at one-loop order reads
\be
\label{JJbar.oneloop}
\begin{split}
\langle   J_a(x_1) \bar  J_b(\bar x_2)\rangle^{(1)}_\text{uhp}=
-\frac{\l}{\pi}\int_S\text{d}^2z\langle   J_a(x_1) \bar  J_b(\bar x_2)  J_c(z)\bar  J_c(\z)\rangle_\text{uhp}\,.
\end{split}
\ee
The above four-point function can be evaluated using Cardy's doubling trick
\be
\begin{split}
\langle   J_a(x_1) & \bar  J_b(\bar x_2)  J_c(z)\bar  J_c(\z)\rangle_\text{uhp}=
\langle   J_a(x_1)  J_b(\bar x_2)  J_c(z)  J_c(\z)\rangle\\
&=\frac{\d_{ab}\,\text{d}_G}{(x_1-\bar x_2)^2(z-\bar z)^2}+\frac{\d_{ab}}{(x_1-z)^2(\bar x_2-\z)^2}+\frac{\d_{ab}}{(x_1-\z)^2(\bar x_2-z)^2}\\
&+\frac{c_G\d_{ab}}{k(z-x_1)(z-\bar x_2)(\z-x_1)(\z-\bar x_2)}\,.
\end{split}
\ee
Inserting the above into \eqref{JJbar.oneloop} we find that the first term is a bubble diagram and therefore it is dismissed.
Moving on to the second and third terms these cancel each other since from \eqref{JJ.int}
\be
\label{JJ.int.text0}
\begin{split}
&J_{1|1}(x_1;x_2)=\int_S\frac{\text{d}^2z}{(z-x_1)^2(\z-\bar x_2)^2}=\frac{\pi}{(x_1-\bar x_2)^2}\, ,\\
&J_{2|1}(x_1;x_2)=\int_S\frac{\text{d}^2z}{(z-\bar x_2)^2(\z- x_1)^2}=-\frac{\pi}{(x_1-\bar x_2)^2}\,,
\end{split}
\ee
 where we have dismissed contact terms of external points.
Therefore, all we are left is the integral corresponding to the last term
\be
\langle   J_a(x_1) \bar J_b(x_2)\rangle^{(1)}_\text{uhp}=
-\frac{\l}{\pi}\frac{c_G\d_{ab}}{k}\int_S\frac{\text{d}^2z}{(z-x_1)(z-\bar x_2)(\z-x_1)(\z-\bar x_2)}\,.
\ee
To evaluate the above expression we use twice the identity \eqref{identity.split}, the integral \eqref{int.I1x1x2} and
the integrals \eqref{JJ.int}
\be
\label{JJ.int.text}
\begin{split}
&J_1(x_1;x_2)=\int_S\frac{\text{d}^2z}{(z-x_1)(\z-\bar x_2)}=\pi\ln\frac{R(\bar x_2- x_1)}{|x_{12}|^2}-\frac{i\pi^2}{2}\,,\\
&J_2(x_1;x_2)=\int_S\frac{\text{d}^2z}{(z-\bar x_2)(\z- x_1)}=\pi\ln\frac{R}{x_1-\bar x_2}+\frac{i\pi^2}{2}\,,\\
&J_3(x_2;x_2)=\int_S\frac{\text{d}^2z}{(z-\bar x_2)(\z-\bar x_2)}=\pi\ln\frac{R}{x_2-\bar x_2}+\frac{i\pi^2}{2}\,,
\end{split}
\ee
where $R$ is a large cut-off radius of a half-disc located at the upper-half plane.
Employing the above we find
\be
\label{JJbar.1loop.solo}
\langle   J_a(x_1)\bar  J_b(\bar x_2)\rangle^{(1)}_\text{uhp}
=-\l\frac{c_G\d_{ab}}{k}\frac{\d_{ab}}{(x_1-\bar x_2)^2}\ln(-\xi)
\,,
\ee
 in terms of the invariant ratio variable \eqref{comparison.xi}. Adding up \eqref{JJbar.1loop.solo} with \eqref{JJbar.CFT} we find
 that $\langle J\bar  J\rangle$ to order $\nicefrac{\l}{k}$ is given by \eqref{JbarJ.oneloop0}. We will not compute the order $\nicefrac{\l^2}{k}$.

\subsection{Two-point composite current-bilinear correlation function}

We will next examine the two-point function of the composite current-bilinear ${\cal O}(x,\bar x)= J_a(x)\bar  J_a(\bar x)$
to order in $\nicefrac{\l}{k^2}$ and show that it takes the form
\be
\boxed{
\small
\label{2pointOFull}
\begin{split}
\langle{\cal O}(x_1,\bar x_1)& {\cal O}(x_2,\bar x_2)\rangle^{(\l)}_\text{uhp}\\
&=\frac{\text{d}_G}{|x_{12}|^4}\left(1-\frac{2\l c_G}{k}\ln\frac{(1-\xi)\varepsilon^2}{|x_{12}|^2}\right)
+\frac{\text{d}_G}{|x_1-\bar x_2|^4}\left(1-\frac{2\l c_G}{k}\ln\frac{-\xi\,\varepsilon^2}{|x_1-\bar x_2|^2}\right)\\
&+\frac{\text{d}_G^2}{(x_1-\bar x_1)^2(x_2-\bar x_2)^2}\left(1-\frac{2\l c_G}{k}\ln\frac{\varepsilon^2}{|(x_1-\bar x_1)(x_2-\bar x_2)|}\right)\\
&+\frac{c_G}{k} \frac{\text{d}_G}{|x_{12}|^2|x_1-\bar x_2|^2} \left(1-4\l-\frac{2\l c_G}{k}\ln\frac{\varepsilon^2{\widetilde F}(\xi)}{|x_{12}(x_1-\bar x_2)|}\right)\ ,
\end{split}}
\ee
where
\be
\label{2pointOFullF}
{\widetilde F}(\xi)=(1-\xi)\left(1-\frac{1}{\xi}\right)^{-\xi}\,,\quad \xi\leqslant0\,,
\ee
in terms of the invariant ratio variable \eqref{comparison.xi}.
The anomalous dimension of the composite current-bilinear, it is then read independently from the four terms in \eqref{2pointOFull}.
All of them at order $\nicefrac{\l}{k}$ give
\be
\g_{\cal O}=-\frac{2c_G}{k}\l+{\cal O}(\l^2)\,.
\ee
This matches the whole plane result, see Eq.(3.7) in~\cite{Georgiou:2015nka}.
Alternatively we could read its anomalous dimension if we evaluate the one-point function of the composite current-bilinear from \eqref{JbarJ.oneloop0},
upon identifying $x_{1,2}=x$ and then summing over $a=b$ 
\begin{equation}
\langle {\cal O}(x,\bar x)\rangle^{(\l)}_\text{uhp}=\frac{\text{d}_G}{(x-\bar x)^2}
\left(1-\l\frac{c_G}{k}\ln\frac{-\varepsilon^2}{(x-\bar x)^2}\right)\,.
\end{equation}

\no
In order to find the exact in $\l$ result using the non-perturbative symmetry \eqref{nonpsym} and demanding
regularity at the non-Abelian and pseudo-dual limits at $\l=\pm 1$ we need to extend the above result to order $\l^2$.
This would involve, after using Cardy's doubling trick, to evaluate an eight-point current correlation function, a quite involved computation.
Instead, we recall that the anomalous dimension is 
a short-distance effect where in the upper-half plane (see footnote~\ref{annulus}) the presence of a boundary is irrelevant.

Matching the whole plane result, see Eq.(3.5) in~\cite{Georgiou:2015nka}, we have that
\be
\g_{\cal O}=-\frac{2c_G\l(1-\l+\l^2)}{k(1-\l)(1+\l)^3} \leqslant 0\,.
\ee

\no
In what follows we will work out the various contributions, that is the conformal result \eqref{2pointOCFT1} and
the one-loop result \eqref{2pointOoneloop}, whose derivation can be found in Appendix~\ref{1loop.append}.

\paragraph{Conformal result:} At the conformal point the correlator is given by \eqref{2pointOCFT} repeated here for convenience
\be
\label{2pointOCFT1}
\begin{split}
&\langle  {\cal O}(x_1,\bar x_1) {\cal O}(x_2,\bar x_2)\rangle_\text{uhp}=\\
&\quad \frac{\text{d}_G}{|x_{12}|^4}+\frac{\text{d}_G}{|x_1-\bar x_2|^4}+\frac{\text{d}_G^2}{(x_1-\bar x_1)^2(x_2-\bar x_2)^2}
+\frac{c_G}{k} \frac{\text{d}_G}{|x_{12}|^2|x_1-\bar x_2|^2}\,.
\end{split}
\ee

\paragraph{One-loop:}
Away from the conformal point, the one-loop contribution to the two-point function is given by
\be
\label{OO.oneloop}
\langle  {\cal O}(x_1,\bar x_1) {\cal O}(x_2,\bar x_2)\rangle^{(1)}_\text{uhp}=
-\frac{\l}{\pi}\int_S\text{d}^2z\langle  {\cal O}(x_1,\bar x_1) {\cal O}(x_2,\bar x_2){\cal O}(z,\z)\rangle_\text{uhp}\,.
\ee
Since even at the conformal point \eqref{2pointOCFT1} there is an explicit $\nicefrac{1}{k}$ dependence  we proceed to
evaluate \eqref{OO.oneloop} up to order $\nicefrac{1}{k^2}$ so that we can read the anomalous dimension of the composite current-bilinear which
scales as $\nicefrac{c_G}{k}$.
After a rather cumbersome computation which is sketched in Appendix~\ref{1loop.append} we reach the order $\l$,
but exact in $\nicefrac1k$ result \eqref{2pointOoneloop.App}, which we repeat here for the reader's convenience
\be
\small
\label{2pointOoneloop}
\begin{split}
\langle  {\cal O}(x_1,\bar x_1) & {\cal O}(x_2,\bar x_2)\rangle^{(1)}_\text{uhp}=
-\frac{2\l c_G}{k}\times\left\{\frac{\text{d}_G}{|x_{12}|^4}\ln\frac{(1-\xi)\varepsilon^2}{|x_{12}|^2}\right.\\
&+\frac{\text{d}_G}{|x_1-\bar x_2|^4}\ln\frac{-\xi\,\varepsilon^2}{|x_1-\bar x_2|^2}
+\left.\frac{\text{d}_G^2}{(x_1-\bar x_1)^2(x_2-\bar x_2)^2}\ln\frac{\varepsilon^2}{|(x_1-\bar x_1)(x_2-\bar x_2)|}\right\}\\
&+\frac{c_G}{k}\frac{\text{d}_G}{|x_{12}|^2|x_1-\bar x_2|^2}\left(-4\l-\frac{2\l c_G}{k}\ln\frac{\varepsilon^2{\widetilde F}(\xi)}{|x_{12}(x_1-\bar x_2)|}\right)\,,
\end{split}
\ee
where $\varepsilon$ is the usual short distance cut-off and the function ${\widetilde F}(\xi)$ was defined in \eqref{2pointOFullF}.
Hence, the two-point function of the operator ${\cal O}(x,\bar x)$ to order $\l$ can be read by adding \eqref{2pointOCFT1} and \eqref{2pointOoneloop} and we find
\eqref{2pointOFull}.

\subsection{One-point primary field correlation function}
\label{pr3.4}

We will next show that the one-point function of the primary field $\Phi_{i,i'}(x,\bar x)$
to order $\nicefrac{\l}{k}$ takes the form
\be
\label{onepoint.oneloop}
\boxed{
\langle \Phi_{i,i'}(x,\bar x)\rangle^{(\l)}_\text{uhp}=\frac{\d_{ii'}}{(x-\bar x)^{2h_R}}\left(1-\l\frac{c_R}{k}\ln\frac{-\varepsilon^2}{(x-\bar x)^2}\right)\,,
}
\ee
thus extending the CFT result obtained in \eqn{onepoint.CFT}.
The anomalous dimension of the primary field to order $\nicefrac{\l}{k}$ can be read from the above expression after
taking into account that $h_R$ is also of order $\nicefrac1k$, yielding
\be
\label{anomalous.Phi.oneloop}
\g_\Phi=\frac{c_R}{k}\left(1-2\l\right)+{\cal O}(\l^2)\,.
\ee
We note that the overall coefficient of the one-point function does not receive a $\l$-dependent contribution 
to order $\nicefrac{\l}{k}$, see also footnote~\ref{norm.1pt}.

Using \eqref{anomalous.Phi.oneloop}, the non-perturbative symmetry \eqref{nonpsym} and demanding
regularity at the non-Abelian and pseudo-dual limits at $\l=\pm 1$, we can constrain the exact in $\l$-dependence
\be
\label{udhf}
\g_\Phi=\frac{1}{k}\frac{c_R+c\l^2+c_R\l^4}{(1-\l)(1+\l)^3}\,,
\ee
up to the constant $c$ which can be fixed from a two-loop computation. The latter would require,
after using Cardy's doubling trick, the evaluation of  a six-point function, which is a quite involved computation.
Alternatively, we recall that the anomalous dimension is determined by the short-distance behavior where in the upper-half plane the presence of a boundary is irrelevant (see footnote~\ref{annulus}).
This can be read from the whole plane result for the two-point function, as the one-point function vanishes identically,
(see Eq. (4.11) of~\cite{Georgiou:2016iom}, with $c_{R'}=c_R$, $N_I=c_R$ and
after setting $c=-2c_R$ in \eqn{udhf})
\be
\label{anomalous.Phi.exact}
\g_\Phi=\frac{c_R}{k}\frac{1-\l}{1+\l} \geqslant 0\,,
\ee
agreeing to order $\l$ with \eqn{anomalous.Phi.oneloop}.

\no
In what follows we shall work out the details in proving \eqref{onepoint.oneloop}.

\paragraph{Conformal result:} At the conformal point the correlator equals to \eqref{onepoint.CFT} which is restated here
\be
\label{onepoint.CFT1}
\langle \Phi_{i,i'}(x,\bar x)\rangle_\text{uhp}=\frac{\d_{ii'}}{(x-\bar x)^{2h_R}}\,.
\ee

\paragraph{One-loop:}
Moving away from the conformal point, we find that the one-loop contribution to the one-point function reads
\be
\langle \Phi_{i,i'}(x,\bar x)\rangle^{(1)}_\text{uhp}=-\frac{\l}{\pi}\int_S\text{d}^2z\langle\Phi_{i,i'}(x,\bar x) J_a(z)\bar J_a(\z)\rangle_\text{uhp}\,.
\ee
We can evaluate the above three-point function using Cardy's doubling trick as follows
\be
\langle\Phi_{i,i'}(x,\bar x) J_a(z)\bar J_a(\z)\rangle_\text{uhp}=\langle\Phi_{i,i'}(x,\bar x)\bar \Phi_{i',i}(\bar x,x) J_a(z) J_a(\z)\rangle_\text{hol}\,.
\ee
The latter four-point function can be easily evaluated using \eqref{J.on.primaries}, \eqref{Casimir.RRp} and the discussion at the end
of  Subsection~\ref{pr2.4}
\be
\begin{split}
&\langle\Phi_{i,i'}(x,\bar x)\bar \Phi_{i',i}(\bar x,x) J_a(z) J_a(\z)\rangle_\text{hol}=\frac{c_R}{k}\frac{\d_{ii'}}{(x-\bar x)^{2h_R}}\left\{\frac{1}{(z-x)(\z-x)}\right.\\
&\left.-\frac{1}{(z-\bar x)(\z-x)}-\frac{1}{(z- x)(\z-\bar x)}+\frac{1}{(z-\bar x)(\z-\bar x)}\right\}+\frac{\text{d}_G}{(x-\bar x)^{2h_R}}\frac{\d_{ii'}}{(z-\z)^2}\,,
\end{split}
\ee
where the last term can be ignored as it corresponds to a bubble diagram. Next, we insert this expression into \eqref{onepoint.oneloop},
using the integrals \eqref{int.I1x1x2}, \eqref{JJ.int.text}, reaching the result
\be
\label{onepoint.oneloop1}
\langle \Phi_{i,i'}(x,\bar x)\rangle^{(1)}_\text{uhp}=-\frac{\l c_R}{k}\frac{\d_{ii'}}{(x-\bar x)^{2h_R}}\ln\frac{-\varepsilon^2}{(x-\bar x)^2}\,.
\ee
Hence, the one-point function to order $\l$ can be read by adding \eqref{onepoint.CFT1} and \eqref{onepoint.oneloop1}, yielding \eqref{onepoint.oneloop}.

\section{Free fields}

\label{freefields.uhp}

In this section we employ the free field approach to  $\l$-deformed $\s$-models~\cite{Georgiou:2019aon}  appropriately adapted
to the upper-half plane. Using this approach we reproduce the three-point function for the current correlators  we found before for Dirichlet
boundary conditions. We also compute the same correlation functions for generalized Neumann boundary conditions, at the free field
level, which do not preserve the current algebra and thus cannot be reproduced form conformal perturbation theory.
In addition, we compute one- and two-point correlation function of primary operators 
and extract their anomalous dimension.

\no
Finally, we point out that the structure constants of the Lie algebra in this section are taken to be real, matching the conventions used in~\cite{Georgiou:2019aon}.

\subsection{The free field expansion of the action}

The expanded action was obtained by parametrizing the group element in terms of normal coordinates as
\be
\label{xphii}
g= e^{i t_a x^a}\ ,\quad x^a = {1\ov \sqrt{k}} \sqrt{1-\l\ov 1+\l}\, \phi^a\ ,
\ee
where the rescaling is introduced so that the kinetic term is canonically normalized.
The action obtained by expanding in the number of fields, which is equivalent to the large $k$-expansion, has terms of the form
$(f^n)_{ab} \del_+\phi^a\del_-\phi^b$, combined in a infinite sum with $n=0,1,2,\dots$, where
\be
\label{fab}
f_{ab} = f_{abc} \phi^c\,.
\ee
The relevant coupling coefficients have a specific dependence on $\l$ dictated by the original $\l$-deformed action~\cite{Sfetsos:2013wia}.
Keeping a few terms, the action is~\cite{Georgiou:2019aon}
\be
\begin{split}
\label{sklinf}
 S_{k,\l} = {1\ov 2\pi} \int_S \text{d}^2\s\, &\Big(\del_+\phi^a\del_-\phi^a +{g_3\ov \sqrt{k}} f_{ab}\del_+\phi^a\del_-\phi^b + {g_4\ov k} f^2_{ab}\del_+\phi^a\del_-\phi^b
 \Big)+\cdots\ ,
\end{split}
\ee
where the couplings are $g_{3,4}$, assume the form
\be
\label{g34}
g_3= -{1\ov 3} {1+4\l+\l^2\ov \sqrt{(1-\l)(1+\l)^3}}\ ,\qquad g_4= {1\ov 12}{1+10\l+\l^2\ov 1-\l^2 }\ .
\ee
The couplings encode the symmetry of \eqref{sklinf} under
\be
\label{invv}
\l\to {1\ov \l} \ , \qquad k\to -k\ , \qquad \phi^a\to -\phi^a\ ,
\ee
which in fact originates from the symmetry of the $\l$-deformed action,
found in~\cite{Itsios:2014lca}. 
Note that, the above symmetry when acting inside square roots is realized as $k\to e^{i\pi}k$ and that $(1-\l)\to e^{i\pi}(\l^{-1}-1)$.
In this paper we will only need the cubic in the fields term with coupling $g_3$.

\subsection{Boundary conditions and computational conventions}

We would like to set up a perturbative expansion around the free theory paying particular attention to the appropriate
boundary condition they obey in conjunction with preservation of integrability. As before we take our boundary to be located at $\s=0$.

\no
Passing from the light-cone coordinates to the world-sheet time and space coordinates as in footnote~\ref{licon},
it is easy to see that in deriving the equations of motion for the action \eqref{sklinf} we get the boundary term
\be
\d \phi^a \Big(\del_\s \phi^a - {g_3\ov\sqrt{k}} f_{ab} \del_\tau \phi^b\Big)\Big |_{\s=0}\ ,
\ee
up to an overall numerical factor and where we keep only the leading interaction term in the action \eqref{sklinf}.
Vanishing of this term is achieved either by Dirichlet boundary conditions
\be
\label{dbc}
 \d\phi^a\big{|}_{\s=0}=0\  \Rightarrow\  \del_\tau\phi^a\big{|}_{\s=0} = 0 \ \ \Leftrightarrow\ \
(\del_++\del_-)\phi^a\big{|}_{\s=0} = 0 \ ,
\ee
for some directions
or by generalized Neumann boundary conditions
\be
\label{nbc}
 \del_\s\phi^a|_{\s=0} = {g_3\ov \sqrt{k}}  f_{ab} \del_\tau \phi^b\big |_{\s=0}
\ \ \Leftrightarrow\ \
\del_+\phi^a|_{\s=0} =  \del_-\phi^a|_{ \s=0} +  {2 g_3\ov \sqrt{k}}  f_{ab} \del_- \phi^b\big |_{\s=0}\ .
\ee
for the rest.
Both boundary conditions preserve the momentum no-flow condition across the boundary given in terms of the
energy momentum tensor by $T_{\tau\s} |_{\s=0}=  (T_{++}-T_{--})|_{\s=0}=0$.\footnote{The above considerations fall into the
general form of possible boundary conditions at $\del S$ for a general $\s$-model with target space coordinates $x^\m$, background metric
$G_{\m\n}$, antisymmetric tensor $B_{\m\n}$ given by
\begin{equation*}
\begin{split}
& {\rm Dirichlet}: \quad \del_\tau x^\m |_{ \del S} = 0\ ,
\\
&
{\rm (Generalized)\ Neumann}: \quad G_{\m\n}\del_\s x^\n |_{ \del S}
=  (B_{\m\n}+ 2\pi F_{\m\n})\del_\tau x^\n |_{ \del S}\ ,
\end{split}
\end{equation*}
where $F=\text{d}A$ is the field strength of the gauge field $A$ ending at some brane.
\label{foot.bc.gen}
} The above boundary conditions are certainly invariant under the symmetry \eqn{invv}.

\no
Among the above boundary conditions the integrable ones form a particularly interesting subset.
 For the case at hand these are given by a condition on the gauge fields $A_\pm$'s involved in the construction of
 the $\l$-deformed $\s$-model~\cite{Sfetsos:2013wia}.
 For a boundary located at $\s=0$, it reads
\be
\label{intbc}
A_+ \big{|}_{\s=0}=A_- \big{|}_{\s=0} \
\ee
and it was found in~\cite{Driezen:2018glg} (see Eqs. (2.8) \& (3.21)).\footnote{In fact a more general
integrability preserving boundary condition is allowed  in which \eqref{intbc} is replaced by
$A_+ |_{ \del S}=\Om A_- |_{ \del S}$, where $\Om$ is a constant Lie algebra inner automorphism
which satisfies $\Omega^2=\mathbb{1}$~\cite{Driezen:2018glg},
(see Eq. (3.22);  in the current work $\eta$ is taken to be the identity matrix.
In the limit $\l\to0$, the above boundary conditions reduce to
the current algebra preserving D-branes (of type-D) of the WZW model~\cite{Alekseev:1998mc,Felder:1999ka,Stanciu:1999id,Figueroa-OFarrill:1999cmq}, as well as those which do not preserve the current algebra (of type-N)~\cite{Stanciu:1999id}.
\label{foot.Omega}
}
Next we would like to investigate the possible boundary conditions for free fields under which the above is satisfied.
In terms of the normal coordinates \eqref{xphii} the expressions for the gauge fields are given by~\cite{Georgiou:2019aon} (see Eqs. (2.6) \& (3.1))
\be
\label{gauge.normal}
A_+ = -i {\l \ov \l e^f- \mathbb{1}}\,   {e^f-\mathbb{1}\ov f}\, \del_+x\ ,\qq
A_- =  -i {\l \ov  e^f- \l\mathbb{1}}\,  {e^f-\mathbb{1}\ov f}\, \del_-x\ ,
\ee
where the matrix $f$ is given in \eqref{fab} but with $x^a$'s in the place of $\phi^a$'s, see also \eqref{xphii}. It is easily seen that
the integrability preserving boundary condition \eqref{intbc} for \eqref{gauge.normal} gives rise to
\be
\frac{\text{e}^f-1}{f}\Big((1-\l)(\text{e}^f+\mathbb{1})\del_\tau x +(1+\l)(\text{e}^f-\mathbb{1})\del_\s x \Big)\Big{|}_{\s=0}=0\,.
\ee
Following~\cite{Alekseev:1998mc,Stanciu:1999id}, we can analyze this boundary condition in directions which stay intact (orthogonal) under the 
adjoint action of $G$ on the algebra, that is $DX^\bot=\text{e}^fX^\bot=X^\bot$, and their perpendicular complement (parallel). The former corresponds
 to Dirichlet boundary conditions and the latter to generalized Neumann ones, that is
\be
\del_\tau x^\bot\big{|}_{\s=0}=0\,,\quad \del_\tau x^\parallel\big{|}_{\s=0}=\frac{1+\l}{1-\l}\frac{\mathbb{1}-\text{e}^f}{\mathbb{1}+\text{e}^f}\del_\s x^\parallel\big{|}_{\s=0}\,.
\ee
Since the $x$'s are proportional to the free fields $\phi$'s -- see \eqref{xphii}, it will be technically beneficiary to have the same Dirichlet boundary
condition for all free fields. Indeed, this is achieved if we set all of them to zero at the boundary $\s=0$ which would consistent with the Dirichlet boundary condition \eqref{dbc}.

\no
 Let's turn now to the boundary condition \eqref{nbc}.
It is not clear that  it preserves integrability as it is not consistent with
\eqref{intbc}. Moreover, even if it can be cast in the more general form described in footnote
\ref{foot.Omega},  integrability issues have to be investigated from scratch.
Leaving  that aside, given \eqref{nbc} one expects that a purely Neumann boundary condition $\del_\s\phi^a\big{|}_{\s=0}$ cannot be imposed.
It turns out that this is indeed the case, for computations giving rise to order $\nicefrac1k$ results such as, the two-point function of currents, but as we will shorty explain it can be used to obtain results of order $\nicefrac{1}{\sqrt{k}}$, e.g. three-point function of currents.

\no
Passing now to the Euclidean regime \eqref{analy} we will use the notation
\be
\label{noott}
j_a(z) = \del \phi^a(z,\z)\ ,\qq \bar j_a(\z) = \bar\del \phi^a(z,\z)\ .
\ee
In order to perform field theory computations using Wick's theorem, we need to determine the basic two-point function
\be
G^{ab}(z,w)= \langle \phi^a(z,\z) \phi^b(w,\w)\rangle_\text{uhp}\  ,
\ee
where $z$ and $w$ are both located at the upper-half plane. The boundary conditions will be imposed at $\s=0$, which in the
Euclidean regime corresponds to $\z=z$. For Dirichlet boundary conditions \eqref{dbc}, we easily find that
\be
{\rm Dirichlet\ b.c.}:\qq  \phi^a(z,\z)|_{\z=z}\ \Rightarrow\ (\del +\bar \del)  G^{ab}  |_{ \z=z} = 0\ .
\ee
For the generalized Neumann boundary conditions we obtain that
\be
{\rm (Generalized)\ Neumann\ b.c.}:\qq   (\del -\bar \del)  G^{ab}  \big|_{ \z=z} ={2g_3\ov \sqrt{k}}\langle f_{ac}\bar \del \phi^c \phi^b
\rangle \big|_{ \z=z}  \ .
\ee
Let us estimate the order of magnitude of the right hand side term responsible for deviating from the standard Neumann boundary.
In order for this term to contribute to the indicated order $\nicefrac{1}{\sqrt{k}}$ the three-point function,
i.e. $\langle f_{ac}\bar \del \phi^c \phi^b
\rangle$, computed in the  $k\to \infty$ limit  has to be non-vanishing. However, since it concerns three Abelian bosons it is clearly zero.
Hence, this term is at best of order $\nicefrac1k$ and can be safely ignored to all computations resulting to order $\nicefrac{1}{\sqrt{k}}$, as are the current three-point functions we will compute using free field methods.

\no
Similarly to Subsection~\ref{subsection 2.2}, the solution to the two-point function with the above
two different boundary conditions can be easily presented in a unified way, including also the case in which the theory is
defined in the entire plane. It reads that\footnote{The two-point function \eqref{gab} is dictated  by varying the free part in the action \eqref{sklinf}. In particular,
it yields in the Euclidean regime \eqref{analy}
$$
\d{\cal L}=\frac{1}{\pi}\left(\d\phi^a\del\bar\del\phi^a-\del_t(\del_t\phi^a\d\phi^a)+\del_\s(\del_\s\phi^a\d\phi^a)\right)\,.
$$
The propagator $G_{ab}$ is read as usual by $\nicefrac{1}{\pi}\,\del\bar\del\, G_{ab}=-\d_{ab}\, \d^{(2)}(z-w)$, having the same
normalization as that in \eqn{olympi}, and it is subjected to the Dirichlet $\phi^a=0$ or to the
Neumann  $\del_\s\phi^a=0$ boundary conditions, at $\s=0$.
}
\be
\label{gab}
\langle \phi^a(z,\z) \phi^b(w,\w)\rangle_\text{uhp} = -\d^{ab}\Big(\ln {|z-w|^2
+\e\ln |z-\bar w|^2}\Big)\ ,
\ee
where, respectively
\be
\e=0 , + 1, -1 \ ,\quad {\rm for\ entire\ plane, \  Neumann\ b.c., Dirichlet\ b.c.}
\ee
We also have that
\be
\label{opeee3}
\begin{split}
&\langle j_a(z) \phi^b(w,\w)\rangle_\text{uhp} =
 -\d^{ab}\Big({1\ov z-w} + {\e\ov z- \bar w}\Big) \  ,
 \\
& \langle j_a(z) j_b(w) \rangle_\text{uhp}= -{\d^{ab}\ov (z-w)^2}\ ,
\\
&\langle \phi^a(z,\z) f_{bc}(w,\w)\rangle_\text{uhp} =-f_{abc}\Big(\ln {|z-w|^2 +\e\,
\ln |z-\bar w|^2}\Big)\ ,
 \\
 &\langle f_{ab}(z,\z)f_{cd}(w,\w)\rangle_\text{uhp} =-f_{abe}f_{cde}\Big(\ln {|z-w|^2 +\e\ln |z-\bar w|^2}\Big) \ ,
\\
&
\langle j_a(z) f_{bc}(w,\w)\rangle_\text{uhp} =-f_{abc}\Big({1\ov z-w} + {\e\ov z- \bar w}\Big) \ ,
\\
&
\langle j_a(z)\bar j_b(\bar w)\rangle_\text{uhp} =\d^{ab}\Big(C\, \d^{(2)}(z-w) - {\e\ov (z-\bar w)^2}
\Big)\ ,
\end{split}
\ee
where $C=\pi$.  We remind the reader that the $\d$-function term arises from the fact that
$\displaystyle \del {1\ov \z} = \bar \del {1\ov z}= \pi \d^{(2)}(z)$. We will keep $C$ as a parameter in the intermediate steps of our
computations in this section, so to keep track of $\d$-term contributions.
Note also that, we have ignored in the second line above the term $\d^{(2)}(z-\bar w)$ since
$z$ located at the upper-half plane and therefore cannot equal to $\w$, which located at the lower-half plane.

\subsection{Correlators with free fields }

In general CFT operators are built using the group element $g$. These operators are expected to get modification in the $\l$-deformed  theory as it was explicitly demonstrated in~\cite{Georgiou:2019jcf}. For the case of free currents
$j^a_\pm=\del_\pm \phi^a$ this dressing amounts to
simply replacing them by the gauge fields $A^a_\pm$ which have a non-trivial
$\l$-dependence and are also expressed in terms of free fields.
In order to have $\mp i  \del_\pm \phi^a$ as the leading term in their free field expansion, one rescales the $A^a_\pm$'s in \eqref{gauge.normal} by
a $\l$-dependent factor and denote them by ${\cal J}^a_\pm$,\footnote{Explicitly, ${\cal J}^a_\pm=-\frac{1}{\l}\sqrt{k(1-\l^2)}A^a_\pm$ with the $A^a_\pm$'s given in \eqref{gauge.normal}, see also~\cite{Georgiou:2019aon}.} in order to distinguish them from the CFT currents $J_\pm$.\footnote{
They are related as 
${\cal J}_\pm\big{|}_{\l=0}=iJ_\pm$, with $J_+=i\sqrt{k}\del_+ gg^{-1}$ and $J_-=-i\sqrt{k}g^{-1}\del_-g$.
In the  Euclidean regime \eqref{analy} the CFT currents $J_+$ and $J_-$ are mapped to 
$J=-\sqrt{k}\del gg^{-1}$ and $\bar J=\sqrt{k}g^{-1}\bar\del g$ respectively, which satisfy the operator product expansion \eqref{KM.OPE}.}

The ${\cal J}_\pm$'s have the following free field expansion~\cite{Georgiou:2019aon}
\be
\label{jjdre}
{\cal J}_\pm =\mp i  \Big(\mathbb{1} \pm  {h_1\ov  \sqrt{k}} f  +\cdots
\Big)\del_\pm\phi\ ,\qquad h_1= \ha \sqrt{1+\l\ov 1-\l}\ ,
\ee
where we have kept only the relevant for this paper terms.
In the Euclidean regime the notation will be $\cal J$ and $\bar {\cal  J}$ in place of
$\cal J_+$ and $ \cal  J_-$, respectively. Taking into account \eqref{analy} we have that
\be
\label{jjdre2}
{\cal J} =\Big(\mathbb{1} +  {h_1\ov  \sqrt{k}} f  +\cdots \Big)\del\phi\ ,
\qquad \bar{\cal J} =-\Big(\mathbb{1} -  {h_1\ov  \sqrt{k}} f  +\cdots \Big)\bar\del\phi\ .
\ee
Let us note that the above dressed currents and the real $f_{abc}$'s are equal to the non-Abelian currents
and the imaginary $f_{abc}$'s in \eqref{KM.OPE}, as
\be
\label{map.sec23.to.sec4}
{\cal J}\big{|}_{\l=0}= i J\,,\quad \bar {\cal J}\big{|}_{\l=0}= i \bar J\,,\quad f_{abc}|_{\rm real} = -i f_{abc}|_{\rm Im}\,.
\ee
These dressed currents obey
\be
{\rm Dirichlet\ b.c. }:\qquad  {\cal J} -   \bar{\cal J} \big |_{\z=z} = 0\ ,
\ee
for Dirichlet boundary conditions, corresponding to the case with $\e=-1$ for the free field propagator \eqref{gab}.
This is precisely the boundary condition \eqref{KM.bnr} for the full non-Abelian currents.

\no
To write an analogous expression for generalized Neumann boundary conditions
we will need another dressed current defined as
\be
\label{jjdre22}
  \bar{\cal J}^\prime=-\Big(\mathbb{1} -  {h^\prime_1\ov  \sqrt{k}} f  +\cdots \Big)\bar\del\phi\ , \qq
 h^\prime_1 = - h_1- 2 g_3= \frac{1+10\l+\l^2}{6\sqrt{(1-\l)(1+\l)^3}}\,.
\ee
Then we have that
\be
\label{nlbc}
{\rm Generalized\ Neumann\ b.c. }:\qquad {\cal J} +   \bar{\cal J}^\prime \big |_{\z=z} = 0\ ,
\ee
where \eqn{nbc} and \eqref{analy} have been also used. 
This condition will be used as a consistency check for the three-point correlation functions that we will compute below.

\subsubsection{The three-point function $\langle{\cal J} {\cal J}{\cal J}\rangle$}

In this section, we will work out the  three-point function $\langle{\cal J} {\cal J}{\cal J}\rangle$ to the
leading result ${\cal O}(\nicefrac{1}{\sqrt{k}})$ and show that it takes the form
\be
\boxed{
\label{JJJ1}
\begin{split}
&\langle{\cal J}_a(x_1,\bar x_1){\cal J}_b(x_2,\bar x_2){\cal J}_c(x_3,\bar x_3)\rangle^{(\l),\text{exact}}_{\text{uhp}}
 = {3\a\ov \sqrt{k}}\, { f_{abc}\ov x_{12} x_{13} x_{23}}
\\
&\qquad
+ \e\, {\b \ov \sqrt{k}}\, f_{abc} \bigg({1\ov x_{23}^2 (x_1-\bar x_2)} -
{1\ov x_{13}^2 (x_2-\bar x_1)}  +\text{cyclic in 1,2,3}\bigg)\ ,
\end{split}
}
\ee
where the coefficients are given by
\be
\label{JJJ1.pre}
\begin{split}
& \a = h_1+\frac{C}{2\pi}\,g_3=
h_1+\frac{g_3}{2}={1\ov 3} \frac{1+\l+\l^2}{\sqrt{(1-\l)(1+\l)^3}}\ ,
\\
&
\b = \a + g_3 =  -  {\l\ov \sqrt{(1-\l)(1+\l)^3}}\ .
\end{split}
\ee
This result is precisely the same as in \eqref{JJJe}, taking also \eqref{map.sec23.to.sec4} into account,
for Dirichlet boundary conditions corresponding to $\e=-1$. In addition,
for $\e=0$ our result coincides with the full plane correlator found in Eq.(3.29) of~\cite{Georgiou:2016iom}, after taking into account  \eqref{map.sec23.to.sec4} and \eqref{JJJ1.pre}.

\no
In what follows, we shall work out the various contributions leading to \eqref{JJJ1}.
For the  $\langle{\cal J} {\cal J}{\cal J}\rangle$ correlator we have that
\be
\small
\langle {\cal J}_a(x_1,\bar x_1) {\cal J}_b(x_2,\bar x_2)  {\cal J}_c(x_3,\bar x_3)\rangle^{(\l),\text{exact}}_{\text{uhp}}
=\langle {\cal J}_a(x_1,\bar x_1) {\cal J}_b(x_2,\bar x_2)  {\cal J}_c(x_3,\bar x_3)
e^{-S_{\rm int}}\rangle_\text{uhp}\ ,
\ee
where the interaction terms in the Euclidean regime can be read from \eqref{sklinf} to be
\be
S_{\text int} =   {g_3\ov 2\pi \sqrt{k}} \int_S \text{d}^2z\, j_af_{ab}\bar j_b
+{\cal O}\left(\frac{1}{k}\right)\ .
\ee
Then, expanding the exponential and keeping terms up to ${\cal O}(\nicefrac{1}{\sqrt{k}})$ we have that
\be
\label{jjj333}
\begin{split}
&\langle {\cal J}_a(x_1,\bar x_1) {\cal J}_b(x_2,\bar x_2)  {\cal J}_c(x_3,\bar x_3)\rangle^{(\l),\text{exact}}_{\text{uhp}}
\\
&
\quad\ ={h_1\ov \sqrt{k}} \langle f_{aa_1}(x_1,\bar x_1) j_{a_1}(x_1) j_b(x_2) j_c(x_3) \rangle_\text{uhp}
+ \big[{\rm cyclic\ in}\ (x_1,a), (x_2,b), (x_3,c)\big]
\\
&  \quad\quad\  -{g_3\ov 2\pi \sqrt{k}}\int_S \text{d}^2z\
\langle j_a(x_1) j_b(x_2) j_c(x_3) j_{a_1}(z) f_{a_1b_1}(z,\z)
\bar j_{b_1}(\z) \rangle_\text{uhp}
\\
& \quad\ ={1\ov \sqrt{k}} \Big(h_1  + { C\ov 2\pi}g_3\Big) \langle f_{aa_1}(x_1,\bar x_1) j_{a_1}(x_1) j_b(x_2) j_c(x_3)\rangle_\text{uhp}
\\
&
\quad\quad\  - \e\, {g_3 \ov 2\pi \sqrt{k}}\int_S {\text{d}^2z\ov (\bar z - x_1)^2} \langle f_{aa_1}(z,\bar z) j_{a_1}(z) j_b(x_2) j_c(x_3)\rangle_\text{uhp}
\\
&
\qq\qq\qq\qq + \big[{\rm cyclic\ in}\ (x_1,a), (x_2,b), (x_3,c)\big] \ .
\end{split}
\ee
Explicitly, the necessary four-point function is given by
\be
\small
\label{EGG1}
 \langle f_{aa_1}(x_1,\bar x_1) j_{a_1}(x_1) j_b(x_2) j_c(x_3) \rangle_\text{uhp}=
 -{f_{abc}\ov x_{12}^2}\bigg({1\ov x_{13}}+{\e\ov \bar x_1 - x_3}\bigg)
 +{f_{abc}\ov x_{13}^2}\bigg({1\ov x_{12}}+{\e\ov \bar x_1 - x_2}\bigg)\ .
\ee
Then we have that
\be
\begin{split}
 & \langle f_{aa_1}(x_1,\bar x_1) j_{a_1}(x_1) j_b(x_2) j_c(x_3)\rangle_\text{uhp}
 + \big[{\rm cyclic\ in}\ (x_1,a), (x_2,b), (x_3,c)\big]
 \\
 &
\qq\qq ={3 f_{abc}\ov x_{12} x_{13} x_{23}}
 + \e f_{abc} \Bigg({1\ov x_{12}^2} \bigg({1\ov \bar x_2 - x_3}-  {1\ov \bar x_1 - x_3}\bigg)
 \\
 &\qq\qq\quad
  +{1\ov x_{23}^2}  \bigg({1\ov \bar x_3 - x_1}-  {1\ov \bar x_2 - x_1}  \bigg)
  +{1\ov x_{31}^2}  \bigg({1\ov \bar x_1 - x_2}-  {1\ov \bar x_3 - x_2}  \bigg)\Bigg)
 \ .
\end{split}
\ee
The remaining integral in \eqref{jjj333} is
\be
\label{jjj3331}
\begin{split}
&\int_S {\text{d}^2 z\ov (\bar z - x_1)^2} \langle f_{aa_1}(z,\bar z) j_{a_1}(z) j_b(x_2) j_c(x_3)\rangle_\text{uhp}
+ \big[{\rm cyclic\ in}\ (x_1,a), (x_2,b), (x_3,c)\big]
\\
& \qq
= f_{abc} \Big(-I_{123} + I_{132}- I_{231}+ I_{213}- I_{312} + I_{321}\Big)
\\
& \qq\quad + \e f_{abc}  \Big(-J_{123} + J_{132}- J_{231}+ J_{213}- J_{312} + J_{321}\Big)\ ,
\end{split}
\ee
where we have defined the integrals
\be
\label{int123}
I_{123}= \int_S {\text{d}^2z\ov (\z-x_1)^2(z-x_2)^2 (z-x_3)}\ ,\quad J_{123}= \int_S {\text{d}^2z\ov (\z-x_1)^2(z-x_2)^2 (\z-x_3)}\ .
\ee
We have evaluated them in~\ref{I123.appendix} and \eqref{J123.appendix}. We recall the result here for the reader's convenience
\be
\label{int123res}
I_{123}=\frac{\pi}{x_{23}^2}\left(\frac{1}{x_1-\bar x_2}-\frac{1}{x_1-\bar x_3}\right)\,,
\quad J_{123}=0\,.
\ee
Upon substitution and after some rearrangements the three-point function for the ${\cal J}$'s equals to \eqref{JJJ1}.

\subsubsection{The three-point function $\langle {\cal J} {\cal J}\bar {\cal J}\rangle $}

Similarly, for the mixed chirality correlator to leading result ${\cal O}(\nicefrac{1}{\sqrt{k}})$ the end result reads
\be
\boxed{
\label{JJJbar.correl}
\begin{split}
&\langle {\cal J}_a(x_1,\bar x_1) {\cal J}_b(x_2,\bar x_2)
\bar{\cal J}_c(x_3,\bar x_3\rangle^{(\l),\text{exact}}_{\text{uhp}}=-\frac{\b}{\sqrt{k}}f_{abc}\times\\
&\left(\frac{\bar x_{12}}{x_{12}^2\bar x_{13}\bar x_{23}}+\frac{\e^2}{(x_1-\bar x_3)^2}\left(\frac{1}{\bar x_1-x_2}-\frac{\e}{x_{23}}\right)+
\frac{\e^2}{(x_2-\bar x_3)^2}\left(\frac{1}{x_1-\bar x_2}+\frac{\e}{x_{13}}\right)\right)\\
&+\frac{\a}{\sqrt{k}} f_{abc}\left(-\frac{\e}{x_{12}(x_1-\bar x_3)(x_2-\bar x_3)}
\right.\\
&\left.+\frac{\e^2}{(x_1-\bar x_3)^2}\left(\frac{1}{x_2-\bar x_3}-\frac{\e}{x_{12}}\right)
-\frac{\e^2}{(x_2-\bar x_3)^2}\left(\frac{1}{x_1-\bar x_3}+\frac{\e}{x_{12}}\right)
\right)\ ,
\end{split}}
\ee
where the coefficients $\a,\b$ can be found in \eqref{JJJ1.pre}.

\no
In what follows, we shall work out the various contributions leading to \eqref{JJJbar.correl}.

\no
Keeping only those terms that potentially contribute to the correlator
up to ${\cal O}(\nicefrac{1}{\sqrt{k}})$, we have that
\be
\small
 \label{corr1}
\begin{split}
&\langle {\cal J}_a(x_1,\bar x_1) {\cal J}_b(x_2,\bar x_2)
\bar{\cal J}_c(x_3,\bar x_3)\rangle^{(\l),\text{exact}}_{\text{uhp}}
=\langle {\cal J}_a(x_1,\bar x_1) {\cal J}_b(x_2,\bar x_2)
\bar{\cal J}_c(x_3,\bar x_3) e^{-S_{\rm int}}\rangle_\text{uhp}
\\
&\quad =-{h_1\ov \sqrt{k}} \langle f_{aa_1}(x_1,\bar x_1) j_{a_1}(x_1) j_b(x_2) \bar j_c(\bar x_3) \rangle_\text{uhp}
+\big[ (x_1,a) \leftrightarrow (x_2,b)\big]
\\
&\qq\ +{h_1\ov \sqrt{k}} \langle f_{c c_1}(x_3,\bar x_3) j_{a}(x_1) j_b(x_2) \bar j_{c_1}(\bar x_3) \rangle_\text{uhp}
\\
 & \qq\  +{g_3\ov 2\pi \sqrt{k}}\int_S \text{d}^2z\
\langle j_a(x_1) j_b(x_2) \bar j_c(\bar x_3) j_{a_1}(z) f_{a_1b_1}(z,\z) \bar j_{b_1}(\z) \rangle_\text{uhp}
\\
&
 \quad = -
 {1\ov \sqrt{k}} \Big(h_1  + { C\ov 2\pi}g_3\Big) \langle f_{aa_1}(x_1,\bar x_1) j_{a_1}(x_1) j_b(x_2) \bar j_c(\bar x_3) \rangle_\text{uhp}
 +\big[ (x_1,a) \leftrightarrow (x_2,b)\big]
 \\
&
\qquad\ +{h_1\ov \sqrt{k}} \langle f_{c c_1}(x_3,\bar x_3) j_{a}(x_1) j_b(x_2) \bar j_{c_1}(\bar x_3) \rangle_\text{uhp}
 \\
 &\qquad\ + {g_3\ov 2\pi \sqrt{k}}\int_S  {\text{d}^2z\ov (\z-\bar x_3)^2}
 \langle f_{ca_1}(z,\z) j_{a_1}(z) j_a(x_1)  j_b( x_2) \rangle_\text{uhp}
 \\
 & \qquad\ +  {\e\, g_3\ov 2\pi \sqrt{k}} \int_S  {\text{d}^2z\ov (\z- x_1)^2}
 \langle f_{aa_1}(z,\z) j_{a_1}(z) j_b(x_2)  \bar j_c(\bar x_3) \rangle_\text{uhp}  +\big[ (x_1,a) \leftrightarrow (x_2,b)\big]\ .
\end{split}
\ee
We need the correlator
\be
\label{EGG0}
\begin{split}
& \langle f_{aa_1}(x_1,\bar x_1) j_{a_1}(x_1) j_b(x_2) \bar j_c(\bar x_3) \rangle_\text{uhp}=
- f_{abc} \Bigg({1\ov x_{12}^2\bar x_{13} } +  {C\, \d^{(2)}(x_{13}) \ov x_{12}}
\\
&\
+ \e\Big({1\ov x_{12}^2 (x_1-\bar x_3)} - {1\ov x_{12} (x_1-\bar x_3)^2} +{C\, \d^{(2)}(x_{13}) \ov \bar x_1 -x_2} \bigg)
- {\e^2 \ov (\bar x_1-x_2)(x_1-\bar x_3)^2}\Bigg)\ ,
\end{split}
\ee
where we have kept the $\d$-function terms since in the last integral above may contribute.
We also need the correlator
\be
\begin{split}
& \langle f_{cc_1}(x_3,\bar x_3) j_{a}(x_1) j_b(x_2)  \bar j_{c_1}(\bar x_3) \rangle_\text{uhp}
\\
& \qq =
-{f_{abc}\ov (x_1-\bar x_3)^2}\Big({1\ov x_{23}}+{\e\ov x_2-\bar x_3}\Big)\Big(C \d^{(2)}(x_{13}) -{\e\ov (x_1-\bar x_3)^2}\Big)
-(x_1 \leftrightarrow x_2)
\\
&\qq =  f_{abc}\Big({\e\ov  (x_1-\bar x_3)^2 x_{23}} - {\e\ov  (x_2-\bar x_3)^2 x_{13}}
\\
&\qq\qq\qq\quad + {\e^2\ov  (x_1-\bar x_3)^2 (x_2-\bar x_3)} - {\e^2\ov  (x_2-\bar x_3)^2 (x_1-\bar x_3)}   \Big)\ ,
 \end{split}
 \ee
 where in the second equality we have neglected contact terms since they will not contribute.
Finally, we need the correlator
\be
\begin{split}
 \langle f_{ca_1}(z,\z) j_{a_1}(z) j_a(x_1)  & j_b( x_2) \rangle_\text{uhp}  = f_{abc}  \Bigg(  {1\ov (z-x_1)(z-x_2)^2}
 -{1\ov (z-x_2)(z-x_1)^2}
 \\
 &
  + \e \bigg({1\ov (\z- x_1)(z-x_2)^2}-{1\ov( \z- x_2)(z-x_1)^2} \bigg)\Bigg)\ ,
\end{split}
\ee
which is the same as \eqref{EGG1} with the necessary relabeling for the points.

\no
Using the above, the first integral  in \eqref{corr1} above can be written as
\be
\int_S  {\text{d}^2z\ov (\z-\bar x_3)^2}
 \langle f_{ca_1}(z,\z) j_{a_1}(z) j_a(x_1)  j_b( x_2) \rangle_\text{uhp} = f_{abc} \big(A_{123}-A_{213}+ \e\, (B_{123}-B_{213})\big)\ ,
\ee
whereas the second integral in \eqref{corr1} as
\be
\begin{split}
& \int_S {\text{d}^2z\ov (\z- x_1)^2}
 \langle f_{aa_1}(z,\z) j_{a_1}(z) j_b(x_2)  \bar j_c(\bar x_3) \rangle_\text{uhp}  +\big[ (x_1,a) \leftrightarrow (x_2,b)\big]
\\
&
\quad = -f_{abc}\big(C_{123}-C_{213}-{C\ov (x_1-\bar x_3)^2 x_{23}} + {C\ov (x_2-\bar x_3)^2 x_{13}}\Big)
\\
&\qq  -\e\, f_{abc} \Big(D_{123}-D_{213}-E_{123}+E_{213}
\\
&\qq\qq\qquad - {C\ov (x_1-\bar x_3)^2 (x_2-\bar x_3)} + {C\ov (x_2-\bar x_3)^2 (x_1-\bar x_3)} \Big)
\\
&\qq + \e^2 f_{abc} (F_{123}-F_{213})\ ,
\end{split}
\ee
where the various  integrals are defined as
\be
\begin{split}
\label{intABCDF123}
&
A_{123}= \int_S {\text{d}^2z\ov (z-x_1)(z-x_2)^2 (\z-\bar x_3)^2}\, ,
\quad B_{123}= \int_S {\text{d}^2z\ov (\z-x_1)(z-x_2)^2 (\z-\bar x_3)^2}\ ,
\\
& C_{123}= \int_S {\text{d}^2z\ov (\z-x_1)^2(z-x_2)^2 (\z-\bar x_3)}\, ,\quad
D_{123}= \int_S {\text{d}^2z\ov (\z-x_1)^2(z-x_2)^2 (z-\bar x_3)}\ ,
\\
& E_{123}= \int_S {\text{d}^2z\ov (\z-x_1)^2(z-x_2) (z-\bar x_3)^2}\, ,\quad
F_{123}= \int_S {\text{d}^2z\ov (\z-x_1)^2(\z-x_2) (z-\bar x_3)^2}\, ,
\end{split}
\ee
whose value is given in \eqref{JJJbar.int}. Summing all the above contributions we find \eqref{JJJbar.correl}.

\no
For the whole plane $\e=0$, \eqref{JJJbar.correl} simplifies to
\be
\langle {\cal J}_a(x_1,\bar x_1) {\cal J}_b(x_2,\bar x_2)
\bar{\cal J}_c(x_3,\bar x_3)\rangle_{\mathbb{R}^2}^{(\l),\text{exact}}=-\frac{\b}{\sqrt{k}}f_{abc}\,\frac{\bar x_{12}}{x_{12}^2\bar x_{13}\bar x_{23}}
\ee
and coincides with the whole plane result in Eq.(3.33) of~\cite{Georgiou:2016iom}, after taking into account 
\eqref{map.sec23.to.sec4} and \eqref{JJJ1.pre}.

\no
For Dirichlet boundary conditions $\e=-1$, \eqref{JJJbar.correl} simplifies to
\be
\label{JJJbar1}
\begin{split}
&\langle {\cal J}_a(x_1,\bar x_1) {\cal J}_b(x_2,\bar x_2)
\bar{\cal J}_c(x_3,\bar x_3)\rangle^{(\l),\text{exact}}_{\text{uhp}}=-\frac{\b}{\sqrt{k}}f_{abc}\times\\
&\left(\frac{\bar x_{12}}{x_{12}^2\bar x_{13}\bar x_{23}}+\frac{1}{(x_1-\bar x_3)^2}\left(\frac{1}{\bar x_1-x_2}+\frac{1}{x_{23}}\right)+
\frac{1}{(x_2-\bar x_3)^2}\left(\frac{1}{x_1-\bar x_2}-\frac{1}{x_{13}}\right)\right)\\
&+\frac{3\a}{\sqrt{k}} f_{abc}\frac{1}{x_{12}(x_1-\bar x_3)(x_2-\bar x_3)}\,,
\end{split}
\ee
which coincides with that in \eqref{jjjb}, after taking into account \eqref{JJJ1.pre} and the map \eqref{map.sec23.to.sec4}.

\no
For Neumann boundary conditions $\e=1$, \eqref{JJJbar.correl} simplifies to
\be
\label{nngf}
\begin{split}
&\langle {\cal J}_a(x_1,\bar x_1) {\cal J}_b(x_2,\bar x_2)
\bar{\cal J}_c(x_3,\bar x_3)\rangle^{(\l),\text{exact}}_{\text{uhp}}=-\frac{\b}{\sqrt{k}}f_{abc}\times\\
&\ \left(\frac{\bar x_{12}}{x_{12}^2\bar x_{13}\bar x_{23}}+\frac{1}{(x_1-\bar x_3)^2}\left(\frac{1}{\bar x_1-x_2}-\frac{1}{x_{23}}\right)+
\frac{1}{(x_2-\bar x_3)^2}\left(\frac{1}{x_1-\bar x_2}+\frac{1}{x_{13}}\right)\right)\\
&\ -\frac{3\a}{\sqrt{k}} f_{abc}\left(\frac{1}{x_{12}(x_1-\bar x_3)(x_2-\bar x_3)}+\frac{2x_{12}}{3(x_1-\bar x_3)^2(x_2-\bar x_3)^2}\right)\,.
\end{split}
\ee

\no
We may independently compute the correlator \eqn{JJJbar.correl} with $\bar {\cal J}$ replaced by  $\bar {\cal J^\prime}$
given in terms of free fields by \eqref{jjdre22}. The end result is
\be
\small
\label{nngfk}
\begin{split}
&\langle {\cal J}_a(x_1,\bar x_1) {\cal J}_b(x_2,\bar x_2)
\bar{\cal J}^\prime_c(x_3,\bar x_3)\rangle^{(\l),\text{exact}}_{\text{uhp}}= -   \frac{\b}{\sqrt{k}}f_{abc}
\times
\\
&\ \left(\frac{\bar x_{12}}{x_{12}^2\bar x_{13}\bar x_{23}}+\frac{\e^2}{(x_1-\bar x_3)^2}\left(\frac{1}{\bar x_1-x_2}-\frac{\e}{x_{23}}\right)+
\frac{\e^2}{(x_2-\bar x_3)^2}\left(\frac{1}{x_1-\bar x_2}+\frac{\e}{x_{13}}\right)\right)\\
&\ +\frac{\a}{\sqrt{k}} f_{abc}\left(-\frac{\e}{x_{12}(x_1-\bar x_3)(x_2-\bar x_3)}
\right.\\
&\ \left.+\frac{\e^2}{(x_1-\bar x_3)^2}\left(\frac{1}{x_2-\bar x_3}-\frac{\e}{x_{12}}\right)
-\frac{\e^2}{(x_2-\bar x_3)^2}\left(\frac{1}{x_1-\bar x_3}+\frac{\e}{x_{12}}\right)
\right)
\\
&
-2\e\frac{h_1+g_3}{\sqrt{k}}f_{abc}\left(\frac{1}{(x_1-\bar x_3)^2}\left(\frac{1}{x_{23}}+\frac{\e}{x_2-\bar x_3}\right)-
\frac{1}{(x_2-\bar x_3)^2}\left(\frac{1}{x_{13}}+\frac{\e}{x_1-\bar x_3}\right)\right)\, ,
\end{split}
\ee
where
\be
h_1+g_3=\frac{1}{6}\bigg(\frac{1-\l}{1+\l}\bigg)^{\!\!\nicefrac32} .
\ee
Then from \eqn{nngfk} with $\e=1$ and \eqref{JJJ1} we may easily check the consistency relation
\be
\label{bc.JJJ.JJJbar2}
{\big\langle} {\cal J}_a(x_1,\bar x_1){\cal J}_b(x_2,\bar x_2)
\Big({\cal J}_c(x_3,\bar x_3)+\bar{\cal J}^\prime_c(x_3,\bar x_3)\Big)\big\rangle^{(\l),\text{exact}}_{\text{uhp}}\Big{|}_{\bar x_3= x_3}=0\, .
\ee

\subsection{Anomalous dimensions of primary fields}

In this subsection we will compute one- and two-point functions of primary operators using free fields with  Dirichlet boundary conditions so that we may directly compare with the results of Subsection~\ref{pr3.4}.
The discussion will be similar to that in Section 3.3 of~\cite{Georgiou:2019aon}, adapted  to the upper-half plane.
Consider the field $D_{ab}$ and its free field expansion
\be
D_{ab}=\text{Tr}(t_agt_bg^{-1})=\d_{ab}+\frac{1}{\sqrt{k}}\sqrt{{1-\l\ov 1+\l}}f_{ab}+\frac{1}{2k}{1-\l\ov 1+\l}f^2_{ab}+\cdots
\ee
We would like to evaluate its one- and two-point correlation function. Starting with the one-point function, it can be easily seen
 that to ${\cal O}(\nicefrac1k)$  the path integral insertions have either vanishing or bubble diagram contribution.
 As a result we are left with
\be
\label{free.onepoint.D}
\begin{split}
\langle D_{ab}(x,\bar x)\rangle^{(\l),\text{exact}}_{\text{uhp}}&=\d_{ab}+\frac{1}{2k}{1-\l\ov 1+\l}\langle f^2_{ab}(x,\bar x)\rangle_\text{uhp}\\
&=\d_{ab}\bigg(1+\frac{c_G}{2k}\frac{1-\l}{1+\l}\ln{-\varepsilon^2\ov (x-\bar x)^2}\bigg)\,.
\end{split}
\ee
Similarly, we can evaluate the one-point function of the group element $g_{ij}(x,\bar x)$
which is a primary field in an irreducible representation $R$, with Hermitian matrices $t_a$.
Using \eqref{xphii} to obtain its free field expansion we find that
\be
\label{free.onepoint.g}
\begin{split}
\langle g_{ij}(x,\bar x)\rangle^{(\l),\text{exact}}_{\text{uhp}}&=\d_{ij}-\frac{1}{2k}{1-\l\ov 1+\l}\left(t_at_b\right)_{ij}
\langle\phi^a(x,\bar x)\phi^b(x,\bar x)\rangle_\text{uhp}\\
&=\d_{ij}\bigg(1+\frac{c_R}{2k}\frac{1-\l}{1+\l}\ln{-\varepsilon^2\ov (x-\bar x)^2}\bigg)\,.
\end{split}
\ee
Hence, we can read out of \eqref{free.onepoint.D} and \eqref{free.onepoint.g} the corresponding anomalous dimensions
\be
\label{gamma.D.g}
\g_D=\frac{c_G}{k}\frac{1-\l}{1+\l}\,,\quad \g_g=\frac{c_R}{k}\frac{1-\l}{1+\l}\,,
\ee
which are in agreement with \eqref{anomalous.Phi.exact} and of course consistent, 
since $D_{ab}$ belongs to the adjoint representation for which $c_R=c_G$.

\no
Similarly, we can evaluate the two-point function for $D_{ab}$ yielding to order $\nicefrac1k$
\be
\begin{split}
\langle D_{ac}(x_1,\bar x_1) D_{b c}(x_2,\bar x_2)\rangle^{(\l),\text{exact}}_{\text{uhp}}
= \d_{ab}\bigg(1 +\frac{c_G}{k}\frac{1-\l}{1+\l}\, \ln\frac{\varepsilon^2 (1-\xi)}{|x_{12}|^2}\bigg)\ ,
\end{split}
\ee
where $\xi$ is the invariant ratio defined in \eqref{comparison.xi}, as well as that for $g_{ij}$
\be
\langle g_{im}(x_1,\bar x_1) g^{-1}_{mj}(x_2,\bar x_2)\rangle^{(\l),\text{exact}}_{\text{uhp}}
= \d_{ij}\bigg( 1 +\frac{c_R}{k}\frac{1-\l}{1+\l}\, \ln\frac{\varepsilon^2 (1-\xi)}{|x_{12}|^2}\bigg)\ ,
\ee
from which we read off  the same anomalous dimensions as above.
Finally, we note that for $\l=0$ and to order $\nicefrac1k$, the above two-point correlation functions are in 
agreement with the generic expression \eqref{2pt.Cardy.double} where
\be
\label{2pt.Prime.Polya}
h_{1,2}=\bar h_{1,2}=h=\frac{c_R}{2k+c_G}\,,\quad F(\xi)=\varepsilon^{4h}(1-\xi)^{4 h}\,,
\ee
where the invariant ratio $\xi$ was given in \eqref{comparison.xi}. 
It is also in agreement with \eqref{Cardy.Osborn}, where
\begin{equation}
\psi(\zeta)=\frac{\varepsilon^{2\D}}{4^{\D}}\bigg(\frac{\zeta+2}{\zeta-2}\bigg)^{\D}\,,\quad \zeta=2(1-2\xi)\geqslant2\,,\quad \D_{1,2}=\D=\frac{2c_R}{2k+c_G}\,,
\end{equation}
which is also consistent with \eqref{2pt.gen.rel} and \eqref{2pt.Prime.Polya}.

\section{Concluding remarks}
\label{concl}

In the present work we studied quantum aspects of $\l$-deformed models in spaces with boundaries
in particular, the model of~\cite{Sfetsos:2013wia} in the upper-half plane.
For Dirichlet type of boundary conditions, preserving the current algebra at the conformal point~\cite{Alekseev:1998mc,Felder:1999ka,Stanciu:1999id,Figueroa-OFarrill:1999cmq} and  the integrability away from it~\cite{Driezen:2019ykp},
we computed exactly in $\l$ and leading order in $\nicefrac1k$,
one-point correlation function of affine primaries, two-point functions of currents and composite current-bilinear and
three-point functions of currents using low order conformal perturbation theory based on current algebras and Cardy's doubling trick, in association with non-trivial symmetries in the coupling space of the models and meromorphicity arguments. Moreover, using standard QFT techniques based on free fields we arrived at the same results and in addition we were able to consider mixed boundary conditions which do not necessarily preserve integrability.
The correlation functions we computed have a rich structure.
We presented our calculations, in particular those involving delicate integrations in the upper-half plane, in full detail, which will be
useful in further related investigations.

The results of this work should be extendable via a conformal mapping to other geometries with boundaries, provided that they share the same current algebra preserving boundary conditions.
It will be interesting to study correlation functions involving primary fields, using conformal perturbation on the upper-half plane, beyond the one-point correlation functions at one-loop order presented in the current work. Combining low order conformal perturbation with meromorphicity arguments and the non-perturbative symmetry on $(\l,k)$ should yield the same correlation functions as these can be found independently using free field techniques on the upper-half plane. In addition, the derived anomalous dimensions should match the full plane result since it is determined by the short-distance behavior, where in the upper-half plane the boundary has no effect (see footnote~\ref{annulus}).

It is also very interesting to consider integrable deformations of coset CFTs on the upper-half plane.
A priori, we still have the two approaches at our disposal, namely conformal perturbation and that using free fields. A posteriori, it is way more difficult to use conformal perturbation when the underlying CFT is a coset one. The reason is that in such CFTs, the building blocks are parafermions which have more complicated operator product expansions than currents and as a result they contain Wilson-like phases in their expressions in terms of target space fields. However, we can still use the free field expansion as it was done in~\cite{Georgiou:2020bpx} for the full plane. In this case one might expect the anomalous dimension of the single parafermion may still stay intact as it is governed by the short-distance behavior. However, the non-local phase played a crucial r\^ole in determining the anomalous dimension of the parafermion in~\cite{Georgiou:2020bpx}. Hence, the effect of the boundary could be significant in the anomalous dimension of the parafermion and a detailed computation should be
done.

A potential extension of the current work is to consider $\l$-deformations of currents algebras
at unequal levels~\cite{Georgiou:2017jfi}. This class of models smoothly
interpolates between a product of current algebras at levels $k_1$ and $k_2$ in the UV towards a product of
current algebras or coset CFTs in the IR~\cite{Georgiou:2017jfi}. To study this class of models on the upper-half plane
we can either use conformal perturbation or the usual QFT perturbation, 
based on free fields adapted to the appropriate boundary conditions.
In the CFT approach, the set of boundary conditions which preserve the current algebras are again given by \eqref{KM.bnr} but for each of the copies separately.
In terms of a D-brane world-volume point of view, this boundary condition simply describes two {\it identical copies} of the conjugacy classes of the group $G$~\cite{Alekseev:1998mc,Felder:1999ka,Stanciu:1999id,Figueroa-OFarrill:1999cmq}. 

A further possible extension includes studying Yang--Baxter deformations of principal chiral models constructed in~\cite{Klimcik:2002zj,Klimcik:2008eq}.
This class of models are related to the $\l$-deformed ones via Poisson--Lie T-duality and analytic continuation of the coordinates and
parameters of the $\s$-model~\cite{Vicedo:2015pna,Hoare:2015gda,Sfetsos:2015nya,Klimcik:2015gba,Klimcik:2016rov}.
In these modes there is no conformal point in contrast to the $\l$-ones. Nevertheless, in studying the Yang--Baxter models on the upper-half plane 
we can still use QFT techniques based on free fields. This study will include deriving their $\b$-function, correlation functions and anomalous dimensions starting with the case with no boundaries and extending it in its presence. Concerning correlation functions, a natural choice for $\eta$-dressed fields to be pursued are the ones appearing in the Lax connection of the Yang--Baxter models~\cite{Klimcik:2008eq}.

\subsection*{Acknowledgements}

We would like to thank C.~Bachas for useful discussions on boundary conditions and S.~Driezen for a useful correspondence related to \cite{Driezen:2018glg}.\\
The research work of K.~Sfetsos was supported by the Hellenic Foundation for
Research and Innovation (H.F.R.I.) under the ``First Call for H.F.R.I.
Research Projects to support Faculty members and Researchers and
the procurement of high-cost research equipment grant'' (MIS 1857, Project Number: 16519).\\
The research work of K.~Siampos has received funding from the Hellenic Foundation
for Research and Innovation (H.F.R.I.) and the General Secretariat for Research and Technology (G.S.R.T.),
under grant agreement No 15425.

\appendix

\section{A quiver of integrals in the upper-half plane}
\label{integrals.quiver}

In this Appendix we present the various integrals appearing on the main text of the current work which we group them
accordingly. The domain of integration is the  half-disc located at the
upper-half plane and will be denoted by $S=\{\text{Im}(z)\geqslant0\}$ and $R$
parameterizes its radius which is taken much
larger (or even infinite) than the modulus of the external points,  that is $|x_{1,2,3}|$.
We keep $R$ large, but finite, when a corresponding integral diverges.

\paragraph{Integrals appearing in the evaluation of the $\langle JJJ\rangle$ and $\langle{\cal J}{\cal J}{\cal J}\rangle$ correlators}
\be
\label{JJJ.int}
\begin{split}
&I_{123}=\int_S {\text{d}^2z\ov (\z-x_1)^2(z-x_2)^2 (z-x_3)}=\frac{\pi}{x_{23}^2}\left(\frac{1}{x_1-\bar x_2}-\frac{1}{x_1-\bar x_3}\right)\ ,\\
&J_{123}=\int_S {\text{d}^2z\ov (\z-x_1)^2(z-x_2)^2 (\z-x_3)}=0\,.
\end{split}
\ee
\paragraph{Integrals appearing in the evaluation of the $\langle JJ\bar J\rangle$ and $\langle{\cal J}{\cal J}\bar{\cal J}\rangle$ correlators}
\small
\ba
\label{JJJbar.int}
&&A_{123}= \int_S {\text{d}^2z\ov (z-x_1)(z-x_2)^2 (\z-\bar x_3)^2}=
-\frac{\pi\bar x_{12}}{x_{12}^2\bar x_{13}\bar x_{23}}-\frac{\pi}{(x_1-\bar x_3)(x_2-\bar x_3)^2}\,,
\nonumber
\\
&&B_{123}= \int_S {\text{d}^2z\ov (\z-x_1)(z-x_2)^2 (\z-\bar x_3)^2}=
\frac{\pi}{x_{23}(x_1-\bar x_3)^2}+\frac{\pi}{x_{12}}\left(\frac{1}{(x_1-\bar x_3)^2}-\frac{1}{(x_2-\bar x_3)^2}\right)\,,
\nonumber
\\
&& C_{123}= \int_S {\text{d}^2z\ov (\z-x_1)^2(z-x_2)^2 (\z-\bar x_3)}=\frac{\pi}{(x_1-\bar x_3)^2}\left(\frac{1}{x_2-\bar x_3}-\frac{1}{x_{23}}\right)\,,
\\
&&D_{123}= \int_S {\text{d}^2z\ov (\z-x_1)^2(z-x_2)^2 (z-\bar x_3)}=\frac{\pi}{(x_2-\bar x_3)^2}\left(\frac{1}{x_1-\bar x_2}-\frac{1}{x_1-\bar x_3}\right)\,,
\nonumber
\\
&&E_{123}= \int_S {\text{d}^2z\ov (\z-x_1)^2(z-x_2) (z-\bar x_3)^2}=-\frac{\pi}{(x_1-\bar x_2)(x_2-\bar x_3)^2}-\frac{\pi}{x_{12}}\left(\frac{1}{(x_1-\bar x_3)^2}-\frac{1}{(x_2-\bar x_3)^2}\right)\,,
\nonumber
\\
&&F_{123}= \int_S {\text{d}^2z\ov (\z-x_1)^2(\z-x_2) (z-\bar x_3)^2}=\frac{\pi}{(x_2-\bar x_3)(x_1-\bar x_3)^2}\,.\nonumber
\ea
\normalsize
\paragraph{Integrals appearing in the evaluation of the $\langle J J\rangle$ and $\langle J\bar J\rangle$ correlators}
\be
\begin{split}
\label{JJ.int}
& I_1(x_1;x_2)=\int_S\frac{\text{d}^2z}{(z-x_1)(\z-x_2)}=\pi\ln\frac{R}{\bar x_1-x_2}+\frac{i\pi^2}{2}\,,
\\
& I_{1|1}(x_1;x_2)=\int_S\frac{\text{d}^2z}{(z-x_1)^2(\z-x_2)^2}=0\,,
\\
& J_1(x_1;x_2)=\int_S\frac{\text{d}^2z}{(z-x_1)(\z-\bar x_2)}=\pi\ln\frac{R(\bar x_2- x_1)}{|x_{12}|^2}-\frac{i\pi^2}{2}\,,
\\
& J_{1|1}(x_1;x_2)=\int_S\frac{\text{d}^2z}{(z-x_1)^2(\z-\bar x_2)^2}=\frac{\pi}{(x_1-\bar x_2)^2} +\pi^2\d^{(2)}(x_{12})\
\,,
\\
& J_2(x_1;x_2)=\int_S\frac{\text{d}^2z}{(z-\bar x_2)(\z- x_1)}=\pi\ln\frac{R}{x_1-\bar x_2}+\frac{i\pi^2}{2}\,,
\\
& J_{2|1}(x_1;x_2)=\int_S\frac{\text{d}^2z}{(z-\bar x_2)^2(\z- x_1)^2}=-\frac{\pi}{(x_1-\bar x_2)^2}\,,
\\
& J_3(x_1;x_2)=\int_S\frac{\text{d}^2z}{(z-\bar x_1)(\z-\bar x_2)}=\pi\ln\frac{R}{x_2-\bar x_1}+\frac{i\pi^2}{2}\,.
\end{split}
\ee
The above integrals are evaluated with heavy use of Stokes' theorem. For a two-dimensional vector with components  $V_{1,2}$,
Stokes' theorem is
\be
\label{Stokes0}
\int_S\text{d}x_1 \text{d}x_2 \left(\del_1 V_2-\del_2 V_1\right)
=\oint_{\del S}\left(V_1\text{d}x_1+V_2 \text{d}x_2\right)\,.
 \ee
 Defining $z=x_1+i x_2$, its complex conjugate $\z=x_1-i x_2$, as well as $A=V_2-i V_1$
 and $B= V_2+i V_1$ we obtain the form of the theorem suitable for the purposes of this paper
\be
\label{Stokes}
\int_S\text{d}^2z\left(\del_z A+\del_{\bar z}B\right)=\frac{i}{2}\oint_{\del S}\left(A\text{d}\bar z-B\text{d}z\right)\, ,
\ee
where $\text{d}^2z= \text{d}x_1\text{d}x_2$.
Subsequently the loop integral, on the right-hand side of the above expression, is evaluated for an appropriate choice of contour.
In what follows, we shall explicitly evaluate some of the above integrals using the above ingredients. Note that the choice of the 
functions $A$ and $B$ is to a certain extent arbitrary and depending on the two-dimensional integral we wish to evaluate, it is chosen to our convenience.

\subsection{The integral $I_{123}$}
\label{I123.appendix}

Let us consider the integral
\be
\label{I123.append}
I_{123}= \int_S {\text{d}^2z\ov (\z-x_1)^2(z-x_2)^2 (z-x_3)}\, .
\ee
In this case we will use Stokes' theorem \eqref{Stokes} with
\be
A=0\,,\quad B=-\frac{1}{(\z-x_1)(z-x_2)^2(z-x_3)}\, ,
\ee
so we have that
\be
\label{I123.appendS}
\begin{split}
I_{123}=&-\frac{i}{2}\oint_{\del S}\text{d}z\,B=-\frac{i}{2}\left(\int_\G\text{d}z\,B
+\int_\text{III} \text{d}z\,B+\oint_{c_{\varepsilon'}}\text{d}z\,B+\int_\text{IV}\text{d}z\,B\right.\\
&\left.+\int_\text{I} \text{d}z\,B+\oint_{c_\varepsilon}\text{d}z\,B+\int_\text{II}\text{d}z\,B
-\int_{-\infty}^{+\infty}\frac{\text{d}x}{(x-x_1)(x-x_2)^2(x-x_3)}\right)\, ,
\end{split}
\ee
where the contour of integration $\del S$ is depicted in figure~\ref{contour1}.
Note that due to the choice of the integration contour we ignore a term
proportional to $\d^{(2)}(z- x_3)$ which arises in evaluating $\del_\z B$, since $x_3$ lies outside the
domain surrounded by the contour of integration, as seen in figure~\ref{contour1}.

\begin{center}
\begin{tikzpicture}[scale=1.0,decoration={markings,
mark=at position 0.8cm with {\arrow[line width=1pt]{>}},
mark=at position 6.0cm with {\arrow[line width=1pt]{>}},
mark=at position 12.0cm with {\arrow[line width=1pt]{>}},
mark=at position 13.6cm with {\arrow[line width=1pt]{>}},
mark=at position 14.5cm with {\arrow[line width=1pt]{>}},
mark=at position 16.2cm with {\arrow[line width=1pt]{>}},
mark=at position 19.2cm with {\arrow[line width=1pt]{>}},
mark=at position 20.2cm with {\arrow[line width=1pt]{>}},
mark=at position 21.5cm with {\arrow[line width=1pt]{>}},
}
]
\draw[help lines,->] (-4,0) -- (4,0) coordinate (xaxis);
\draw[help lines,->] (0,-0.3) -- (0,4) coordinate (yaxis);

\node at (1.28,1.38) {$\bullet$};
\node at (-1.15,1.38) {$\bullet$};
\node at (-0.2,-0.2) {O};

\path[draw,line width=1.0pt,postaction=decorate] (1.3,0) node[below] {} -- (3,0) node[below] {$R$} arc
(0:180:3) node[below] {$-R$} -- (-1.2,0) node[below] {} -- (-1.2,1.2) node[below] {} arc (250:-80:.2) --
(-1.1,0) node[below] {} -- (1.2,0) node[below] {} -- (1.2,1.2) node[below] {} arc (250:-80:.2) -- (1.3,0) node[below] {};

\node[below] at (xaxis) {$\text{Re}(z)$};
\node[left] at (yaxis) {$\text{Im}(z)$};
\node at (1.5,0.6) {II};
\node at (1.0,0.6) {I};
\node at (-1.5,0.6) {III};
\node at (-0.8,0.6) {IV};
\node at (1.2,1.8) {$c_\varepsilon$};
\node at (-1.2,1.8) {$c_{\varepsilon'}$};
\node at (2,2.6) {$\Gamma$};
\draw[<-,shorten <=2mm] (-1.0,1.4) -- ++ (40:0.5) node[right] {$x_3$};
\draw[<-,shorten <=2mm] (1.4,1.4) -- ++ (40:0.5) node[right] {$x_2$};
\end{tikzpicture}
\vskip -.4 cm
\captionof{figure}{Contour of integration $\del S$ for the integral \eqref{I123.append}  with $R\to\infty$.
\label{contour1}}
\end{center}
The integrals on I and II cancel each other and similarly
for the integrals of III and IV. In addition, the integral on $\G$ vanishes as well as.
Finally, using Cauchy's theorem on the lower-half plane,
the last integral on the real line is also vanishing since $\text{Im}(x_{1,2,3})>0$.
Hence, we are left with the integrals on $c_\varepsilon$
and $c_{\varepsilon'}$ which can be easily evaluated
\be
\oint_{c_{\varepsilon}}\text{d}z\,B=\frac{2\pi i}{(x_1-\bar x_2)x_{23}^2}\, ,\qquad
\oint_{c_{\varepsilon'}}\text{d}z\,B=-\frac{2\pi i}{(x_1-\bar x_3)x_{23}^2}\, .
\ee
Using the above in \eqref{I123.appendS} we find the result
\be
\label{I123.result}
I_{123}=\frac{\pi}{x_{23}^2}\left(\frac{1}{x_1-\bar x_2}-\frac{1}{x_1-\bar x_3}\right)\,.
\ee

\subsection{The integral $J_{123}$}
\label{J123.appendix}

Let us now consider the integral
\be
\label{J123.appen}
J_{123}= \int_S {\text{d}^2z\ov (\z-x_1)^2(z-x_2)^2 (\z-x_3)}\,.
\ee
To evaluate it we will use again the Stokes' theorem \eqref{Stokes} with
\be
A=-\frac{1}{(\z-x_1)^2(z-x_2)(\z-x_3)}\,  , \qquad B=0\,,
\ee
so that we have that
\be
\label{J123.appenS}
\begin{split}
J_{123}=&\frac{i}{2}\oint_{\del S}\text{d}\z A=\frac{i}{2}\left(\int_\Gamma\text{d}\z A+
\int_\text{I}\text{d}\z A+\oint_{c_\varepsilon}\text{d}\z A+\int_\text{II}\text{d}\z A\right.\\
&\left.-\int_{-\infty}^{+\infty}\frac{\text{d}x}{(x-x_1)^2(x-x_2)(x-x_3)}\right)\,,
\end{split}
\ee
where the contour of integration $\del S$ is depicted in figure~\ref{contour2}.  Note that a
term proportional to $\d^{(2)}(z-\bar x_3)$ resulting from evaluating $\del_z A$ has been
ignored since $z$ and $\bar x_3$ are located the upper- and lower-halves of the plane, respectively.
\begin{center}
\begin{tikzpicture}[scale=1.0,decoration={markings,
mark=at position 1.3cm with {\arrow[line width=1pt]{>}},
mark=at position 7.0cm with {\arrow[line width=1pt]{>}},
mark=at position 12.0cm with {\arrow[line width=1pt]{>}},
mark=at position 15.9cm with {\arrow[line width=1pt]{>}},
mark=at position 17.0cm with {\arrow[line width=1pt]{>}},
mark=at position 18.1cm with {\arrow[line width=1pt]{>}},
}
]
\draw[help lines,->] (-4,0) -- (4,0) coordinate (xaxis);
\draw[help lines,->] (0,-0.3) -- (0,4) coordinate (yaxis);

\node at (1.28,1.38) {$\bullet$};
\node at (-0.2,-0.2) {O};

\path[draw,line width=1.0pt,postaction=decorate] (1.3,0) node[below] {} -- (3,0) node[below] {$R$} arc
(0:180:3) node[below] {$-R$} -- (1.2,0) node[below] {} -- (1.2,1.2) node[below] {} arc (250:-80:.2) -- (1.3,0) node[below] {};

\node[below] at (xaxis) {$\text{Re}(z)$};
\node[left] at (yaxis) {$\text{Im}(z)$};
\node at (1.6,0.6) {II};
\node at (1.0,0.6) {I};
\node at (1.2,1.8) {$c_\varepsilon$};
\node at (2,2.6) {$\Gamma$};
\draw[<-,shorten <=2mm] (1.4,1.4) -- ++ (40:0.5) node[right] {$x_2$};
\end{tikzpicture}
\vskip -.4 cm
\captionof{figure}{Contour of integration $\del S$ for the integral \eqref{J123.appen}  with $R\to\infty$.
\label{contour2}}
\end{center}
The integrals on I and II cancel each other and the integrals over the contours $\G$ and $c_\varepsilon$ vanish. Finally, the last integral
on the real line which can be easily seen to vanish using Cauchy's theorem on the lower-half plane, with $\text{Im}(x_{1,2,3})>0$.
Using the above in \eqref{J123.appenS} we find that
\be
J_{123}=0\, .
\ee

 \subsection{The integral $I_1(x_1,x_2)$}
 \label{I1x1x2.sect}

Let us now consider the integral
\be
\label{Ix1x2.append}
I_1(x_1,x_2)=\int_S\frac{\text{d}^2z}{(z-x_1)(\z-x_2)}\,.
\ee
To evaluate it we use Stokes' theorem \eqref{Stokes} with
\be
 A=0\,,\quad B=\frac{\ln(\z-x_2)}{z-x_1}\, ,
 \ee
so that we have
\be
\label{Ix1x2.appendp}
\begin{split}
I_1(x_1,x_2)&=-\frac{i}{2}\oint_{\del S}\text{d}z\,B=-\frac{i}{2}\left(\int_\G\text{d}z\,B+\int_\text{I}\text{d}z\,B
+\oint_{c_\varepsilon}\text{d}z B+\int_\text{II}\text{d}z\,B\right.\\
&\left.+\int_{-R}^{+R}\text{d}x\,\frac{\ln(x-x_2)}{x-x_1}\right)\,,
\end{split}
\ee
with the contour of integration as in figure~\ref{contour3}. Note that, similarly to before,
we have ignored a $\d^{(2)}(z-x_1)$ term in evaluating $\del_\z B$ since it vanishes for our choice of integration contour.
\begin{center}
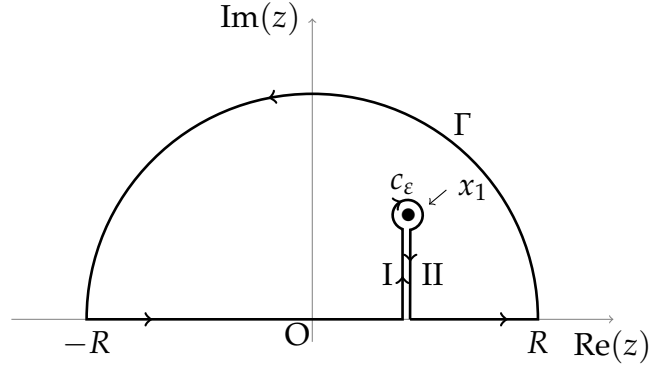

\begin{tikzpicture}[scale=1.0,decoration={markings,
mark=at position 1.3cm with {\arrow[line width=1pt]{>}},
mark=at position 7.0cm with {\arrow[line width=1pt]{>}},
mark=at position 12.0cm with {\arrow[line width=1pt]{>}},
mark=at position 15.9cm with {\arrow[line width=1pt]{>}},
mark=at position 17.0cm with {\arrow[line width=1pt]{>}},
mark=at position 18.1cm with {\arrow[line width=1pt]{>}},
}
]
\draw[help lines,->] (-4,0) -- (4,0) coordinate (xaxis);
\draw[help lines,->] (0,-0.3) -- (0,4) coordinate (yaxis);

\node at (1.28,1.38) {$\bullet$};
\node at (-0.2,-0.2) {O};

\path[draw,line width=1.0pt,postaction=decorate] (1.3,0) node[below] {} -- (3,0) node[below] {$R$} arc
(0:180:3) node[below] {$-R$} -- (1.2,0) node[below] {} -- (1.2,1.2) node[below] {} arc (250:-80:.2) -- (1.3,0) node[below] {};

\node[below] at (xaxis) {$\text{Re}(z)$};
\node[left] at (yaxis) {$\text{Im}(z)$};
\node at (1.6,0.6) {II};
\node at (1.0,0.6) {I};
\node at (1.2,1.8) {$c_\varepsilon$};
\node at (2,2.6) {$\Gamma$};
\draw[<-,shorten <=2mm] (1.4,1.4) -- ++ (40:0.5) node[right] {$x_1$};
\end{tikzpicture}
\vskip -.4 cm
\captionof{figure}{Contour of integration $\del S$ for the integral \eqref{Ix1x2.appendp}.
\label{contour3}}
\end{center}
The integrals on I and II cancel and the loop integral on $c_\varepsilon$ equals to
\be
\oint_{c_\varepsilon}\text{d}z\,B=-2\pi i\ln(\bar x_1-x_2)\, .
\ee
Also the integral on $\G$ equals to
\be
\int_\G\text{d}z\,B=i\pi\ln R+\frac{\pi^2}{2}\ .
\ee
To evaluate the integral on the real line in \eqref{Ix1x2.appendp} we use Cauchy's theorem for the following integral and
and the indicated contour in the lower-half plane is depicted in figure~\ref{contour4} so that we avoid branch cuts,
yielding
\be
\label{Ix1x2.appendp1}
\oint_{\del S'}\text{d}z \frac{\ln (z-x_2)}{z-x_1} =0\quad\Longrightarrow\quad
\int_{\G'}\text{d}z\,\frac{\ln (z-x_2)}{z-x_1}+\int_{-R}^{+R}\text{d}x\,\frac{\ln(x-x_2)}{x-x_1}=0\,.
\ee
\begin{center}
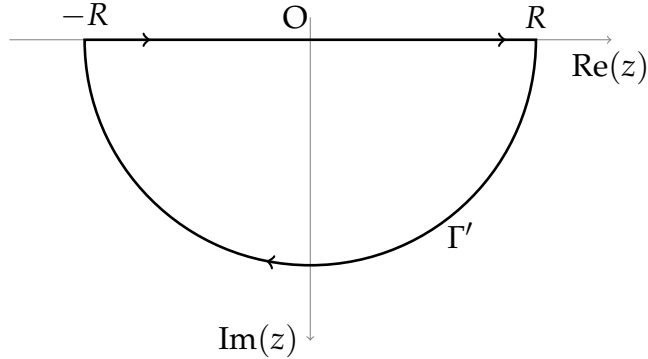

\begin{tikzpicture}[scale=1.0,decoration={markings,
mark=at position 1.3cm with {\arrow[line width=1pt]{>}},
mark=at position 7.0cm with {\arrow[line width=1pt]{>}},
mark=at position 12.0cm with {\arrow[line width=1pt]{>}},
}
]
\draw[help lines,->] (-4,0) -- (4,0) coordinate (xaxis);
\draw[help lines,->] (0,0.3) -- (0,-4) coordinate (yaxis);

\node at (-0.2,0.3) {O};

\path[draw,line width=1.0pt,postaction=decorate] (1.3,0) node[below] {} -- (3,0) node[above] {$R$} arc
(360:180:3) node[above] {$-R$}  -- (1.3,0) node[below] {};

\node[below] at (xaxis) {$\text{Re}(z)$};
\node[left] at (yaxis) {$\text{Im}(z)$};
\node at (2,2.-4.6) {$\Gamma'$};
\end{tikzpicture}
\vskip -.4 cm
\captionof{figure}{Contour of integration $\del S'$ for the integral \eqref{Ix1x2.appendp1}.
\label{contour4}}
\end{center}
We may evaluate the line integral on $\G'$ by letting $z=R e^{i \phi}$, $R\gg 1$ and $\phi\in [2\pi ,\pi]$. The choice
of the range for the angle $\phi$ is dictated by the fact that we should restrict to the lower-half plane in the transversing the curve $\G'$.
We find that
\be
\int_{\G'}\text{d}z\,\frac{\ln (z-x_2)}{z-x_1}=-i\pi\ln R+\frac{3\pi^2}{2}\, .
\ee
Employing the above in \eqref{Ix1x2.appendp} we find the result
\be
\label{ix1x2}
I_1(x_1,x_2)=\pi\ln\frac{R}{\bar x_1-x_2}+\frac{i\pi^2}{2}\, .
\ee

 \subsection{The integral $J_1(x_1,x_2)$}

 \label{J1.basic.int}

Let us now consider the integral
\be
J_1(x_1;x_2)=\int_S\frac{\text{d}^2z}{(z-x_1)(\z-\bar x_2)}\,.
\ee
To evaluate it we use Stokes' theorem \eqref{Stokes} with
\be
A=\frac{\ln|z-x_1|^2}{\z-\bar x_2}\,,\quad B=0\,,
\ee
so we have that
\be
\begin{split}
J_1(x_1;x_2)=&\frac{i}{2}\oint_{\del S}\text{d}\z A=\frac{i}{2}\left(\int_\Gamma\text{d}\z A+
\int_\text{I}\text{d}\z A+\oint_{c_\varepsilon}\text{d}\z A+\int_\text{II}\text{d}\z A\right.\\
&\left.+\int_{-R}^{+R}\text{d}x\frac{\ln|x-x_1|^2}{x-\bar x_2}\right)\,,
\end{split}
\ee
where the contour of integration $\del S$ is depicted in figure~\ref{contour2} with the radius $R$ kept finite.
The integrals on I and II cancel each other and the loop integral on $c_\varepsilon$ equals to
\be
\oint_{c_\varepsilon}\text{d}\z\,A=2\pi i\ln|x_{12}|^2\,
\ee
and the integral along $\G$ equals to
\be
\int_\Gamma\text{d}\z A=-i\pi\ln R^2\,.
\ee
Hence, we are left with the integrals on the real line
\be
\int_{-R}^{+R}\text{d}x\frac{\ln|x-x_1|^2}{x-\bar x_2}=\int_{-R}^{+R}\text{d}x\frac{\ln(x-x_1)}{x-\bar x_2}+
\int_{-R}^{+R}\text{d}x\frac{\ln(x-\bar x_1)}{x-\bar x_2}\, .
\ee
These can be evaluated via Cauchy's theorem on the lower- and upper-half plane, respectively.
Doing so one easily finds
\be
\begin{split}
&\int_{-R}^{+R}\text{d}x\frac{\ln(x-x_1)}{x-\bar x_2}=-2i\pi\ln(\bar x_2-x_1)+i\pi\ln R-\frac{3\pi^2}{2}\,,\\
&\int_{-R}^{+R}\text{d}x\frac{\ln(x-\bar x_1)}{x-\bar x_2}=-i\pi\ln R+\frac{\pi^2}{2}\,.
\end{split}
\ee
Combining all the above, one finds
\be
J_1(x_1;x_2)=\pi\ln\frac{R(\bar x_2-x_1)}{|x_{12}|^2}-\frac{i\pi^2}{2}\,.
\ee

\section{Two-point function $\langle J J\rangle$ at two-loop}
\label{2loop.append}

In this Appendix we sketch the proof of \eqref{JJ.2loop.solo}. Our starting point will be
\eqref{JJ.2loop} which using Cardy's doubling trick we rewrite as
\be
\small
\label{JJ.2loop.Cardy}
\langle  J_a(x_1)  J_b(x_2)\rangle_\text{uhp}^{(2)}=\frac{\l^2}{2\pi^2}\int_S\text{d}^2z_{12}
\langle  J_a(x_1)  J_b(x_2) J_c(z_1)  J_c(\z_1) J_d(z_2)  J_d(\z_2)\rangle\,.
\ee
To evaluate the above six-point current correlation function on the full plane we will use Ward identity 
and the Kac--Moody current algebra \eqref{KM.OPE}
in order to reduce it to four- and five-point functions
\be
\small
\begin{split}
\langle  J_a(x_1) &  J_b(x_2) J_c(z_1)  J_c(\z_1) J_d(z_2)  J_d(\z_2)\rangle=
\\
&\frac{\langle  J_b(x_2)  J_a(\z_1) J_d(z_2)  J_d(\z_2)\rangle}{(x_1-z_1)^2}+
\frac{f_{ace}\langle  J_b(x_2) J_e(z_1)  J_c(\z_1) J_d(z_2)  J_d(\z_2)\rangle}{\sqrt{k}(x_1-z_1)}\\
&+\frac{\langle  J_b(x_2)  J_a(z_1) J_d(z_2)  J_d(\z_2)\rangle}{(x_1-\z_1)^2}+
\frac{f_{ace}\langle  J_b(x_2) J_c(z_1) J_e(\z_1)  J_d(z_2)  J_d(\z_2)\rangle}{\sqrt{k}(x_1-\z_1)}\\
&+\frac{\langle  J_b(x_2)  J_c(z_1) J_c(\z_1) J_a(\z_2)\rangle}{(x_1-z_2)^2}+
\frac{f_{ade}\langle  J_b(x_2) J_c(z_1)  J_c(\z_1) J_e(z_2)  J_d(\z_2)\rangle}{\sqrt{k}(x_1-z_2)}\\
&+\frac{\langle  J_b(x_2)  J_c(z_1) J_c(\z_1)  J_a(z_2)\rangle}{(x_1-\z_2)^2}+
\frac{f_{ade}\langle  J_b(x_2) J_c(z_1) J_e(\z_1)  J_d(z_2)  J_e(\z_2)\rangle}{\sqrt{k}(x_1-\z_2)}
\end{split}
\ee
and we have dismissed two terms corresponding to bubble diagrams.
This expression can be written schematically as
\be
\small
\label{six.JJ}
\begin{split}
\langle  J_a(x_1) &  J_b(x_2) J_c(z_1)  J_c(\z_1) J_d(z_2)  J_d(\z_2)\rangle=\\
&\frac{\langle  J_b(x_2)  J_a(\z_1) J_d(z_2)  J_d(\z_2)\rangle}{(x_1-z_1)^2}+
\frac{f_{ace}\langle  J_b(x_2) J_e(z_1)  J_c(\z_1) J_d(z_2)  J_d(\z_2)\rangle}{\sqrt{k}(x_1-z_1)}\\
&+(z_1\leftrightarrow\z_1)+(z_1\leftrightarrow z_2)+(z_1\leftrightarrow \z_2)\,,
\end{split}
\ee
where the replacements address to the four- and five-point correlation functions as well.
Note that the latter two replacements, namely $z_1\leftrightarrow z_2$ and $z_1\leftrightarrow \z_2$ can be obtained from
the second line of \eqref{six.JJ} in conjuction with the replacement $z_1\leftrightarrow \z_1$ upon relabeling the integration variables.
Hence, upon inserting in \eqref{JJ.2loop.Cardy} we obtain the following simplified expression
\be
\label{six.JJ.int}
\begin{split}
\langle  J_a(x_1) &  J_b(x_2)\rangle_\text{uhp}^{(2)}=
2\times\frac{\l^2}{2\pi^2}\int_S\text{d}^2z_{12}\left\{\frac{\langle  J_b(x_2)  J_a(\z_1) J_d(z_2)  J_d(\z_2)\rangle}{(x_1-z_1)^2}\right.\\
&\left.+\frac{f_{ace}\langle  J_b(x_2) J_e(z_1)  J_c(\z_1) J_d(z_2)  J_d(\z_2)\rangle}{\sqrt{k}(x_1-z_1)}
+(z_1\leftrightarrow\z_1)\right\}\,,
\end{split}
\ee
where again the replacement $z_1\leftrightarrow\z_1$ addresses to the four- and five-point correlation functions.
The four-point function contribution in \eqref{six.JJ} is
\be
\label{four.JJ}
\begin{split}
&\frac{\langle  J_b(x_2)  J_a(\z_1) J_d(z_2)  J_d(\z_2)\rangle}{(x_1-z_1)^2}=\frac{\d_{ab}\text{d}_G}{(x_1-z_1)^2(x_2-\z_1)^2(z_2-\z_2)^2}\\
&\qq +\frac{\d_{ab}}{(x_1-z_1)^2(x_2-z_2)^2\z_{12}^2}
+\frac{\d_{ab}}{(x_1-z_1)^2(x_2-\z_2)^2(\z_1-z_2)^2}\\
&\qq +\frac{c_G\d_{ab}}{k}\frac{1}{(x_1-z_1)^2(x_2-z_2)(x_2-\z_2)\z_{12}(\z_1-z_2)}\,.
\end{split}
\ee
Also, the five-point function contribution in \eqref{six.JJ} to order $\nicefrac1k$ reads
\be
\small
\label{five.JJ}
\begin{split}
&\frac{f_{ace}\langle  J_b(x_2) J_e(z_1)  J_c(\z_1) J_d(z_2)  J_d(\z_2)\rangle}{\sqrt{k}(x_1-z_1)}
\\
&\ = \frac{c_G\d_{ab}}{k}\frac{1}{(x_1-z_1)(z_2-x_2)^2(z_1-\z_1)(z_1-\z_2)\z_{12}}\\
&\quad +\frac{c_G\d_{ab}}{k}\frac{1}{(x_1-z_1)(\z_2-x_2)^2z_{12}(\z_1-z_2)(z_1-\z_1)}\\
&\quad -\frac{c_G\d_{ab}}{k}\frac{1}{(x_1-z_1)z_{12}^2\z_{12}^2}\left(\frac{1}{x_2-\z_1}-\frac{1}{x_2-\z_2}-\frac{1}{x_2-z_1}+\frac{1}{x_2-z_2}\right)\\
&\quad -\frac{c_G\d_{ab}}{k}\frac{1}{(x_1-z_1)(z_1-\z_2)^2(\z_1-z_2)^2}\left(\frac{1}{x_2-\z_1}+\frac{1}{x_2-\z_2}-\frac{1}{x_2-z_1}-\frac{1}{x_2-z_2}\right)\\
&\quad -\frac{c_G\d_{ab}}{k}\frac{\text{d}_G}{(x_1-z_1)(z_1-\z_1)^2(z_2-\z_2)^2}\left(\frac{1}{x_2-\z_1}-\frac{1}{x_2-z_1}\right)\,.
\end{split}
\ee
Next, we insert \eqref{four.JJ}, \eqref{five.JJ} into  \eqref{six.JJ.int} and we perform the double integrals
in the upper-half plane. We can organize the various terms in $k$-independent Abelian ones and $\nicefrac1k$-terms as they appear in \eqref{four.JJ} and \eqref{five.JJ}.
Doing so, we find that \eqref{six.JJ.int} takes schematically the following form
\be
\label{2loop.solo.pre}
\langle  J_a(x_1)  J_b(x_2)\rangle_\text{uhp}^{(2)}=\frac{\l^2}{\pi^2}\left(\text{Abelian terms}+\text{$\nicefrac1k$-terms}\right)\, .
\ee
The various contributions are listed below.

\paragraph{Abelian terms:} These terms appear in \eqref{four.JJ}, and upon integration over $z_1$ and $z_2$
they can be written as
\be
\label{JJ.Abel.sum}
\text{Abelian terms}=\d_{ab}\sum_{i=1}^3(P_i+\widetilde P_i)\,,
\ee
where  the corresponding integrals are defined as they appear in \eqref{four.JJ}
\be
\label{JJ.Abel.Table}
\begin{split}
&P_1=\int_S\frac{\text{d}^2z_{12}}{(x_1-z_1)^2(x_2-\z_1)^2(z_2-\z_2)^2}=0\,,\\
&P_2=\int_S\frac{\text{d}^2z_{12}}{(x_1-z_1)^2(x_2-z_2)^2\z_{12}^2}=\frac{\pi^2}{x_{12}^2}\,,\\
&P_3=\int_S\frac{\text{d}^2z_{12}}{(x_1-z_1)^2(x_2-\z_2)^2(\z_1-z_2)^2}=0\ .
\end{split}
\ee
The $\widetilde P_i$'s which are related to the $P_i$'s upon the replacement  $z_1\leftrightarrow\bar z_1$ in the corresponding integrands
\be
\label{JJ.Abel.Table2}
\begin{split}
&\widetilde P_1=\int_S\frac{\text{d}^2z_{12}}{(x_1-\z_1)^2(x_2-z_1)^2(z_2-\z_2)^2}=0\,,\\
&\widetilde P_2=\int_S\frac{\text{d}^2z_{12}}{(x_1-\z_1)^2(x_2-z_2)^2(z_1-\z_2)^2}=0\,,\\
&\widetilde P_3=\int_S\frac{\text{d}^2z_{12}}{(x_1-\z_1)^2(x_2-\z_2)^2z_{12}^2}=0\,.
\end{split}
\ee
The above integrals were evaluated using the results of \eqref{JJ.int}. Inserting the above into \eqref{JJ.Abel.sum},
we easily find that
\be
\label{JJ.Abel.sum.res}
\text{Abelian terms}=\pi^2\frac{\d_{ab}}{x_{12}^2}\,.
\ee

\paragraph{$\nicefrac1k$-terms:} These terms appearing in \eqref{four.JJ} and \eqref{five.JJ} upon integration over $z_1$ and $z_2$
can be written as
\be
\label{JJ.1ovk.sum}
\text{$\nicefrac1k$-terms}=\d_{ab}\frac{c_G}{k}\sum_{i=1}^6(Q_i+\widetilde Q_i)\, ,
\ee
where the  integrals in order of appearance  are
\be
\small
\label{JJ.1ovk.Table}
\begin{split}
&Q_1=\int_S\frac{\text{d}^2z_{12}}{(z_1-x_1)^2(z_2-x_2)(\z_2-x_2)\z_{12}(\z_1-z_2)}\\
&\phantom{xx}=\frac{\pi^2}{x_{12}^2}\ln\frac{|x_1-\bar x_2|^2}{(x_1-\bar x_1)(\bar x_2- x_2)}+\frac{\pi^2}{x_{12}(\bar x_1-x_2)}\,,\\
&Q_2=\int_S\frac{\text{d}^2z_{12}}{(x_1-z_1)(z_2-x_2)^2(z_1-\z_1)(z_1-\z_2)\z_{12}}=K+\frac{\pi^2}{x_{12}(x_2-\bar x_2)}\,,\\
&Q_3=\int_S\frac{\text{d}^2z_{12}}{(x_1-z_1)(\z_2-x_2)^2z_{12}(\z_1-z_2)(z_1-\z_1)}=0\,,\\
&Q_4=-\int_S\frac{\text{d}^2z_{12}}{(x_1-z_1)z_{12}^2\z_{12}^2}\left(\frac{1}{x_2-\z_1}-\frac{1}{x_2-\z_2}-\frac{1}{x_2-z_1}+\frac{1}{x_2-z_2}\right)\\
&\phantom{xx}=\frac{\pi^2}{x_{12}^2}\left(\ln\frac{|x_1-\bar x_2|^2}{(x_1-\bar x_1)(\bar x_2- x_2)}+\ln\frac{\bar x_1-x_2}{\bar x_1-x_1}+\ln\frac{\varepsilon^2}{|x_{12}|^2}-\frac{x_{12}}{\bar x_1-x_2}\right)\,,\\
&Q_5=-\int_S\frac{\text{d}^2z_{12}}{(x_1-z_1)(z_1-\z_2)^2(\z_1-z_2)^2}\left(\frac{1}{x_2-\z_1}+\frac{1}{x_2-\z_2}-\frac{1}{x_2-z_1}-\frac{1}{x_2-z_2}\right)\\
&\phantom{xx}=-2K+\frac{\pi^2\bar x_{12}}{x_{12}|x_1-\bar x_2|^2}\,,\\
&Q_6=\int_S\frac{\text{d}_G\,\text{d}^2z_{12}}{(x_1-z_1)(x_2-z_1)(x_2-\z_1)(z_1-\z_1)(z_2-\z_2)^2}\,,
\end{split}
\ee
where in $Q_2$ and $Q_5$ we have introduced
\be
\label{helpK}
K=\pi\int_S\frac{\text{d}^2z}{(x_1-z)(z-x_2)(\z-x_2)(z-\z)}\,.
\ee
Furthermore, the $\widetilde Q_i$'s appearing in \eqref{JJ.1ovk.sum} are related to the $Q_i$'s in \eqref{JJ.1ovk.Table}
upon the replacement  $z_1\leftrightarrow\bar z_1$ in the corresponding integrands
\small
\ba
\label{JJ.1ovk.Table2}
&&\widetilde Q_1=\int_S\frac{\text{d}^2z_{12}}{(\z_1-x_1)^2(z_2-x_2)(\z_2-x_2)(z_1-\z_2)z_{12}}=0\,,\nonumber\\
&&\widetilde Q_2=-\int_S\frac{\text{d}^2z_{12}}{(x_1-\z_1)(z_2-x_2)^2(z_1-\z_1)\z_{12}(z_1-\z_2)}=\widetilde{K}-\frac{\pi^2}{(x_1-\bar x_2)(x_2-\bar x_2)}\,,\nonumber\\
&&\widetilde Q_3=-\int_S\frac{\text{d}^2z_{12}}{(x_1-\z_1)(\z_2-x_2)^2(\z_1-z_2)z_{12}(z_1-\z_1)}=0\,,\\
&&\widetilde Q_4=-\int_S\frac{\text{d}^2z_{12}}{(x_1-\z_1)(\z_1-z_2)^2(z_1-\bar z_2)^2}\left(\frac{1}{x_2-z_1}-\frac{1}{x_2-\z_2}-\frac{1}{x_2-\z_1}+\frac{1}{x_2-z_2}\right)=-2\widetilde{K}\,,\nonumber\\
&&\widetilde Q_5=-\int_S\frac{\text{d}^2z_{12}}{(x_1-\z_1)z_{12}^2\z_{12}^2}\left(\frac{1}{x_2-z_1}+\frac{1}{x_2-\z_2}-\frac{1}{x_2-\z_1}-\frac{1}{x_2-z_2}\right)\nonumber\\
&&\phantom{xx}=-\frac{\pi^2}{x_{12}^2}\ln\frac{\bar x_1-x_2}{\bar x_1-x_1}-\frac{\pi^2}{(x_1-\bar x_2)x_{12}}
+\frac{\pi^2}{x_{12}^2}\ln\frac{|x_1-\bar x_2|^2}{(x_1-\bar x_1)(\bar x_2- x_2)}\,,\nonumber\\
&&\widetilde Q_6=-\int_S\frac{\text{d}_G\,\text{d}^2z_{12}}{(x_1-\z_1)(x_2-\z_1)(x_2-z_1)(z_1-\z_1)(z_2-\z_2)^2}\,,\nonumber
\ea
\normalsize
where in $\widetilde Q_{2}$ and  $\widetilde Q_{4}$ we have introduced
\be
\label{helpKt}
\widetilde{K}=-\pi\int_S\frac{\text{d}^2z}{(x_1-\z)(z-x_2)(\z-x_2)(z-\z)}\,.
\ee
The above integrals \eqref{JJ.1ovk.Table} and \eqref{JJ.1ovk.Table2}
can be evaluated by an extensive use of the Stokes's theorem \eqref{Stokes} and of \eqref{JJ.int}.
Before inserting the various pieces into \eqref{JJ.1ovk.sum}, we focus on the contributions of $Q_6$ and $\widetilde Q_6$
which we add them up
\be
\label{JJ.help2}
\begin{split}
Q_6+\widetilde Q_6&=\text{d}_G\int_S\frac{\text{d}^2z_{12}}{(x_1-z_1)(x_2-z_1)(x_1-\z_1)(x_2-\z_1)(z_2-\z_2)^2}\\
&=\text{d}_G\frac{\pi}{x_{12}^2}\ln\frac{|x_1-\bar x_2|^2}{(x_1-\bar x_1)(\bar x_2- x_2)}\int_S\frac{\text{d}^2z_2}{(z_2-\z_2)^2}\,.
\end{split}
\ee
Hence we dismiss it since it corresponds to a bubble diagram.
In addition, we note that the integrals $K$ and $\widetilde{K}$, defined in \eqref{helpK} and \eqref{helpKt} respectively, are divergent but
their sum is finite and equals to
\be
\label{JJ.help1}
K+\widetilde{K}=\pi\int_S\frac{\text{d}^2z}{(z-x_1)(z-x_2)(\z-x_1)(\z-x_2)}
=\frac{\pi^2}{x_{12}^2}\ln\frac{|x_1-\bar x_2|^2}{(x_1-\bar x_1)(\bar x_2- x_2)}\,,
\ee
where in the last step we have used the integral \eqref{JJ.lego.int}.

Inserting \eqref{JJ.1ovk.Table}, \eqref{JJ.1ovk.Table2}
into \eqref{JJ.1ovk.sum}, we find that
\be
\label{JJ.1ovk.sum.res}
\text{$\nicefrac1k$-terms}=\pi^2\frac{c_G}{k}\frac{\d_{ab}}{x_{12}^2}
\left(-\frac{\xi}{1-\xi}+\ln\frac{(1-\xi)^2\varepsilon^2}{|x_{12}|^2}\right)\,,
\ee
where we have also used \eqref{JJ.help1}, ignored the bubble diagram \eqref{JJ.help2}
and expressed the various quantities in terms of the invariant ratio \eqref{comparison.xi}.
Finally, we insert \eqref{JJ.Abel.sum.res} and \eqref{JJ.1ovk.sum.res} into \eqref{2loop.solo.pre} and
we find \eqref{JJ.2loop.solo} which is repeated here for the reader's convenience
\be
\label{JJ.2loop.solo.App}
\langle  J_a(x_1)  J_b(x_2)\rangle^{(2)}_\text{uhp}
=\l^2\left(1-\frac{c_G}{k}\frac{\xi}{1-\xi}\right)\frac{\d_{ab}}{x_{12}^2}+
\frac{c_G\l^2}{k}\frac{\d_{ab}}{x_{12}^2}\ln\frac{(1-\xi)^2\varepsilon^2}{|x_{12}|^2}\,.
\ee

\section{Two-point function $\langle{\cal O}{\cal O}\rangle$ at one-loop}
\label{1loop.append}

In this Appendix we sketch the proof of \eqref{2pointOoneloop}.
We consider the integrand of \eqref{OO.oneloop} which corresponds to a six-point current correlation function and can be evaluated
using Cardy's doubling trick and the Kac--Moody current algebra \eqref{KM.OPE}
\be
\small
\label{six.JJ.comp}
\begin{split}
&\langle  {\cal O}(x_1,\bar x_1) {\cal O}(x_2,\bar x_2){\cal O}(z,\z)\rangle_\text{uhp}=
\langle  J_a(x_1) J_a(\bar x_1) J_b(x_2) J_b(\bar x_2) J_c(z) J_c(\z) \rangle\\
&\quad =\frac{\langle J_a(\bar x_1) J_b(x_2) J_b(\bar x_2) J_a(\z)\rangle}{(z-x_1)^2}
+\frac{1}{\sqrt{k}}\frac{f_{cae}}{z-x_1}\langle J_e(x_1) J_a(\bar x_1) J_b(x_2) J_b(\bar x_2) J_c(\z)\rangle\\
&\qq +(\text{second line}: x_1\leftrightarrow\bar x_1)\\
&\qq +(\text{second and third line}: x_1\leftrightarrow x_2)\,,
\end{split}
\ee
where we have ignored a term which would correspond to a bubble diagram. Next, we focus on the four-point function in the second line of \eqref{six.JJ.comp}
\be
\small
\label{four.JJ.comp}
\begin{split}
&\frac{\langle J_a(\bar x_1) J_b(x_2) J_b(\bar x_2) J_a(\z)\rangle}{(z-x_1)^2}=
\frac{\text{d}_G}{(\bar x_1-x_2)^2(\bar x_2-\z)^2(z-x_1)^2}+\frac{\text{d}_G}{\bar x_{12}^2(x_2-\z)^2(z-x_1)^2}\\
&\quad +\frac{\text{d}^2_G}{(\bar x_1-\z)^2(x_2-\bar x_2)^2(z-x_1)^2}+\frac{c_G}{k}\frac{\text{d}_G}{(\bar x_1-x_2)\bar x_{12}(x_2-\z)(\bar x_2-\z)(z-x_1)^2}\,.
\end{split}
\ee
Then we move on to the five-point function in the second line of \eqref{six.JJ.comp}, which can be organized in $\nicefrac1k$ and $\nicefrac{1}{k^2}$ terms, namely
\be
\small
\label{five1.JJ.comp}
\begin{split}
&\frac{1}{\sqrt{k}}\frac{f_{cae}}{z-x_1}\langle J_e(x_1) J_a(\bar x_1) J_b(x_2) J_b(\bar x_2) J_c(\z)\rangle\bigg{|}_\text{$\nicefrac1k$-terms}=\\
&\quad +\frac{c_G\text{d}_G}{k}\frac{1}{\bar x_{12}x_{12}^2|z-x_1|^2(\bar x_2-\z)}\\
&\quad +\frac{c_G\text{d}_G}{k}\frac{1}{(\bar x_1-x_2)(x_1-\bar x_2)^2|z-x_1|^2(x_2-\z)}\\
&\quad+\frac{c_G\text{d}_G}{k}\frac{1}{(z-x_1)(\bar x_1-x_2)^2(\bar x_2-\z)^2}\left(\frac{1}{x_1-\bar x_1}-\frac{1}{x_{12}}+\frac{1}{x_1-\bar x_2}-\frac{1}{x_1-\z}\right)\\
&\quad +\frac{c_G\text{d}_G}{k}\frac{1}{(z-x_1)\bar x_{12}^2(x_2-\z)^2}\left(\frac{1}{x_1-\bar x_1}+\frac{1}{x_{12}}-\frac{1}{x_1-\bar x_2}-\frac{1}{x_1-\z}\right)\\
&\quad +\frac{c_G\text{d}_G^2}{k}\frac{1}{(z-x_1)(\bar x_1-\z)^2(x_2-\bar x_2)^2}\left(\frac{1}{x_1-\bar x_1}-\frac{1}{x_1-\z}\right)
\end{split}
\ee
and
\small
\ba
\small
\label{five2.JJ.comp}
&&\frac{1}{\sqrt{k}}\frac{f_{cae}}{z-x_1}\langle J_e(x_1) J_a(\bar x_1) J_b(x_2) J_b(\bar x_2) J_c(\z)\rangle\bigg{|}_\text{$\nicefrac{1}{k^2}$ terms}=
\nonumber
\\
& &+\frac{c_G^2\text{d}_G}{k^2}\frac{1}{(x_1-\bar x_1)\bar x_{12}(\bar x_1-x_2)(z-x_1)(\z-x_2)(\z-\bar x_2)}
\nonumber
\\
&&+\frac{c_G^2\text{d}_G}{2k^2}\frac{1}{|x_{12}|^2(\bar x_1-x_2)(z-x_1)(\z-x_2)(\z-\bar x_2)}
-\frac{c_G^2\text{d}_G}{k^2}\frac{1}{x_{12}(x_2-\bar x_2)(\bar x_1-x_2)|z-x_1|^2(\z-\bar x_2)}
\nonumber
\\
&&+\frac{c_G^2\text{d}_G}{2k^2}\frac{1}{|x_1-\bar x_2|^2\bar x_{12}(z-x_1)(\z-x_2)(\z-\bar x_2)}
+\frac{c_G^2\text{d}_G}{k^2}\frac{1}{(x_1-\bar x_2)(x_2-\bar x_2)\bar x_{12}|z-x_1|^2(\z-x_2)}
\nonumber
\\
&&+\frac{c_G^2\text{d}_G}{k^2}\frac{1}{\bar x_{12}(\bar x_1-x_2)(z-x_1)(\z-x_1)(\z-x_2)(\z-\bar x_2)}\,.
\ea
\normalsize
Inserting \eqref{six.JJ.comp}, \eqref{four.JJ.comp}, \eqref{five1.JJ.comp} and \eqref{five2.JJ.comp} into
\eqref{OO.oneloop} we find schematically
\be
\label{OO.1loop.schem}
\langle  {\cal O}(x_1,\bar x_1) {\cal O}(x_2,\bar x_2)\rangle^{(1)}_\text{uhp}=-\frac{\l}{\pi}\left(\text{Abelian terms}
+\text{$\nicefrac1k$-terms}
+\text{$\nicefrac{1}{k^2}$-terms}\right)\,,
\ee
where the various terms are listed below.

\paragraph{Abelian terms:}
There are three such terms appear in \eqref{four.JJ.comp} and upon integration over $z$  they can be written as
\be
\label{OO.k.ind}
\text{Abelian terms}=\text{d}_G\sum_{i=1}^3(R_i+\widetilde R_i)+(x_1\leftrightarrow x_2)\,,
\ee
where the corresponding integrals are defined in order of appearance in \eqref{four.JJ.comp} and read
\be
\small
\label{OO.Abel.Table}
\begin{split}
&R_1=\int_S\frac{\text{d}^2z}{(\bar x_1-x_2)^2(\bar x_2-\z)^2(z-x_1)^2}=\frac{\pi}{|x_1-\bar x_2|^4}\,,\\
&R_2=\int_S\frac{\text{d}^2z}{\bar x_{12}^2(x_2-\z)^2(z-x_1)^2}=0\,,\\
&R_3=\text{d}_G\int_S\frac{\text{d}^2z}{(\bar x_1-\z)^2(x_2-\bar x_2)^2(z-x_1)^2}=\frac{\pi\,\text{d}_G}{(x_1-\bar x_1)^2(x_2-\bar x_2)^2}
\end{split}
\ee
and we have dismissed terms involving contact terms of external points.
The integrals denoted by $\widetilde R_i$'s  are related to the $R_i$'s upon the replacement
$x_1\leftrightarrow\bar x_1$ in the corresponding integrands
\be
\small
\label{OO.Abel.Table2}
\begin{split}
&\widetilde R_1=\int_S\frac{\text{d}^2z}{x_{12}^2(\bar x_2-\z)^2(z-\bar x_1)^2}=0\,,\\
&\widetilde R_2=\int_S\frac{\text{d}^2z}{(x_1-\bar x_2)^2(x_2-\z)^2(z-\bar x_1)^2}=-\frac{\pi}{|x_1-\bar x_2|^4}\,,\\
&\widetilde R_3=\text{d}_G\int_S\frac{\text{d}^2z}{(x_1-\z)^2(x_2-\bar x_2)^2(z-\bar x_1)^2}=-\frac{\pi\,\text{d}_G}{(x_1-\bar x_1)^2(x_2-\bar x_2)^2}\,.
\end{split}
\ee
The above integrals \eqref{OO.Abel.Table} and \eqref{OO.Abel.Table2} can be easily evaluated
using the results of \eqref{JJ.int}. Next we insert  \eqref{OO.Abel.Table} and \eqref{OO.Abel.Table2} into \eqref{OO.k.ind} and we find 
that there is no contribution
\be
\label{OO.k.ind.res}
\text{Abelian terms}=0\,.
\ee

\paragraph{$\nicefrac1k$-terms:}
There are six such terms appear in \eqref{four.JJ.comp} and \eqref{five1.JJ.comp}, and upon integration over $z$ they can be written as
\be
\label{OO.1ovk}
\text{$\nicefrac1k$-terms}=\frac{c_G\text{d}_G}{k}\sum_{i=1}^6(S_i+\widetilde S_i)+(x_1\leftrightarrow x_2)\,,
\ee
where the  corresponding integrals are as usual defined  in order of appearance in  \eqref{four.JJ.comp}, \eqref{five1.JJ.comp}
and read
\be
\small
\label{OO.1ovk.Table}
\begin{split}
&S_1=\int_S\frac{\text{d}^2z}{(\bar x_1-x_2)\bar x_{12}(x_2-\z)(\bar x_2-\z)(z-x_1)^2}=\frac{\pi}{|x_{12}|^2|x_1-\bar x_2|^2}\,,\\
&S_2=\int_S\frac{\text{d}^2z}{\bar x_{12}x_{12}^2|z-x_1|^2(\bar x_2-\z)}=\frac{\pi}{|x_{12}|^4}
\ln\left(\frac{\varepsilon^2}{|x_{12}|^2}\frac{x_1-\bar x_2}{x_1-\bar x_1}\right)\,,\\
&S_3=\int_S\frac{\text{d}^2z}{(\bar x_1-x_2)(x_1-\bar x_2)^2|z-x_1|^2(x_2-\z)}=
\frac{\pi}{|x_1-\bar x_2|^4}\ln\frac{\varepsilon^2}{(x_1-\bar x_1)(\bar x_1-x_2)}\,,\\
&S_4=\int_S\frac{\text{d}^2z}{(z-x_1)(\bar x_1-x_2)^2(\bar x_2-\z)^2}\left(\frac{1}{x_1-\bar x_1}-\frac{1}{x_{12}}+\frac{1}{x_1-\bar x_2}-\frac{1}{x_1-\z}\right)\\
&\phantom{xx}=\frac{\pi}{|x_{12}|^2|x_1-\bar x_2|^2}+\frac{\pi}{|x_1-\bar x_2|^4}\ln\frac{|x_{12}|^2}{(\bar x_1-x_1)(x_1-\bar x_2)}\,,\\
&S_5=\int_S\frac{\text{d}^2z}{(z-x_1)\bar x_{12}^2(x_2-\z)^2}\left(\frac{1}{x_1-\bar x_1}+\frac{1}{x_{12}}-\frac{1}{x_1-\bar x_2}-\frac{1}{x_1-\z}\right)\\
&\phantom{xx}=\frac{\pi}{(x_1-\bar x_1)\bar x_{12}|x_1-\bar x_2|^2}+\frac{\pi}{|x_{12}|^4}\ln\frac{\bar x_1-x_2}{\bar x_1-x_1}\,,\\
&S_6=\text{d}_G\int_S\frac{\text{d}^2z}{(z-x_1)(\bar x_1-\z)^2(x_2-\bar x_2)^2}\left(\frac{1}{x_1-\bar x_1}-\frac{1}{x_1-\z}\right)
\\
&\phantom{xx} =\frac{\pi\,\text{d}_G}{(x_1-\bar x_1)^2(x_2-\bar x_2)^2}\ln\frac{\varepsilon^2}{|x_1-\bar x_1|^2}\, .
\end{split}
\ee
The $\widetilde S_i$'s are related to the $S_i$'s
upon the replacement $x_1\leftrightarrow\bar x_1$ in the corresponding integrands
\be
\small
\label{OO.1ovk.Table2}
\begin{split}
&\widetilde S_1=\int_S\frac{\text{d}^2z}{x_{12}(x_1-\bar x_2)(x_2-\z)(\bar x_2-\z)(z-\bar x_1)^2}=0\,,\\
&\widetilde S_2=\int_S\frac{\text{d}^2z}{(x_1-\bar x_2)(\bar x_1-x_2)^2|z-\bar x_1|^2(\bar x_2-\z)}=
\frac{\pi}{|x_1-\bar x_2|^4}\ln\frac{x_1-\bar x_1}{x_2-\bar x_1}\,,\\
&\widetilde S_3=\int_S\frac{\text{d}^2z}{x_{12}\bar x_{12}^2|z-\bar x_1|^2(x_2-\z)}=\frac{\pi}{|x_{12}|^4}\ln\frac{x_1-\bar x_1}{x_2-\bar x_1}\,,\\
&\widetilde S_4=\int_S\frac{\text{d}^2z}{(z-\bar x_1)x_{12}^2(\bar x_2-\z)^2}\left(\frac{1}{\bar x_1-x_1}-\frac{1}{\bar x_1-x_2}
\qq +\frac{1}{\bar x_{12}}-\frac{1}{\bar x_1-\z}\right)
\\
&\phantom{xx} =\frac{\pi}{|x_{12}|^4}\ln\frac{x_2-\bar x_1}{x_1-\bar x_1}\,,
\\
&\widetilde S_5=\int_S\frac{\text{d}^2z}{(z-\bar x_1)(x_1-\bar x_2)^2(x_2-\z)^2}\left(\frac{1}{\bar x_1- x_1}+\frac{1}{\bar x_1-x_2}-\frac{1}{\bar x_{12}}-
\frac{1}{\bar x_1-\z}\right)\\
&\phantom{xx}=\frac{\pi}{|x_1-\bar x_2|^4}\ln\frac{\bar x_1-x_2}{\bar x_1-x_1}-\frac{\pi\, }{|x_1-\bar x_2|^2(x_1-\bar x_1)\bar x_{12}}\,,\\
&\widetilde S_6=\text{d}_G\int_S\frac{\text{d}^2z}{(z-\bar x_1)( x_1-\z)^2(x_2-\bar x_2)^2}\left(\frac{1}{\bar x_1-\bar x_1}-\frac{1}{\bar x_1-\z}\right)=0\,.
\end{split}
\ee
The above integrals \eqref{OO.1ovk.Table} and \eqref{OO.1ovk.Table2} can be easily evaluated using the results of \eqref{JJ.int}.
Inserting \eqref{OO.1ovk.Table} and \eqref{OO.1ovk.Table2} into \eqref{OO.1ovk}, we find that
\be
\label{OO.1ovk.res}
\begin{split}
&\text{$\nicefrac1k$-terms}=\frac{c_G\text{d}_G}{k}\left(\frac{2\pi}{|x_{12}|^4}\ln\left(\frac{\varepsilon^2}{|x_{12}|^2}(1-\xi)\right)+
\frac{2\pi}{|x_1-\bar x_2|^4}\ln\left(\frac{-\varepsilon^2\xi}{|x_1-\bar x_2|^2}\right)\right.\\
&\qq \left.\frac{2\pi\text{d}_G}{(x_1-\bar x_1)^2(x_2-\bar x_2)^2}\ln\frac{\varepsilon^2}{|(x_1-\bar x_1)||(x_2-\bar x_2)|}+\frac{4\pi}{|x_{12}|^2|x_1-\bar x_2|^2}\right)\,.
\end{split}
\ee

\paragraph{$\nicefrac{1}{k^2}$-terms:}
There are four such terms appearing in \eqref{five2.JJ.comp}, one for each line, and upon integration over $z$ they can be written as
\be
\label{OO.1ovk2}
\text{$\nicefrac{1}{k^2}$-terms}=\frac{c^2_G\text{d}_G}{k^2}\sum_{i=1}^4(T_i+\widetilde T_i)+(x_1\leftrightarrow x_2)\,,
\ee
where in order of appearance in \eqref{five2.JJ.comp} the corresponding integrals read
\small
\ba
\small
\label{OO.1ovk2.Table}
&& T_1=\int_S\frac{\text{d}^2z}{(x_1-\bar x_1)\bar x_{12}(\bar x_1-x_2)(z-x_1)(\z-x_2)(\z-\bar x_2)}
\nonumber
\\
&&\phantom{xx} =\frac{\pi}{(x_1-\bar x_1)(x_2-\bar x_2)(x_1-\bar x_2)\bar x_{12}}\ln\frac{|x_{12}|^2}{|x_1-\bar x_2|^2}\,,
\nonumber
\\
&&T_2=\int_S\text{d}^2z\left(\frac{1}{2|x_{12}|^2(\bar x_1-x_2)(z-x_1)(\z-x_2)(\z-\bar x_2)}
-\frac{1}{x_{12}(x_2-\bar x_2)(\bar x_1-x_2)|z-x_1|^2(\z-\bar x_2)}\right)
\nonumber
\\
&&\phantom{xx}=\frac{\pi}{2|x_{12}|^2(x_2-\bar x_2)(\bar x_1-x_2)}\ln\frac{(x_1-\bar x_2)^2\varepsilon^4}{|x_1-\bar x_2|^2|x_{12}|^2(x_1-\bar x_1)^2}\,,
\nonumber
\\
&&T_3=\int_S\text{d}^2z\left(\frac{1}{2|x_1-\bar x_2|^2\bar x_{12}(z-x_1)(\z-x_2)(\z-\bar x_2)}
+\frac{1}{(x_1-\bar x_2)(x_2-\bar x_2)\bar x_{12}|z-x_1|^2(\z-x_2)}\right)
\nonumber
\\
&& \phantom{xx}=\frac{\pi}{2|x_1-\bar x_2|^2\bar x_{12}(x_2-\bar x_2)}\ln\frac{|x_{12}|^2(x_1-\bar x_1)^2(\bar x_1-x_2)^2}{|x_1-\bar x_2|^2\varepsilon^4}\,,
\\
&&T_4=\int_S\frac{\text{d}^2z}{\bar x_{12}(\bar x_1-x_2)(z-x_1)(\z-x_1)(\z-x_2)(\z-\bar x_2)}
\nonumber
\\
&& \phantom{xx}=\frac{\pi}{|x_{12}|^2|x_1-\bar x_2|^2}\ln\frac{|x_{12}|^2}{(x_1-\bar x_2)(\bar x_1- x_1)}-
\frac{\pi}{|x_{12}|^2(\bar x_1-x_2)(x_2-\bar x_2)}\ln\frac{|x_{12}|^2}{|x_1-\bar x_2|^2}\, .
\nonumber
\ea
\normalsize
The $\widetilde T_i$'s are related to the $T_i$'s upon the replacement $x_1\leftrightarrow\bar x_1$ in the corresponding integrands
\small
\ba
\small
\label{OO.1ovk2.Table2}
&& \widetilde T_1=\int_S\frac{\text{d}^2z}{(\bar x_1- x_1)(x_1-\bar x_2)x_{12}(z-\bar x_1)(\z-x_2)(\z-\bar x_2)}=0\,,
\nonumber
\\
&&\widetilde T_2=\int_S\text{d}^2z\left(\frac{1}{2|x_1-\bar x_2|^2x_{12}(z-\bar x_1)(\z-x_2)(\z-\bar x_2)}
-\frac{1}{(\bar x_1-x_2)(x_2-\bar x_2)x_{12}|z-\bar x_1|^2(\z-\bar x_2)}\right)
\nonumber
\\
&&\phantom{xx}=-\frac{\pi}{|x_1-\bar x_2|^2(x_2-\bar x_2)x_{12}}\ln\frac{x_2-\bar x_1}{x_1-\bar x_1}\,,
\nonumber
\\
&&\widetilde T_3=\int_S\text{d}^2z\left(\frac{1}{2|x_{12}|^2(x_1-\bar x_2)(z-\bar x_1)(\z-x_2)(\z-\bar x_2)}
+\frac{1}{\bar x_{12}(x_2-\bar x_2)(x_1-\bar x_2)|z-\bar x_1|^2(\z-x_2)}\right)
\nonumber
\\
&&\phantom{xx}=\frac{\pi}{|x_{12}|^2(x_2-\bar x_2)(x_1-\bar x_2)}\ln\frac{x_2-\bar x_1}{x_1-\bar x_1}\,,
\\
&&\widetilde T_4=\int_S\frac{\text{d}^2z}{(x_1-\bar x_2)x_{12}(z-\bar x_1)(\z-\bar x_1)(\z-x_2)(\z-\bar x_2)}
=\frac{\pi}{|x_{12}|^2|x_1-\bar x_2|^2}\ln\frac{x_2-\bar x_1}{x_1-\bar x_1}\,.
\nonumber
\ea
\normalsize
The above integrals \eqref{OO.1ovk2.Table} and \eqref{OO.1ovk2.Table2} can be easily evaluated using the results of \eqref{JJ.int}.
Inserting  \eqref{OO.1ovk2.Table} and \eqref{OO.1ovk2.Table2} into \eqref{OO.1ovk2}, we find
\be
\label{OO.1ovk2.res}
\text{$\nicefrac{1}{k^2}$-terms}=\frac{2\pi}{|x_{12}|^2|x_1-\bar x_2|^2}\ln\frac{\varepsilon^2{\widetilde F}(\xi)}{|x_{12}(x_1-\bar x_2)|}\,,
\ee
where
\be
\label{function.H.OO}
{\widetilde F}(\xi)=(1-\xi)\left(1-\frac{1}{\xi}\right)^{-\xi}\,,\quad \xi\leqslant0\,.
\ee
Finally, inserting \eqref{OO.k.ind.res}, \eqref{OO.1ovk.res} and \eqref{OO.1ovk2.res} into \eqref{OO.1loop.schem} we find
\be
\label{2pointOoneloop.App}
\begin{split}
\langle  {\cal O}&(x_1,\bar x_1)  {\cal O}(x_2,\bar x_2)\rangle^{(1)}_\text{uhp}=
-\frac{2\l c_G}{k}\times\left\{\frac{\text{d}_G}{|x_{12}|^4}\ln\frac{(1-\xi)\varepsilon^2}{|x_{12}|^2}\right.\\
&+\frac{\text{d}_G}{|x_1-\bar x_2|^4}\ln\frac{-\xi\,\varepsilon^2}{|x_1-\bar x_2|^2}
+\left.\frac{\text{d}_G^2}{(x_1-\bar x_1)^2(x_2-\bar x_2)^2}\ln\frac{\varepsilon^2}{|(x_1-\bar x_1)(x_2-\bar x_2)|}\right\}\\
&+\frac{c_G}{k}\frac{\text{d}_G}{|x_{12}|^2|x_1-\bar x_2|^2}\left(-4\l-\frac{2\l c_G}{k}\ln\frac{\varepsilon^2{\widetilde F}(\xi)}{|x_{12}(x_1-\bar x_2)|}\right)\,,
\end{split}
\ee
where the function ${\widetilde F}(\xi)$ was defined in \eqref{function.H.OO}.


\begin{thebibliography}{1}

 \bibitem{Dai:1989ua}
J.~Dai, R.~G.~Leigh and J.~Polchinski,
{\it New Connections Between String Theories},\\
\href{https://www.worldscientific.com/doi/abs/10.1142/S0217732389002331}{Mod. Phys. Lett. \textbf{A4} (1989), 2073-2083}.

\bibitem{Leigh:1989jq}
R.~G.~Leigh,
{\it Dirac-Born-Infeld Action from Dirichlet Sigma Model},\\
\href{https://www.worldscientific.com/doi/abs/10.1142/S0217732389003099}{Mod. Phys. Lett. \textbf{A4} (1989), 2767-2772}.

\bibitem{Horava:1989ga}
P.~Ho\v{r}ava,
{\it Background Duality of Open String Models},\\
\href{https://www.sciencedirect.com/science/article/pii/0370269389902098?via\%3Dihub}{Phys. Lett. \textbf{B231} (1989), 251-257}.



\bibitem{Green:1991et}
M.~B.~Green,
{\it Space-time duality and Dirichlet string theory},\\
\href{https://www.sciencedirect.com/science/article/pii/037026939191048Z?via\%3Dihub}{Phys. Lett. \textbf{B266} (1991), 325-336}.

\bibitem{Bachas:1995kx}
C.~Bachas, {\it D-brane dynamics},
\href{https://www.sciencedirect.com/science/article/pii/0370269396002389?via\%3Dihub}{Phys. Lett.  \textbf{B374} (1996), 37-42},
[\href{https://arxiv.org/abs/hep-th/9511043}{\ttfamily hep-th/9511043}].


\bibitem{Hull:1994ys}
C.~M.~Hull and P.~K.~Townsend,
{\it Unity of superstring dualities},\\
\href{https://www.sciencedirect.com/science/article/pii/055032139400559W?via\%3Dihub}{Nucl. Phys. \textbf{B438} (1995), 109-137},
[\href{https://arxiv.org/abs/hep-th/9410167}{\ttfamily hep-th/9410167}].

\bibitem{Witten:1995ex}
E.~Witten,
{\it String theory dynamics in various dimensions},\\
\href{https://www.sciencedirect.com/science/article/pii/055032139500158O?via\%3Dihub}{Nucl. Phys. \textbf{B443} (1995), 85-126},
[\href{https://arxiv.org/abs/hep-th/9503124}{\ttfamily hep-th/9503124}].

\bibitem{Maldacena:1997re}
J.~M.~Maldacena,
{\it The Large N limit of superconformal field theories and supergravity},
\href{https://link.springer.com/article/10.1023\%2FA\%3A1026654312961}{Int. J. Theor. Phys. \textbf{38} (1999), 1113-1133},
\href{https://www.intlpress.com/site/pub/pages/journals/items/atmp/content/vols/0002/0002/a001/}{Adv. Theor. Math. Phys. {\bf 2} (1998), 231-252},
[\href{https://arxiv.org/abs/hep-th/9711200}{\ttfamily hep-th/9711200}].



\bibitem{Gubser:1998bc}
S.~S.~Gubser, I.~R.~Klebanov and A.~M.~Polyakov,
{\it Gauge theory correlators from noncritical string theory},
\href{https://www.sciencedirect.com/science/article/abs/pii/S0370269398003773?via\%3Dihub}{Phys. Lett. \textbf{B428} (1998), 105-114},
[\href{https://arxiv.org/abs/hep-th/9802109}{\ttfamily hep-th/9802109}].

\bibitem{Witten:1998qj}
E.~Witten,
{\it Anti-de Sitter space and holography},\\
\href{https://www.intlpress.com/site/pub/pages/journals/items/atmp/content/vols/0002/0002/a002/}{Adv. Theor. Math. Phys. \textbf{2} (1998), 253-291},
[\href{https://arxiv.org/abs/hep-th/9802150}{\ttfamily hep-th/9802150}].

\bibitem{Witten:1998zw}
E.~Witten,
{\it Anti-de Sitter space, thermal phase transition, and confinement in gauge theories},
\href{https://www.intlpress.com/site/pub/pages/journals/items/atmp/content/vols/0002/0003/a003/}{Adv. Theor. Math. Phys. \textbf{2} (1998), 505-532},
[\href{https://arxiv.org/abs/hep-th/9803131}{\ttfamily hep-th/9803131}].

\bibitem{Cardy:1984bb}
J.~L.~Cardy,
{\it Conformal Invariance and Surface Critical Behavior},\\
\href{https://www.sciencedirect.com/science/article/pii/0550321384902414?via\%3Dihub}{Nucl. Phys. \textbf{B240} (1984), 514-532}.

\bibitem{McAvity:1995zd}
D.~M.~McAvity and H.~Osborn,
{\it Conformal field theories near a boundary in general dimensions},
\href{https://www.sciencedirect.com/science/article/pii/0550321395004769?via\%3Dihub}{Nucl. Phys. \textbf{B455} (1995), 522-576},
[\href{https://arxiv.org/abs/cond-mat/9505127}{\ttfamily cond-mat/9505127}].

 \bibitem{Sfetsos:2013wia}
  K.~Sfetsos, {\it Integrable interpolations: From exact CFTs to non-Abelian T-duals},\hfill\break
  \href{https://www.sciencedirect.com/science/article/pii/S0550321314000066?via\%3Dihub}{Nucl. Phys. {\bf B880} (2014), 225-246},
  [\href{http://arxiv.org/abs/arXiv:1312.4560}{\ttfamily 1312.4560}].

   \bibitem{Hollowood:2014rla}
  T.~J.~Hollowood, J.~L.~Miramontes and D.~M.~Schmidtt,
 {\it Integrable Deformations of Strings on Symmetric Spaces},
  \href{https://link.springer.com/article/10.1007\%2FJHEP11\%282014\%29009}{JHEP {\bf 1411} (2014), 009},
  [\href{http://arxiv.org/abs/1407.2840}{\ttfamily 1407.2840}].

\bibitem{Hollowood:2014qma}
  T.~J.~Hollowood, J.~L.~Miramontes and D.~Schmidtt,
{\it An Integrable Deformation of the $AdS_5 \times S^5$ Superstring},
\href{https://iopscience.iop.org/article/10.1088/1751-8113/47/49/495402}{J.\ Phys.\ {\bf A47} (2014) 49,  495402},
 [\href{http://arxiv.org/abs/1409.1538}{\ttfamily 1409.1538}].

  \bibitem{Georgiou:2016urf}
  G.~Georgiou and K.~Sfetsos,
  {\it A new class of integrable deformations of CFTs},\\
   \href{https://link.springer.com/article/10.1007\%2FJHEP03\%282017\%29083}{JHEP {\bf 1703} (2017), 083},
  [\href{https://arxiv.org/abs/1612.05012}{\ttfamily 1612.05012}].

  \bibitem{Georgiou:2017jfi}
  G.~Georgiou and K.~Sfetsos,
  {\it Integrable flows between exact CFTs},\\
  \href{https://link.springer.com/article/10.1007\%2FJHEP11\%282017\%29078}{JHEP {\bf 1711}  (2017), 078},
   [\href{https://arxiv.org/abs/1707.05149}{\ttfamily 1707.05149}].

   \bibitem{Sfetsos:2017sep}
  K.~Sfetsos and K.~Siampos,
  {\it Integrable deformations of the $G_{k_1} \times G_{k_2}/G_{k_1+k_2}$ coset CFTs},
  \href{https://linkinghub.elsevier.com/retrieve/pii/S0550321317303966}{Nucl. Phys. {\bf B927} (2018), 124-139},
  [\href{https://arxiv.org/abs/1710.02515}{\ttfamily 1710.02515}].

   \bibitem{Sfetsos:2015nya}
  K.~Sfetsos, K.~Siampos and D.~C.~Thompson,
 {\it Generalised integrable $\lambda$- and $\eta$-deformations and their relation},
  \href{https://linkinghub.elsevier.com/retrieve/pii/S0550321315003004}{Nucl. Phys. {\bf B899} (2015), 489-512},
  [\href{http://arxiv.org/abs/1506.05784}{\ttfamily 1506.05784}].


\bibitem{Georgiou:2018hpd}
  G.~Georgiou and K.~Sfetsos,
  {\it Novel all loop actions of interacting CFTs: Construction, integrability and RG flows},
  \href{https://www.sciencedirect.com/science/article/pii/S0550321318302992?via\%3Dihub}{Nucl. Phys. {\bf B937} (2018), 371-393},
 [\href{https://arxiv.org/abs/1809.03522}{\ttfamily 1809.03522}].

\bibitem{Georgiou:2018gpe}
  G.~Georgiou and K.~Sfetsos,
  {\it The most general $\lambda$-deformation of CFTs and integrability},
    \href{https://link.springer.com/article/10.1007\%2FJHEP03\%282019\%29094}{JHEP {\bf 1903} (2019), 094},
  [\href{https://arxiv.org/abs/1812.04033} {\ttfamily 1812.04033}].

\bibitem{Driezen:2019ykp}
  S.~Driezen, A.~Sevrin and D.~C.~Thompson,
  {\it Integrable asymmetric $\lambda$-deformations},
  \href{https://link.springer.com/article/10.1007\%2FJHEP04\%282019\%29094}{JHEP {\bf 1904} (2019), 094},
  [\href{https://arxiv.org/abs/1902.04142}{\ttfamily 1902.04142}].

  \bibitem{Witten:1983ar}
  E.~Witten,
  {\it Nonabelian Bosonization in Two-Dimensions},\\
  \href{https://link.springer.com/article/10.1007\%2FBF01215276}{Commun.\ Math.\ Phys.\  {\bf 92} (1984), 455-472}.



    \bibitem{Karabali:1988au}
D.~Karabali, Q.~H.~Park, H.~J.~Schnitzer and Z.~Yang,
{\it A GKO Construction Based on a Path Integral Formulation of Gauged Wess-Zumino-Witten Actions},\\
\href{https://www.sciencedirect.com/science/article/pii/0370269389911209?via\%3Dihub}{Phys. Lett. \textbf{B216} (1989), 307-312}.

  \bibitem{Karabali:1989dk}
D.~Karabali and H.~J.~Schnitzer,
{\it BRST Quantization of the Gauged WZW Action and Coset Conformal Field Theories},
\href{https://www.sciencedirect.com/science/article/pii/055032139090075O}{Nucl. Phys. \textbf{B329} (1990), 649-666}.

\bibitem{DiVecchia:1984df}
P.~Di Vecchia and P.~Rossi,
{\it On the Equivalence Between the \{Wess-Zumino\} Action and the Free Fermi Theory in Two-dimensions},
\href{https://www.sciencedirect.com/science/article/pii/0370269384907688?via\%3Dihub}{Phys. Lett. \textbf{B140} (1984), 344-348}.

\bibitem{DiVecchia:1984ksr}
P.~Di Vecchia, B.~Durhuus and J.~L.~Petersen,
{\it The Wess-Zumino Action in Two-Dimensions and Nonabelian Bosonization},
\href{https://www.sciencedirect.com/science/article/pii/0370269384918136?via\%3Dihub}{Phys. Lett. \textbf{B144} (1984), 245-249}.


  \bibitem{Itsios:2014lca}
G.~Itsios, K.~Sfetsos and K.~Siampos,
{\it The all-loop non-Abelian Thirring model and its RG flow},
\href{https://www.sciencedirect.com/science/article/pii/S0370269314003037?via\%3Dihub}{Phys. Lett. \textbf{B733} (2014), 265-269},
 [\href{https://arxiv.org/abs/1404.3748}{\ttfamily 1404.3748}].



  \bibitem{Kutasov:1989aw}
  D.~Kutasov, {\it Duality Off the Critical Point in Two-dimensional Systems With Nonabelian Symmetries},
  \href{http://www.sciencedirect.com/science/article/pii/0370269389913257}{Phys. Lett. {\bf B233} (1989), 369-373}.

   \bibitem{Georgiou:2016iom}
G.~Georgiou, K.~Sfetsos and K.~Siampos,
{\it All-loop correlators of integrable \ensuremath{\lambda}-deformed \ensuremath{\sigma}-models},
\href{https://www.sciencedirect.com/science/article/pii/S0550321316301298?via\%3Dihub}{Nucl. Phys. \textbf{B909} (2016), 360-393},
[\href{https://arxiv.org/abs/1604.08212}{\ttfamily 1604.08212}].


\bibitem{Kutasov:1989dt}
  D.~Kutasov,
  {\it String Theory and the Nonabelian Thirring Model},\\
 \href{https://www.sciencedirect.com/science/article/abs/pii/0370269389912859}{Phys. Lett. {\bf B227} (1989), 68-72}.



 \bibitem{Sfetsos:2014jfa}
  K.~Sfetsos and K.~Siampos,
  {\it Gauged WZW-type theories and the all-loop anisotropic non-Abelian Thirring model},
  \href{https://www.sciencedirect.com/science/article/pii/S0550321314001953?via\%3Dihub}{Nucl. Phys. {\bf B885} (2014), 583-599},
 [\href{http://arxiv.org/abs/arXiv:1405.7803}{\ttfamily 1405.7803}].
 

   \bibitem{Gerganov:2000mt}
  B.~Gerganov, A.~LeClair and M.~Moriconi,
 {\it On the beta function for anisotropic current interactions in 2-D},
  \href{https://journals.aps.org/prl/abstract/10.1103/PhysRevLett.86.4753}{Phys. Rev. Lett. {\bf 86} (2001) 4753-4756},
 [\href{http://arxiv.org/abs/hep-th/0011189}{\ttfamily hep-th/0011189}].


\bibitem{Balog:1993es}
  J.~Balog, P.~Forgacs, Z.~Horvath and L.~Palla,
 {\it A New family of su(2) symmetric integrable sigma models},
  \href{https://www.sciencedirect.com/science/article/pii/0370269394902135?via\%3Dihub}{Phys. Lett. {\bf B324} (1994), 403-408},
  [\href{http://arxiv.org/abs/hep-th/9307030}{\ttfamily hep-th/9307030}].


\bibitem{Georgiou:2015nka}
G.~Georgiou, K.~Sfetsos and K.~Siampos,
{\it All-loop anomalous dimensions in integrable \ensuremath{\lambda}-deformed \ensuremath{\sigma}-models},
\href{https://www.sciencedirect.com/science/article/pii/S055032131500351X?via\%3Dihub}{Nucl. Phys. \textbf{B901} (2015), 40-58},
[\href{https://arxiv.org/abs/1509.02946}{\ttfamily 1509.02946}].





\bibitem{Georgiou:2019jcf}
G.~Georgiou, P.~Panopoulos, E.~Sagkrioti and K.~Sfetsos,
{\it Exact results from the geometry of couplings and the effective action},\\
\href{https://www.sciencedirect.com/science/article/pii/S0550321319302652?via\%3Dihub}{Nucl. Phys. \textbf{B948} (2019), 114779},
[\href{https://arxiv.org/abs/1906.00984}{\ttfamily1906.00984}].

\bibitem{Sagkrioti:2020mkw}
E.~Sagkrioti,
{\it A generalized method for all-loop results in $\lambda$-models},\\
\href{https://link.springer.com/article/10.1007\%2FJHEP08\%282020\%29050}{JHEP \textbf{08} (2020), 050},
[\href{https://arxiv.org/abs/2007.00034}{\ttfamily 2007.00034}].




\bibitem{Georgiou:2019aon}
G.~Georgiou and K.~Sfetsos,
{\it Field theory and $\lambda$-deformations: Expanding around the identity},
\href{https://www.sciencedirect.com/science/article/pii/S0550321319303414?via\%3Dihub}{Nucl. Phys. \textbf{B950} (2020), 114855},
[\href{https://arxiv.org/abs/1910.01056}{\ttfamily 1910.01056}].

\bibitem{Georgiou:2020bpx}
G.~Georgiou, K.~Sfetsos and K.~Siampos,
{\it A free field perspective of \ensuremath{\lambda}-deformed coset CFT\textquoteright{}s},
\href{https://link.springer.com/article/10.1007\%2FJHEP07\%282020\%29187}{JHEP \textbf{07} (2020), 187},
[\href{https://arxiv.org/abs/2004.10216}{\ttfamily 2004.10216}].

\bibitem{Driezen:2018glg}
S.~Driezen, A.~Sevrin and D.~C.~Thompson,
{\it D-branes in $\lambda$-deformations},\\
\href{https://link.springer.com/article/10.1007\%2FJHEP09\%282018\%29015}{JHEP \textbf{09} (2018), 015},
[\href{https://arxiv.org/abs/1806.10712}{\ttfamily 1806.10712}].






\bibitem{Alekseev:1998mc}
A.~Y.~Alekseev and V.~Schomerus,
{\it D-branes in the WZW model},\\
\href{https://journals.aps.org/prd/abstract/10.1103/PhysRevD.60.061901}{Phys. Rev. \textbf{D60} (1999), 061901},
[\href{https://arxiv.org/abs/hep-th/9812193}{\ttfamily hep-th/9812193}].

\bibitem{Felder:1999ka}
G.~Felder, J.~Frohlich, J.~Fuchs and C.~Schweigert,
{\it The Geometry of WZW branes},\\
\href{https://www.sciencedirect.com/science/article/pii/S0393044099000613?via\%3Dihub}{J. Geom. Phys. \textbf{34} (2000), 162-190},
[\href{https://arxiv.org/abs/hep-th/9909030}{\ttfamily hep-th/9909030}].


\bibitem{Stanciu:1999id}
S.~Stanciu,
{\it D-branes in group manifolds},
\href{https://iopscience.iop.org/article/10.1088/1126-6708/2000/01/025}{JHEP \textbf{01} (2000), 025},
[\href{https://arxiv.org/abs/hep-th/9909163}{\ttfamily hep-th/9909163}].


\bibitem{Figueroa-OFarrill:1999cmq}
J.~M.~Figueroa-O'Farrill and S.~Stanciu,
{\it More D-branes in the Nappi-Witten background},
\href{https://iopscience.iop.org/article/10.1088/1126-6708/2000/01/024}{JHEP \textbf{01} (2000), 024},
[\href{https://arxiv.org/abs/hep-th/9909164}{\ttfamily hep-th/9909164}].


\bibitem{CFT.book}
P.~Di Francesco, P.~Mathieu and D. S\'en\'echal,
{\it Conformal Field Theory},\\ 
Graduate texts in contemporary physics. Springer, New York, NY, 1997,\\
\href{https://link.springer.com/book/10.1007\%2F978-1-4612-2256-9}{DOI: 10.1007/978-1-4612-2256-9}.



\bibitem{Fredenhagen:2006dn}
S.~Fredenhagen, M.~R.~Gaberdiel and C.~A.~Keller,
{\it Bulk induced boundary perturbations},
\href{https://iopscience.iop.org/article/10.1088/1751-8113/40/1/F03}{J. Phys. \textbf{A40} (2007), F17},
[\href{https://arxiv.org/abs/hep-th/0609034}{\ttfamily hep-th/0609034}].


\bibitem{Fischler:1986ci}
W.~Fischler and L.~Susskind,
{\it Dilaton Tadpoles, String Condensates and Scale Invariance},
\href{https://www.sciencedirect.com/science/article/pii/0370269386914255?via\%3Dihub}{Phys. Lett. \textbf{B171} (1986), 383-389}.

\bibitem{Fischler:1986tb}
W.~Fischler and L.~Susskind,
{\it Dilaton Tadpoles, String Condensates and Scale Invariance II},
\href{https://www.sciencedirect.com/science/article/pii/0370269386905149?via\%3Dihub}{Phys. Lett. \textbf{B173} (1986), 262-264}.

\bibitem{Keller:2007nd}
C.~A.~Keller,
{\it Brane backreactions and the Fischler-Susskind mechanism in conformal field theory},
\href{https://iopscience.iop.org/article/10.1088/1126-6708/2007/12/046}{JHEP \textbf{12} (2007), 046},
[\href{https://arxiv.org/abs/0709.1076}{\ttfamily 0709.1076}].




  \bibitem{Knizhnik:1984nr}
  V.G.~Knizhnik and A.B.~Zamolodchikov,
  {\it Current Algebra and Wess--Zumino Model in Two-Dimensions},
  \href{http://www.sciencedirect.com/science/article/pii/0550321384903742}{Nucl. Phys. {\bf B247} (1984), 83-103}.

\bibitem{Cardy:1989ir}
J.~L.~Cardy,
{\it Boundary Conditions, Fusion Rules and the Verlinde Formula},\hfill\break
\href{https://www.sciencedirect.com/science/article/pii/055032138990521X?via\%3Dihub}{Nucl. Phys. \textbf{B324} (1989), 581-596}.


\bibitem{Schomerus:2002dc}
V.~Schomerus,
{\it Lectures on branes in curved backgrounds},\hfill\break
\href{https://iopscience.iop.org/article/10.1088/0264-9381/19/22/305}{Class. Quant. Grav. \textbf{19} (2002), 5781-5847}.



\bibitem{Bachas:2000ik}
C.~Bachas, M.~R.~Douglas and C.~Schweigert,
{\it Flux stabilization of D-branes},\hfill\break
\href{https://iopscience.iop.org/article/10.1088/1126-6708/2000/05/048}{JHEP \textbf{05} (2000), 048},
[\href{https://arxiv.org/abs/hep-th/0003037v2}{\ttfamily hep-th/0003037}].


\bibitem{Bachas:2001id}
C.~Bachas, N.~Couchoud and P.~Windey,
{\it Orientifolds of the three sphere},\hfill\break
\href{https://iopscience.iop.org/article/10.1088/1126-6708/2001/12/003}{JHEP \textbf{12} (2001), 003},
[\href{https://arxiv.org/abs/hep-th/0111002v2}{\ttfamily hep-th/0111002}].


 \bibitem{Polyakov:1970xd}
A.~M.~Polyakov,
{\it Conformal symmetry of critical fluctuations},\\
\href{http://www.jetpletters.ac.ru/ps/1737/article_26381.shtml}{JETP Lett. \textbf{12} (1970), 381-383}.

\bibitem{Georgiou:2016zyo}
G.~Georgiou, K.~Sfetsos and K.~Siampos,
{\it $\lambda$-deformations of left\textendash{}right asymmetric CFTs},
\href{https://www.sciencedirect.com/science/article/pii/S0550321316303832?via\%3Dihub}{Nucl. Phys. \textbf{B914} (2017), 623-641},
[\href{https://arxiv.org/abs/1610.05314}{\ttfamily 1610.05314}].


\bibitem{Klimcik:2002zj}
C. Klim\v c\'\i k,
  {\it YB sigma models and dS/AdS T-duality}, \hfill\break
  \href{https://iopscience.iop.org/article/10.1088/1126-6708/2002/12/051}{JHEP {\bf 0212} (2002) 051},
[\href{https://arxiv.org/abs/hep-th/0210095}{\ttfamily hep-th/0210095}].

\bibitem{Klimcik:2008eq}
  C. Klim\v c\'\i k,
  {\it On integrability of the YB sigma-model}, \hfill\break
  \href{https://aip.scitation.org/doi/10.1063/1.3116242}{J.\ Math.\ Phys.\  {\bf 50} (2009) 043508},
  [\href{http://arxiv.org/abs/0802.3518}{\ttfamily 0802.3518}].
  
  
  \bibitem{Vicedo:2015pna}
  B.~Vicedo,
  {\it Deformed integrable $\sigma$-models, classical $R$-matrices and classical exchange algebra on Drinfel'd doubles},
  \href{https://iopscience.iop.org/article/10.1088/1751-8113/48/35/355203}{J. Phys. A: Math. Theor. {\bf 48} (2015) 355203},
 [\href{http://arxiv.org/abs/1504.06303}{\ttfamily 1504.06303}].

\bibitem{Hoare:2015gda}
  B.~Hoare and A.~A.~Tseytlin,
  {\it On integrable deformations of superstring sigma models related to $AdS_n \times S^n$ supercosets},
  \href{https://www.sciencedirect.com/science/article/pii/S0550321315002035?via\%3Dihub}{Nucl.\ Phys.\ {\bf B897} (2015), 448-478},
    [\href{http://arxiv.org/abs/1504.07213}{\ttfamily 1504.07213}].


\bibitem{Klimcik:2015gba}
C. Klim\v c\'\i k,
  {\it $\eta$ and $\lambda$ deformations as ${\cal E}$-models}, \hfill\break
\href{https://www.sciencedirect.com/science/article/pii/S0550321315003302?via\%3Dihub}{Nucl.\ Phys.\ {\bf B900} (2015), 259-272},
  [\href{http://arxiv.org/abs/1508.05832}{\ttfamily 1508.05832}].


\bibitem{Klimcik:2016rov}
C.~Klim\v{c}\'\i{}k,
{\it Poisson\textendash{}Lie T-duals of the bi-Yang\textendash{}Baxter models},\\
\href{https://www.sciencedirect.com/science/article/pii/S0370269316303380?via\%3Dihub}{Phys. Lett. \textbf{B760} (2016), 345-349},
[\href{https://arxiv.org/abs/1606.03016}{\ttfamily 1606.03016}].



\end{thebibliography}
\end{document}